# Systematic electronic structure in the cuprate parent state from quantum many-body simulations

Zhi-Hao Cui,[1] Huanchen Zhai,[1] Xing Zhang,[1] Garnet Kin-Lic Chan,[1*]

[1]Division of Chemistry and Chemical Engineering, California Institute of Technology, Pasadena, California 91125 USA

*To whom correspondence should be addressed; E-mail: gkc1000@gmail.com.

**The quantitative description of correlated electron materials remains a modern computational challenge. We demonstrate a numerical strategy to simulate correlated materials at the fully ab initio level beyond the solution of effective low-energy models, and apply it to gain a detailed microscopic understanding across a family of cuprate superconducting materials in their parent undoped states. We uncover microscopic trends in the electron correlations and reveal the link between the material composition and magnetic energy scales via a many-body picture of excitation processes involving the buffer layers. Our work illustrates a path towards a quantitative and reliable understanding of more complex states of correlated materials at the ab initio many-body level.**

# Background

Currently, we have a qualitative theoretical understanding of many electronic phases of matter. However, there remains a deficit in the quantitative understanding of correlated electron materials (*1, 2*). This limits our ability to connect the atomic structure and composition to the electronic phenomena, as well to answer fundamental physical questions related to microscopic mechanisms. Here, we describe and apply a strategy to precisely simulate properties of a prototypical family of correlated electronic materials, the high-temperature superconducting cuprates, in their undoped, parent, electronic state. We directly approximate the solution of the *ab initio* many-electron Schrödinger equation instead of solving a low-energy effective model, within an approach that is numerically improvable without adjustable parameters. Using this strategy, we show that we can reveal the systematics of the cuprate parent state across a family of layered cuprate materials, connecting the observed low-energy physics to specific microscopic processes governed by the atomic and structural composition.

Among correlated quantum materials, the high temperature ($T_c$) superconductors remain a fertile source of new physics (*3–6*). We focus on the cuprates, where one finds the highest superconducting $T_c$ in the mercury-barium cuprate family (*7*). Although progress has been made in understanding the universal phase diagram through numerical calculations on lattice models, the understanding of properties of individual compounds remains largely empirical, with substantial difficulties in linking the observed trends to model parameters.

In principle, a quantitative understanding is simply a matter of many-electron quantum mechanics, but solving the Schrödinger equation beyond lattice models involves three challenges: the quantum many-body correlations, the thermodynamic limit (TDL), and the high-energy degrees of freedom / long-range interactions of real materials. We here adopt a pragmatic computational framework where the challenges can be tackled simultaneously: *ab initio* solvers for the many-body problem beyond models (*8, 9*); self-consistent quantum embedding to develop phases in the TDL (*10–12*); and periodic quantum chemistry using local bases (*13, 14*) to efficiently treat long-range interactions and high-energy degrees of freedom. Each component has been individually tested in prior work, but the important feature of our combined strategy is that the solution process bypasses models with uncontrolled parameters; the only remaining parameters are the size of the computational cell, the basis size, and the level of the many-body solver. Thus, all aspects of the calculation can in principle be controlled towards exactness.

In this work, we describe the full application of this strategy to the *ab initio* simulation of a family of cuprates in their parent phase at zero temperature. Although the parent phase is qualitatively simple, and elements of our framework have been used to understand exotic physics in simplified models (*15*), obtaining quantitative material systematics and functional relationships even in the parent phase is a major challenge, which serves as a litmus test of the promise of our overall *ab initio* strategy. As we shall describe, our detailed simulations bring a new level of resolution to the electronic structure, with which we uncover direct links between the material specific physics and composition.

## Cuprates and the parent state

**Structure**. The main structural feature of the cuprates is the two-dimensional $CuO_2$ (formally $[CuO_2]^{2-}$) square lattice plane [Fig. 1 (a)]. In different cuprates, the copper-oxygen plane is surrounded by other atoms and buffer layers in the vertical direction. We consider three specific compounds, in addition to layer-stacked idealized $CuO_2$ planes [geometries in Table S1 in (*16*)]. The first is infinite layer $CaCuO_2$ (CCO) [Fig. 1 (d)], where calcium counterions intercalate between the $CuO_2$ planes in an infinitely repeating structure. CCO does not itself superconduct, due to difficulties in doping the material. However, high $T_c$s are observed in the related mercury-barium cuprates (the Hg-Ba-Ca-Cu-O family). Here, the $CuO_2$ plane is decorated by apical oxygens, which connect to buffers of Hg and Ba ions. Unlike in CCO, the buffer layers form large spacers between the copper-oxygen layers. Different mercury-barium cuprates can be synthesized with different numbers of $CuO_2$ planes between each buffer layer, leading to single-layer, double-layer, etc. cuprates. We consider two members in this family: $HgBa_2CuO_4$ (Hg-1201, single-layer, $T_c$ = 97 K) and $HgBa_2CaCu_2O_6$ (Hg-1212, double-layer, $T_c = 127$ K). Hg-1201 exhibits distorted octahedral Cu-O coordination [2 apical oxygens per Cu, Fig. 1 (b)], while each layer of Hg-1212 contains pyramidal Cu-O coordination [1 apical oxygen per Cu, Fig. 1 (c)]. Hg-1201, Hg-1212, and CCO are compositionally related by replacing Hg-Ba-apical O layers by Ca layers.

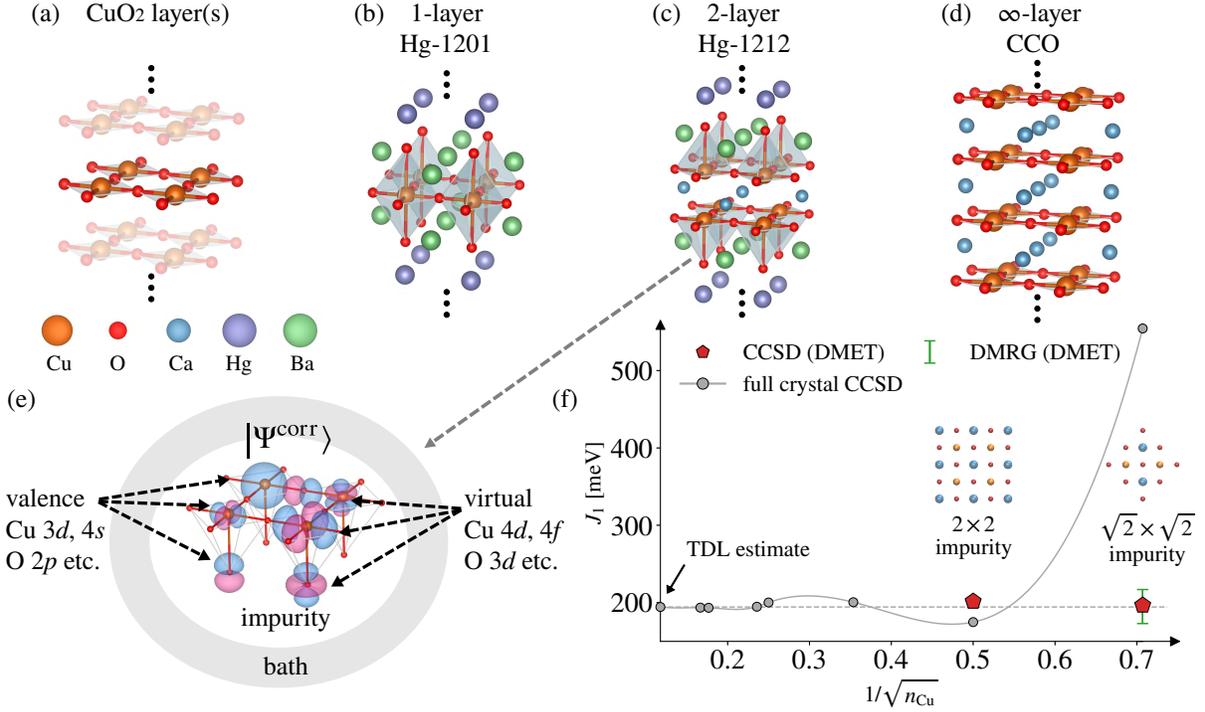

Figure 1: **Structures and computational strategy.** (a) $[CuO_2]^{2-}$ plane(s) in cuprates. (b) → (c) → (d): Relationship between single-layer Hg-1201 ($HgBa_2CuO_4$), double-layer Hg-1212 ($HgBa_2CaCu_2O_6$), infinite layer CCO ($CaCuO_2$); Ca layers replace the Hg-Ba-apical-O layers. (e) The *ab initio* density matrix embedding framework. The Hg-1212 lattice is divided into an impurity (e.g., the 2×2 cell) with the environment replaced by a bath; the atoms are represented by local valence and virtual orbitals, and the impurity problem is solved for the many-body wavefunction $\Psi^{corr}$. (f) Correlation and finite size effects in the nearest-neighbor Heisenberg exchange coupling $J_1$. We compare the exchange coupling ($y$ axis) from a full crystal CCSD calculation as a function of CCO crystal size [plane side-length in units of Cu atoms, $n_{Cu}$ ($x$ axis)] in a small basis, to embedded calculations with two impurity sizes and two solvers DMRG, CCSD. The embedded $2 \times 2$ impurity is already close to the TDL, while the DMRG and CCSD impurity solvers agree well in the smallest impurity.

**Parent state**. Unlike conventional superconductors, the parent state of the cuprates is an antifer-romagnetic (AFM) insulator with long-range order, due to the strong Cu $d$-$d$ electron interaction. Typical Néel temperatures for the AFM state range from about 250 K (in $Nd_2CuO_4$) to 450 K (in $YBa_2Cu_3O_6$) (*5*), and only after doping does the ground-state enter the superconducting phase. It is generally thought that the antiferromagnetism is to first order approximated by 2D nearest-neighbor (NN) Heisenberg-like physics. However, the 2D NN Heisenberg model does not reproduce the dispersion of the experimental spin-wave spectrum and questions remain as to the magnitude, sign, and material specific origin of corrections to the nearest-neighbor picture.

There have been many attempts to correlate properties of the cuprates in the superconducting phase (such as $T_c$) with structure, composition, and band structure (*17–22*). However, without a direct ability to simulate the material $T_c$ with different parameters it is difficult to distinguish correlation from causation. Although there has been less focus on correlating properties of the parent state with physical features, many proposals relate the high Néel temperatures and strong exchange coupling in the parent state to the superconducting mechanism and other exotic physics under doping. Below, we establish causal, quantitative relationships between the magnetic features of the parent state and the atomic-scale structural and electronic features of the materials.

## Theoretical techniques

**Strategy**. Previous numerical work on cuprate electronic structure [with a few exceptions e.g., (*23, 24*)] falls in two classes: (i) *ab initio* all electron simulations with a modest treatment of electron correlation (*25–27*), often used to derive low-energy effective models, and (ii) accurate many-body methods applied to low-energy effective models, to obtain phase diagrams and more exotic orders (*28–34*).

Our strategy is to use families of methods associated with the model studies of (ii), but technologically elevated to the fully *ab initio* Hamiltonians of (i). This bypasses the ambiguities of intermediate downfolded models, while allowing correlated physics to emerge. The three numerical components are the quantum embedding, the *ab initio* all-electron infrastructure, and the many-body solvers. Our technical setup uses density matrix embedding theory (DMET) to self-consistently embed a $2 \times 2$ supercell (impurity) of the cuprate material within an all-electron description, and we solve the resulting embedded impurity with an *ab initio* many-body approximation [coupled cluster (CC) theory]. To do so feasibly and reliably relies on recent advances and new techniques specific to this work, such as a sub-impurity formalism and improved DMET self-consistency algorithms for large impurities; improved *ab initio* matrix element generation; and careful solver benchmarking against a massively parallel *ab initio* density matrix renormalization group (DMRG) implementation. Below we describe the quantum embedding and many body solvers; the *ab initio* infrastructure is discussed in Sec. 1.1 of (*16*).

*Quantum embedding*. This provides a framework for phases that emerge due to interactions (*35*), and includes dynamical mean-field theory and its relations (*36–38*), and the DMET (*39, 40*) used in this work. The material is separated into an impurity region and a bath that describes fluctuations out of the impurity, and their self-consistency yields emergent phases. The embedding becomes exact with increasing impurity size.

In previous work on the 1-band and 3-band Hubbard models, DMET has been extensively benchmarked against other methods, and for example, accurately resolves exotic order in the underdoped region (*15, 41*). (The ability of DMET to capture exotic physics in doped lattices shines a light on the path from the *ab initio* studies of the parent state here to the physics of the doped materials). To move beyond models to the *ab initio* physics, we start from our recently introduced all-electron, full cell approach (*10–12*). Here, the impurity is a supercell of the cuprate containing *all* atoms and orbitals, with all quartic interactions between the orbitals. In contrast to downfolded approaches with a handful of impurity orbitals and possibly simplified interactions (*37, 42*), our largest impurity (in Hg-1212) contains 48 atoms and close to 900 orbitals [Fig. 1(e)]. These orbitals include many "virtual" bands, which capture quantitative electron correlation effects and screening. Part of the reason why these large impurities are feasible is the DMET formulation itself, which bypasses expensive frequency dependent quantities. The other critical factors are the choice of solvers discussed below, and the periodic quantum chemistry infrastructure based on local atomic basis sets, which compactly discretize the virtual bands for electron correlation.

*Ab initio many-body solvers*. The quantum impurity problem in the full-cell approach is a many-body problem with hundreds of orbitals. This can be solved because many orbitals do not display strongly correlated physics. We use two impurity solvers in this work. The majority of the results are obtained using *ab initio* coupled cluster singles and doubles (CCSD) (*43*) solvers. Although approximate, they exactly treat clusters of (arbitrarily) strongly correlated particles, and have previously been shown to yield accurate results in various quantum impurity problems (*9–12,44*), particularly in ordered phases. To verify the accuracy of the CC approximation, we use a second solver, the quantum chemistry DMRG (*8, 45*) to benchmark a subset of problems.

**Computational setup**. The $2 \times 2$ supercell impurities are shown in Fig. 1 for the different mercury-barium cuprates. [In Sec. 2.1.5 of (*16*) we also discuss a benchmark study of lanthanum copper oxide]. Every atom is represented in a valence double-zeta with polarization basis [def2-SVP (*46*)] e.g., each Cu is represented by $[5s3p2d1f]$ shells and each O by $[3s2p1d]$ shells, and the embedding lattice is chosen to be an $8 \times 8 \times 2$ lattice of the primitive cell. Large impurities (e.g., in Hg-1212) were further fragmented into smaller sub-impurities with up to 364 orbitals (280 impurity orbitals and 84 valence bath orbitals), and impurity solutions were obtained using CCSD or DMRG. [Unless otherwise indicated, data is from the CCSD solver; DMRG data is in Sec. 2.1.4 of (*16*)]. The DMET equations were then solved with self-consistency and valence-shell lattice-impurity density matrix fitting.

**Benchmarks**. Within the above strategy, the only sources of error are from the finite size of the impurity (and embedding lattice), the approximate nature of the impurity solver, and the finite size of the local atomic basis. We have carried out extensive benchmarking to verify the specific approximations. In Fig. 1(f) we compare results from finite impurities to the TDL (which can be estimated from a full crystal calculation within a small local atomic basis) for the energy difference between the ferromagnetic (FM) and the AFM state ($\propto$ the NN exchange coupling $J_1$). We also show the deviation between this energy difference estimate from the *ab initio* DMRG and CCSD solvers in a small impurity where DMRG is tractable. Both sets of data illustrate that the TDL and many-body character of the physics is well-captured within the approximations in this work. Additional benchmarks (e.g. basis set convergence) can be found in Secs. 2.1 and 3.3 of (*16*) (see e.g. Figs. S4, S5 and Tables S15-S18).

# Results

## Multi-orbital electronic structure

We start with general electronic trends across the series Hg-1201, Hg-1212, CCO, and $[CuO_2]^{2-}$ as a baseline to understand trends in the physics in later sections.

**Order parameters and bonding**. We first extract order parameters from the $2\times 2$ computational supercell: charge, local moment, bond orders (from the off-diagonal elements of $\gamma_{ij} = \left\langle a_j^\dagger a_i \right\rangle$ where $i$, $j$ label local atomic orbitals in the cell); and the spin correlation function $\langle S_z(0) S_z(r) \rangle$ measured across the full crystal [Figs. 2 (a) - (e)].

The key features are: (i) The ground-state is AFM with long-range order, with the moment in the Cu half-filled $3d_{x^2-y^2}$ orbital. Cu $4s/4p$ occupancy reduces the total moment by about 10%. The unit cell moment ranges from 0.71 in Hg-1201 to 0.55 in $[CuO_2]^{2-}$. (ii) Charge is transferred from in-plane O orbitals to the other ions, with the degree of transfer increasing across the series. There is significant charge transfer to the Cu minority spin orbitals (as much as 0.3 electrons in $[CuO_2]^{2-}$). (iii) Ca and Ba buffer atoms in CCO, Hg-1201, and Hg-1212 are ionic, with Hg covalently bonded to the apical oxygen via the O $2p_z$-Hg $6s$, $5d_{z^2}$ bonds. Hg-1201 and Hg-1212 do not differ much with respect to the out-of-plane observables, but do differ for their observables in the CuO$_2$ plane. (iv) In-plane $\sigma$-bonding [Fig. 2(f)] is predominantly Cu $4p$-O $2p$ and does not differ much across the compounds. However, Cu $3d$-O $2p$ bonding and out-of-plane $\pi$ bonding increase across the series, reflecting increasing in-plane $3d/4p$ hybridization. The change in bonding is not (solely) due to the structural changes (e.g., CCO and [CuO$_2$]$^{2-}$ have the same Cu-O bond-length but different bond orders) but instead reflects redistribution of charge from the buffer layers. (v) The apical oxygen bond order [Fig. 2(g)] decreases from Hg-1201 to Hg-1212, with the oxygen only weakly bound to Cu. Cu $4s$ and $4p_z$ contribute to apical bonding, with little $3d_{z^2}$ participation.

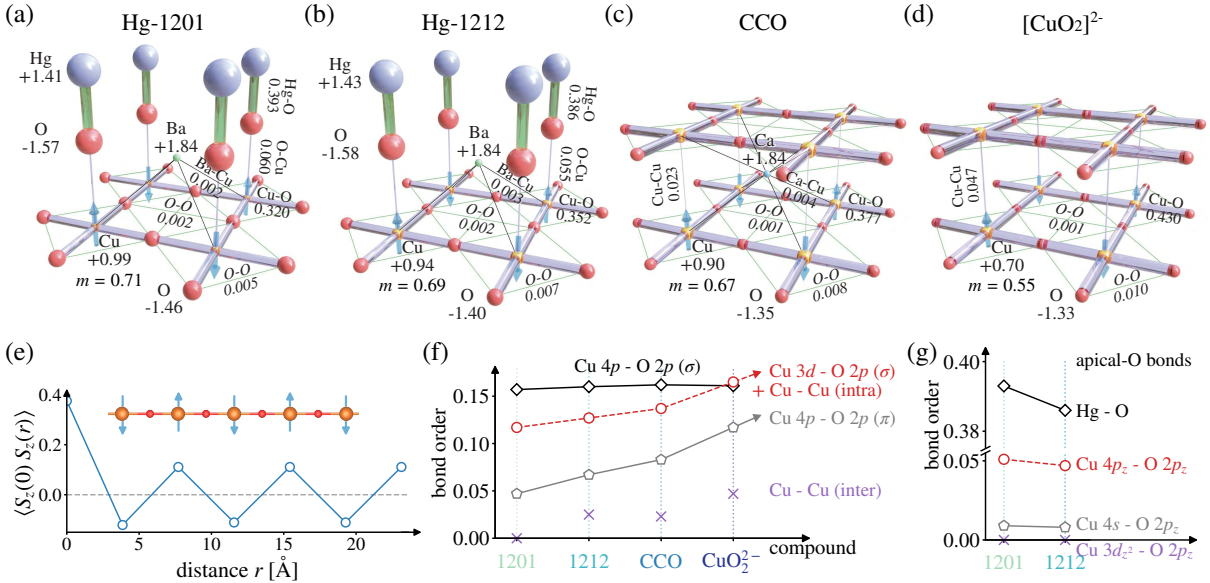

Figure 2: **Charge, spin and bond orders.** (a)-(d): Charge $c$, magnetic moment $m$ and bond order $b$ of different cuprates. Cu: yellow; O: red; Hg: violet; Ba: green; Ca: blue. Atomic sphere radius - number of electrons $n$ (local charge $Z-n$ is labelled, $Z$: nuclear charge); arrow length - magnitude of local moment $m = n^\uparrow - n^\downarrow$; bond width - bond order $b$. (e) Spin-spin correlation function $\langle S_z(0) S_z(r)\rangle$ in CCO. (f) Cu orbital-resolved bond orders. (g) Apical O orbital-resolved bond orders. For more details, see Sec. 2.2 of (*16*).

**Natural occupancy distributions and effects of correlation.** We obtain additional insight from the spin-resolved ($\gamma^\sigma$) and spin-traced ($\gamma = \sum_\sigma \gamma^\sigma$) single-particle density matrices (equal-time Green's functions) evaluated in the full crystal. These provide non-local and **k**-space information on correlations. We first discuss the spin-traced single-particle density matrix. The eigenvalues (i.e., the natural occupancy distribution, sometimes called the momentum distribution function) and eigenvectors (natural orbitals) illustrate the degree of symmetry breaking and highlight the important degrees of freedom near the Fermi level. The spin-traced natural occupancy distribution together with the projected atomic character of the eigenvectors is shown in Figs. 3(a), (b). We see that the most important orbitals near the Fermi level are the classic 3-band orbitals - Cu $3d_{x^2-y^2}$ and O $2p_x$, $2p_y$. We also find no single next most important orbital: Cu $4s$, $4p$, $3d_{z^2}$, as well as the apical oxygen and Hg orbitals all contribute to a similar degree.

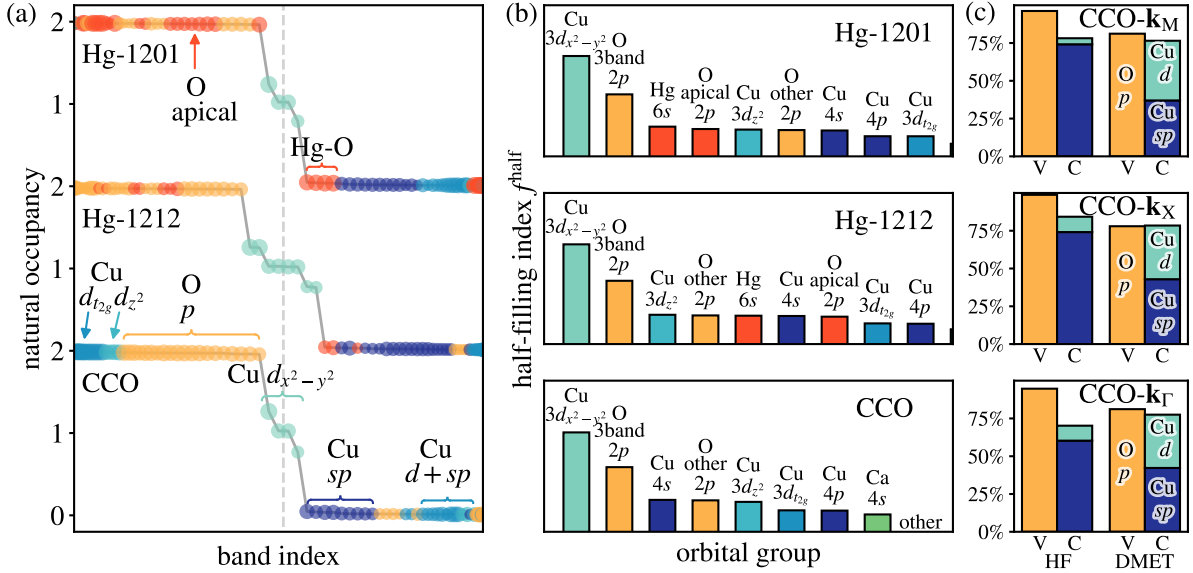

Figure 3: **Natural occupancy distribution (eigenvalues of the single-particle density matrix) and quasiparticle character.** (a) Occupancy of natural orbitals around the Fermi level (dashed line), from the spin-traced density matrix ($\gamma^\alpha + \gamma^\beta$) in Hg-1201, Hg-1212 and CCO. Orbital character denoted by colors and labels. (b) Half-filling index of the different local orbitals [see Eqn. (S50) in (*16*) for definition], measuring their importance in the most correlated orbitals of the calculation. (c) Orbital component analysis of the spin-resolved mean-field (HF) and correlated (DMET) top valence (V) and bottom conduction (C) bands of CCO at different **k** points (averaged from the 8 bands near the Fermi level), $\Gamma$: $(0, 0)$; X: $(\frac{1}{2}, 0)$; M: $(\frac{1}{2}, \frac{1}{2})$.

The spin-resolved natural occupancies and eigenvectors indirectly reflect the nature of the quasiparticles and the importance of dynamical effects [discussion in Sec. 1.3.4 of (*16*)]. The eigenvectors with natural occupancies closest to the jump across the Fermi level can be viewed as "pseudo"-valence band maximum (VBM)/conduction band minimum (CBM) states. Defined in this way, from Fig. 3(c), we see that the pseudo-VBM is dominated by O $2p_{x(y)}$, while the pseudo-CBM is dominated by Cu $3d_{x^2-y^2}$ (and the apical O and Hg bands in the Hg-Ba compounds). This classifies all the compounds as charge-transfer insulators.

We can further untangle the effect of interactions from pure single-particle physics by comparing the spin-resolved natural occupancies of the correlated calculation with that of a spin-polarized Hartree-Fock (HF) reference. The correlated spin-resolved natural occupancies are all quite close to 0 and 1 (Fig. S15), i.e., the mean-field values, thus dynamical effects are small. However, the orbital components of the eigenvectors are very different between the mean-field and correlated distributions [Fig. 3(c)], indicating strong static effects. It appears in the AFM state, the effect of interactions on the quasiparticles is mainly static rather than dynamical, and can be largely captured via static screening of the interactions, correlation driven rehybridization of the orbitals, and renormalization of their energies.

## Magnetic trends across the cuprates

We next characterize the low-lying magnetic excitations across the series of cuprates. To do so compactly, we introduce a magnetic model (not to solve for the electronic structure, but for interpretation) and extract exchange couplings from our correlated calculations of different spin-configurations: the AFM state, the FM state, and a spin-density wave state [Fig. S3 in (*16*)]. From these we derive parameters for the NN Heisenberg model ($J$) and a multi-$J$ Heisenberg model where the exchange couplings $J_1$, $J_2$, $J_3$ and $J_c$ are related via the perturbation expansion of the 1-band Hubbard model (with only 3 free parameters). [A 3$J^{\text{eff}}$ model where $J_c$ is renormalized into the $J_1$, $J_2$, $J_3$ parameters can also be derived. In CCO, we also derive an interlayer $J_\perp$ using two AFM layer configurations. For a full discussion of all models and the spin-wave calculation see Sec. 1.5.2 of (*16*)]. The parameters are illustrated in Fig. 4(b) and tabulated in Tables S10-S13. We display the corresponding spin-wave spectrum from linear spin wave theory in Fig. 4 (a).

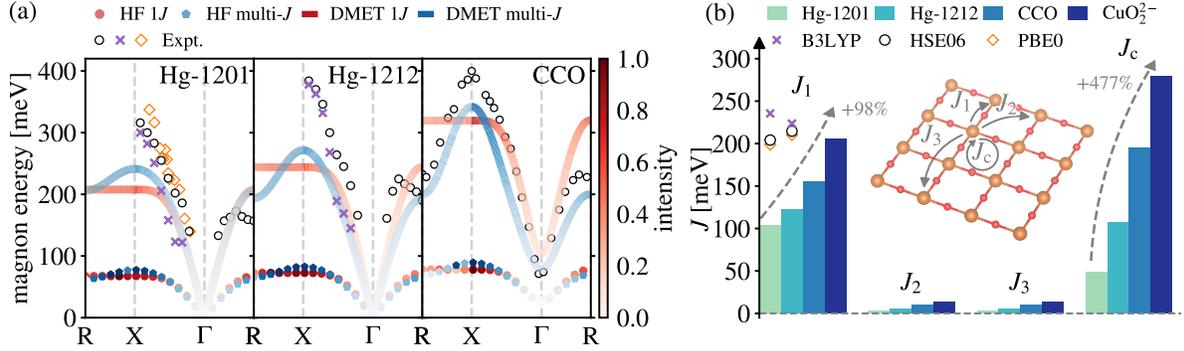

Figure 4: **Spin wave dispersion of Hg-1201, Hg-1212 and CCO.** (a) The 2D magnetic Brillouin zone is sampled along Γ: (0, 0), X: ($\frac{1}{2}$, 0), R: ($\frac{1}{4}$, $\frac{1}{4}$), $k_z$ is fixed at 0.46 to match the experimental conditions in CCO, and fixed at 0 for Hg-1201 and Hg-1212. NN Heisenberg (1$J$) and multi-$J$ model curves are shown. The multi-$J$ model includes a quantum renormalization factor of $Z_c = 1.219$ (*47*). Experimental RIXS data is extracted from (*48*), (*49*) for Hg-1201 and Hg-1212; (*50*) for CCO. (b) Trends in the multi-$J$ model parameters across the cuprate family. Hybrid density functional (PBE0, HSE06, B3LYP) results for the first two Hg compounds are also shown with symbols. For details, see Sec. 2.3 of (*16*).

**Spin-wave spectrum**. In CCO the full experimental dispersion is available, while for Hg-1201 and Hg-1212 only part of the dispersion near the Γ point has been measured. As is well-known the NN Heisenberg model does not capture dispersion away from the Γ point, but the derived NN $J$ agrees well with that derived from experiment by fitting near the Γ point; for example, in CCO, the NN $J$ fit to DMET data yields $J = 155$ meV, compared to $J = 142, 158$ meV (the two numbers are from different experiments) (*50, 51*). The multi-$J$ model with *ab initio* parameters yields improved agreement across the experimental dispersion, illustrating the importance of long-range exchange. The discrepancies are largest near the X point ($\frac{1}{4}, \frac{1}{4}$), likely due to finite size effects in the embedding, although there are also confounding factors from the experimental setting in Hg-1201, Hg-1212 [Sec. 2.3.5 of (*16*)]. Compared to CCO, the Hg-Ba compounds display flatter dispersions, and we capture this in our derived spin-wave spectrum.

**Magnetic parameters**. Trends in the magnetic couplings of the multi-$J$ Heisenberg model among the four compounds are shown in Fig. 4 (b). Across the series Hg-1201, Hg-1212, CCO, $CuO_2^{2-}$, all couplings $J_1$, $J_2$, $J_3$ and $J_c$ increase significantly. $J_1$ roughly doubles and $J_c$ increases by a factor of 5, illustrating (i) the importance of the buffer layers in the long-range exchange coupling and (ii) the increasing "delocalization" across the series of compounds. A recent resonant inelastic X-ray scattering (RIXS) experiment (*48*) suggests that $J_1$ increases significantly (by about 20 % - 30 %) from the single-layer Hg-1201 to the double-layer Hg-1212, similar to the increase in $T_c$. We find quantitative agreement with our correlated calculations, where Hg-1212 shows an increase in $J_1$ by about 18 %.

**Effect of interactions**. To understand the effect of interactions, we can compare to the mean-field HF results. These give almost flat dispersion curves, since the $J$ couplings are very small (e.g., $J_1 \sim 40$ meV), while the magnon energy at the $\Gamma$ point is also lower than the experimental value. Thus, the observed magnetic energy scales require a careful treatment of electron correlation. As suggested in the last section, a large part of the effect of interactions can be captured by a renormalization of the low-energy band structure and interaction. Choosing a density functional treatment or Hubbard $U$ parameter can mimic this, however, we do not find a single choice of functional or Hubbard $U$ consistently or accurately reproduces the material trends. For example, moving from Hg-1201 to Hg-1212 should yield a significant increase in the exchange couplings, but from Fig. 4 (b) (symbol data), one finds $J_1$ decreases with the B3LYP functional, and increases only marginally with HSE06 (5%) and PBE0 (6%). In addition, $J_1$ is significantly overestimated by all the above functionals.

## Untangling layer effects

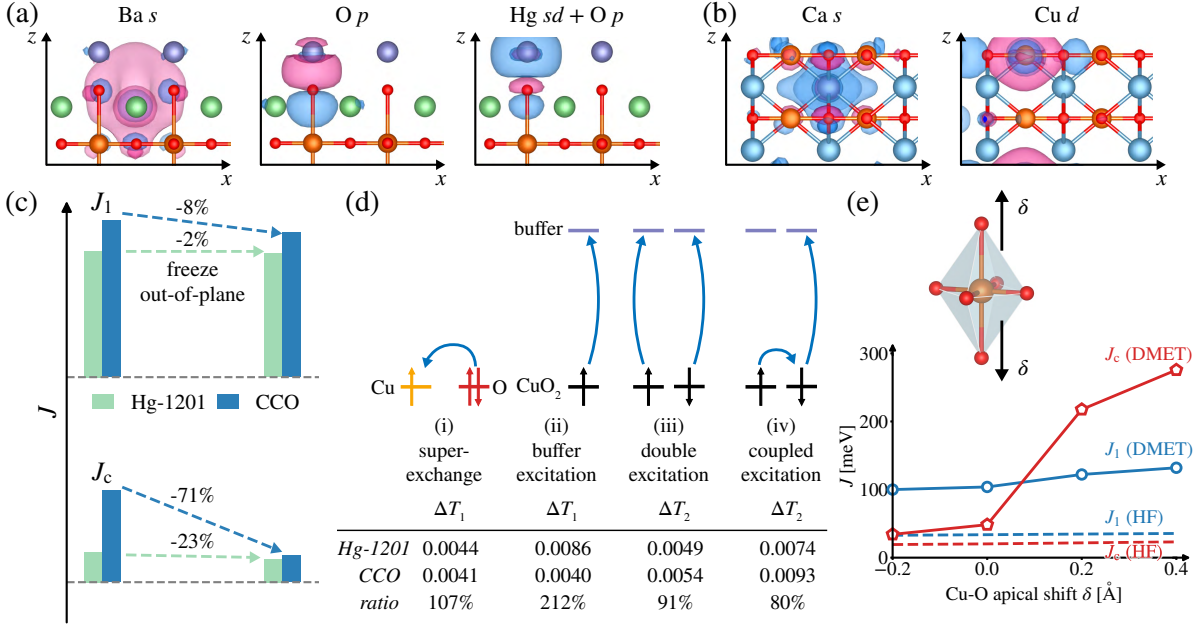

Figure 5: **Effects of buffer layers.** Representative out-of-plane orbitals (isosurfaces) in (a) Hg-1201 and in (b) CCO. (c) The effect of freezing fluctuations to out-of-plane orbitals on the NN magnetic coupling $J_1$ and cyclic exchange coupling $J_c$. (d) Excitations relevant to exchange pathways in cuprates: super-exchange is facilitated by excitations from in-plane oxygen orbitals to empty copper states (i); in Hg-1201, substantial excitations from the copper-oxygen plane to the buffer layer (ii) reduce super-exchange. The numbers ($\Delta T_1$, $\Delta T_2$) reflect the change in excitation weight upon unfreezing the buffer orbitals. (e) Influence of apical Cu-O distance on exchange coupling $J_1$ and $J_c$ at the mean-field (HF) and correlated (DMET) level.

We now connect the microscopic correlated electronic structure with the trends in the magnetic physics observed above to derive mechanistic insights. As seen above, changing the buffer layer leads to large changes in the exchange couplings (particularly for the non-local terms). However, this effect does not appear at the HF mean-field level. To verify that it originates due to fluctuations (electron correlation) with the buffer layers (and not simply via the effect of the electrostatic potential of the buffer layer on electron correlation within the cuprate plane), we first devise a procedure that allows us to switch electron correlation with the buffer layer orbitals on and off. To do so, we explicitly freeze excitations involving out-of-plane orbitals in the correlated impurity solver calculations [i.e. the impurity wavefunction excludes configurations with such excitations relative to the HF determinant, Sec. 2.4.1 of (*16*)]. Any changes from freezing and unfreezing these fluctuations therefore directly reflect the influence of electron correlation with the orbitals of the buffer plane.

Representative out-of-plane orbitals of Hg-1201 and CCO are shown in Fig. 5(a) and (b). The out-of-plane impurity orbitals consist of empty outer valence shells on Ca, Hg, and Ba, apical oxygen orbitals, and other orbitals that originate from the adjacent copper-oxygen plane. The Ca and Ba centered localized orbitals (4*s* and 6*s*) are similar in CCO and Hg-1201.

The changes in the $J_1$ and $J_c$ from unfreezing the out-of-layer orbitals are shown in Fig. 5(c). In both compounds, the exchange couplings are decreased by freezing, but in CCO, the effect is stronger and $J_c$ is especially strongly influenced by freezing, decreasing by as much as 71% in CCO. To understand this, we analyze the correlated impurity wavefunctions in CCO and Hg-1201. Shown in Fig. 5(d) are the changes in the weights of single-particle excitations $\Delta T_1$ and connected two-particle excitations $\Delta T_2$ upon unfreezing the buffer layer in the two compounds. Generally speaking, when the buffer layer is unfrozen, the increased excitation manifold increases screening and decreases the energetic penalty to excite from filled to empty states, such as the empty Cu and buffer layer states. In CCO and Hg-1201, we find that this increases the O → Cu excitation associated with superexchange [process (i) in Fig. 5(d)], increasing the exchange couplings. However, in Hg-1201, we see in addition a significant increase in excitations from in-plane Cu, O orbitals to the empty Hg, apical O states [process (ii)]. This change in the copper-plane to buffer excitation is more than twice as large in Hg-1201 than in CCO, and it depletes the ground configuration associated with in-plane exchange and reduces the effective non-local hopping by rehybridizing the Cu empty states (*17*), cancelling the enhancement of in-plane O → Cu excitations, and yielding an aggregate small change in exchange coupling upon unfreezing the buffer orbitals [Sec. 2.4.2 of (*16*)]. Note that this also explains why the exchange couplings of Hg-1212 lie in between those of Hg-1201 and CCO, as the buffer suppression of in-plane super-exchange occurs via a single buffer layer in Hg-1212 versus two on either side in Hg-1201. The analysis also reveals (smaller) differences between the compounds in the connected two-particle fluctuations involving the buffer [processes (iii), (iv)]; these are material specific effects that cannot be folded into a static renormalization. Finally, in Fig. 5(e) we show the effect on the exchange coupling of increasing the apical oxygen distance in Hg-1201, both at the mean-field level and at the correlated level. Consistent with the above mechanism, we find that increasing apical oxygen distance removes the buffer suppression effect in the correlated calculation (increasing the exchange coupling), but makes little difference in the mean-field calculation, as fluctuations must first renormalize the energies of the empty states for them to be accessible.

## Discussion

We have demonstrated that through a numerical strategy combining quantum embedding, *ab initio* quantum solvers, and periodic quantum chemistry, we can determine at the many-body level, material specific correlated electron structure in the parent state of the cuprates. This reveals trends in the multi-orbital bonding, correlation effects in the Fermi distribution and quasiparticles, and gives a quantitative description of the low-energy magnetic excitations. Across a series of homologous mercury-barium and calcium cuprates, the systematic trends in the nature of the magnetic exchange can be explained through the analysis of the many-body state, which uncovers a competition between super-exchange and plane-to-buffer excitation processes.

A general observation is that while the interactions are strong, many of their effects in the parent state can be renormalized into a static low-energy theory. This supports the long-standing practice of interpreting physics in this region through simple band-structures and static interaction parameters. However, we also find that empirical approaches to determine this renormalization do not have the accuracy to capture the trends amongst the materials, unlike the controlled many-body approaches used here.

A strength of the many-body approach is that we can interrogate individual electronic processes, and our *ab initio* formulation allows us to trace these processes beyond models to the individual atomic orbital level. We use this capability to untangle the links between layer composition and magnetic exchange. In prescient work, it was conjectured that the range of magnetic exchange is related to electronic processes involving an effective apical conduction band, and that this further correlates with the superconducting transition temperature (*17*). We now have a direct picture of the first part of this conjecture, with rich atomic-scale and many-body resolution.

Components of the numerical strategy in this work have previously been used to describe exotic phases in models. The success of the current *ab initio* realization for cuprate parent states thus extrapolates to the exciting prospect that a similar approach may eventually yield a quantitative picture of more complex cuprate phases. If that is the case, we may be able to answer the second part of the above and similar conjectures about superconducting properties, through a direct *ab initio* simulation of the superconducting orders and the energy scales of the cuprates in their doped states.

We thank Tianyu Zhu, Linqing Peng, Yuan Li, Patrick Lee, Andrew Millis, and Steve White for helpful discussions. Funding: This work was primarily supported by the US Department of Energy, Office of Science, via grant no. DE-SC0018140. The DMRG calculations were performed using the BLOCK2 code which was developed with funding from the US National Science Foundation, via CHE-2102505. This work also relied on improvements to the PYSCF density fitting and integral transformation modules, carried out as part of work supported by the Center for Molecular Magnetic Quantum Materials, an Energy Frontier Research Center funded by the U.S. Department of Energy, Office of Science, Basic Energy Sciences under award no. DE-SC0019330. G.K.-L.C. is a Simons Investigator in Physics and is part of the Simons Collaboration on the Many-Electron Problem. Z.-H.C. acknowledges support from the Eddleman Quantum Institute through a graduate fellowship. Calculations were conducted in the Resnick High Performance Computing Center, supported by the Resnick Sustainability Institute at Caltech, and National Energy Research Scientific Computing Center (NERSC), a U.S. Department of Energy Office of Science User Facility located at Lawrence Berkeley National Laboratory.

Data used in this work are in the supplementary materials and online at [github.com/zhcui/cuprate_parent_state_data](github.com/zhcui/cuprate_parent_state_data). The LIBDMET code is available at [github.com/gkclab/libdmet_preview](github.com/gkclab/libdmet_preview). The BLOCK2 code is available at [github.com/block-hczhai/block2-preview](github.com/block-hczhai/block2-preview). PYSCF is available from [pyscf.org](pyscf.org).

**Supplementary materials**

Supplementary text

Figures S1-S24, Tables S1-S18, References 52-111.

# Supplementary Materials for "Systematic electronic structure in the cuprate parent state from quantum many-body simulations"

Zhi-Hao Cui,[1] Huanchen Zhai,[1] Xing Zhang,[1] Garnet Kin-Lic Chan,[1*]

[1]Division of Chemistry and Chemical Engineering, California Institute of Technology,
Pasadena, California 91125 USA

*To whom correspondence should be addressed; E-mail: gkc1000@gmail.com.

# Contents

# 1 Methods

## 1.1 Periodic quantum chemistry formalism

### 1.1.1 Periodic Gaussian bases

We will use a quantum chemistry formalism based on crystalline Gaussian bases, i.e., translational-symmetry-adapted linear combinations of Gaussian atomic orbitals (AO) (*52*),

$$\chi_p^{\mathbf{k}}(\mathbf{r}) = \sum_{\mathbf{T}} e^{i\mathbf{k}\cdot\mathbf{T}} \chi_p(\mathbf{r}-\mathbf{T}), \tag{S1}$$

where $\mathbf{T}$ denotes a lattice vector and $\mathbf{k}$ is a crystal momentum vector in the first Brillouin zone (FBZ). To formulate the *ab initio* calculation, it is necessary to express the Hamiltonian matrix elements (integrals) in this basis. The one-electron integrals, namely the overlap $S$, kinetic $T$ and electron-nuclear interaction integrals $V^{\text{N-el}}$ are (*13, 53*),

$$S_{pq}^{\mathbf{k}} = \frac{1}{N}\langle\chi_p|\chi_q\rangle = \sum_{\mathbf{T}} e^{i\mathbf{k}\cdot\mathbf{T}} \int d\mathbf{r}\, \chi_p^*(\mathbf{r})\chi_q(\mathbf{r}-\mathbf{T}), \tag{S2}$$

$$T_{pq}^{\mathbf{k}} = \frac{1}{N}\langle\chi_p^{\mathbf{k}}|\nabla_{\mathbf{r}}^2|\chi_q^{\mathbf{k}}\rangle = \sum_{\mathbf{T}} e^{i\mathbf{k}\cdot\mathbf{T}} \int d\mathbf{r}\, \chi_p^*(\mathbf{r})\nabla_{\mathbf{r}}^2\chi_q(\mathbf{r}-\mathbf{T}), \tag{S3}$$

$$V_{pq}^{\text{N-el},\mathbf{k}} = \frac{1}{N}\langle\chi_p^{\mathbf{k}}|v^{\text{N-el}}|\chi_q^{\mathbf{k}}\rangle = \sum_{\mathbf{T}} e^{i\mathbf{k}\cdot\mathbf{T}} \int d\mathbf{r}\, \chi_p^*(\mathbf{r})v^{\text{N-el}}(\mathbf{r})\chi_q(\mathbf{r}-\mathbf{T}), \tag{S4}$$

where the divergent part ($\mathbf{G} = \mathbf{0}$) in the electron-nuclear interaction $v^{\text{N-el}}(\mathbf{r})$ is removed. The total one-electron Hamiltonian integral (core Hamiltonian matrix element) is then,

$$h_{pq}^{\text{core},\mathbf{k}} = T_{pq}^{\mathbf{k}} + V_{pq}^{\text{N-el},\mathbf{k}}. \tag{S5}$$

We also define the matrix elements of the 2-electron Coulomb interaction. This leads to electron repulsion integrals (ERI) involving 4 crystalline Gaussian AOs (4 "centers"),

$$V_{pqrs}^{\mathbf{k}_p\mathbf{k}_q\mathbf{k}_r\mathbf{k}_s} = \int d\mathbf{r}_1\, d\mathbf{r}_2\, \chi_p^{\mathbf{k}_p*}(\mathbf{r}_1)\chi_q^{\mathbf{k}_q}(\mathbf{r}_1)\frac{1}{r_{12}}\chi_r^{\mathbf{k}_r*}(\mathbf{r}_2)\chi_s^{\mathbf{k}_s}(\mathbf{r}_2). \tag{S6}$$

Note that crystal momentum conservation means that the ERI vanishes unless $\mathbf{k}_p + \mathbf{k}_r - \mathbf{k}_q - \mathbf{k}_s = n\mathbf{b}$, where $n\mathbf{b}$ is an integer multiple of the reciprocal lattice vectors.

With all the matrix elements evaluated, standard molecular quantum chemistry techniques can be applied in the periodic setting. This lays the foundation for the efficient *ab initio* implementation of quantum embedding theories below.

### 1.1.2 Density fitting

There are a large number of ERIs in the above formulation. To reduce the cost of evaluating them, we employ density fitting (DF) which factorizes the 4-center electron repulsion integral (ERI) into a product of 3-center ERIs and a metric matrix. Using the Coulomb metric (*13, 14*) and for auxiliary basis functions labelled $\{P, Q, \cdots\}$, we obtain

$$V_{pqrs}^{\mathbf{k}_p\mathbf{k}_q\mathbf{k}_r\mathbf{k}_s} = \sum_{PQ}(p\mathbf{k}_p q\mathbf{k}_q|P)\mathcal{J}_{PQ}^{-1}(Q|r\mathbf{k}_r s\mathbf{k}_s), \tag{S7}$$

with the 3-center ERI,

$$(P|p\mathbf{k}_p q\mathbf{k}_q) = \frac{1}{N}\sum_{\mathbf{T}_p\mathbf{T}_q}\int d\mathbf{r}_1 d\mathbf{r}_2\, e^{i\mathbf{k}_q\cdot\mathbf{T}_q - i\mathbf{k}_p\cdot\mathbf{T}_p}\chi_P(\mathbf{r}_1)\frac{1}{r_{12}}\chi_p^*(\mathbf{r}_2-\mathbf{T}_p)\chi_q(\mathbf{r}_2-\mathbf{T}_q), \tag{S8}$$

and the Coulomb metric,

$$\mathcal{J}_{PQ} = \int d\mathbf{r}_1 d\mathbf{r}_2 \, \chi_P^*(\mathbf{r}_1) \frac{1}{r_{12}} \chi_Q(\mathbf{r}_2). \tag{S9}$$

It is computationally convenient to absorb the Coulomb metric symmetrically into the definition of the 3-center integrals,

$$\begin{aligned} V_{pqrs}^{\mathbf{k}_p\mathbf{k}_q\mathbf{k}_r\mathbf{k}_s} &= \sum_L \sum_{PQ} \left[ (p\mathbf{k}_p q\mathbf{k}_q|P) \mathcal{J}_{PL}^{-\frac{1}{2}} \right] \left[ \mathcal{J}_{LQ}^{-\frac{1}{2}} (Q|r\mathbf{k}_r s\mathbf{k}_s) \right] \\ &= \sum_L W_{Lpq}^{\mathbf{k}_p\mathbf{k}_q*} W_{Lrs}^{\mathbf{k}_r\mathbf{k}_s}. \end{aligned} \tag{S10}$$

where the symmetrical decomposition can be carried out using the eigenvalue decomposition of $\mathcal{J}$ and linear dependence is handled by discarding small eigenvalues. In the following, we use the symmetric DF form and use $L$ to label auxiliary basis functions. We choose the auxiliary basis to be also a crystalline Gaussian basis, thus the above formulae correspond to (crystalline) Gaussian density fitting (GDF).

The 3-center integral $W$ obeys several useful relations, which we use later to derive some of formulae. Similar to the 4-center integral, there is momentum conservation,

$$\mathbf{k}_L = \mathbf{k}_p - \mathbf{k}_q + n\mathbf{b}. \tag{S11}$$

From Eq. (S8) and Eq. (S9), one can also verify the following complex conjugation relation,

$$W_{Lqp}^{\mathbf{k}_q\mathbf{k}_p*} = W_{Lpq}^{\mathbf{k}_p\mathbf{k}_q}. \tag{S12}$$

## 1.2 Ab initio quantum embedding

### 1.2.1 Density matrix embedding theory

Density matrix embedding theory (DMET) (*39, 40*) is a quantum embedding theory designed to handle strongly correlated systems. There are passing similarities to dynamical mean-field theory (DMFT) as both relate the solution of a bulk problem to that of a self-consistent quantum impurity problem (*39*). However, DMET differs substantially in its physical interpretation and mathematical formulation, as frequency dependent quantities do not appear. The latter makes the approach computationally more tractable, allowing the treatment of the complicated impurity supercells in this work.

In the cuprate setting, DMET has been benchmarked and applied to various phases of the 1-band and 3-band Hubbard models. It was extensively benchmarked in the 1-band Hubbard model as part of the Simons Collaboration (*54*). In particular, accurate large-scale DMET calculations were used to resolve existing uncertainty in the half-filling magnetic moment in the 1-band model (*41*). DMET has further been used to character the $d$-wave pairing order and to discover magnetic and inhomogeneous orders as a function of 1-band and 3-band doping (*55, 56*). For example, DMET played an important role in resolving the stripe order and wavelength in the underdoped region of the 1-band Hubbard model (*15*). DMET has been generalized to finite temperatures (*57*), spectral quantities (*58*), and time-dependent settings (*59*), but the current work uses only the original zero-temperature ground-state formalism, which we briefly describe below.

The DMET computation self-consistent solves two problems: a fragment (quantum impurity) problem, solved at the many-body level, and the full crystal, solved at the mean-field level. The former is referred to as the high-level calculation and the latter as the low-level calculation. Consider partitioning the crystal into fragments. If the fragments are chosen with translational symmetry, i.e., they are a crystal unit cell, it is sufficient to consider a single fragment $x$ as other solutions are related by symmetry. The fragment Hilbert space is spanned by a set of fragment orbitals $\{\phi^x\}$, and the remaining part of the crystal is termed the environment. Then, we consider a mean-field wavefunction approximating the crystal electronic ground-state. As we are concerned only with non-superconducting parent states, we will assume this is a Slater determinant $|\Phi\rangle$. We can write any such $|\Phi\rangle$ as

$$|\Phi\rangle = |\Psi^{x,\text{emb}}\rangle \otimes |\Phi^{x,\text{core}}\rangle, \tag{S13}$$

where $|\Phi^{x,\text{core}}\rangle$ is a Slater determinant with no support in $\{\phi^x\}$, and $|\Psi^{x,\text{emb}}\rangle$ is a Slater determinant in the embedded Hilbert space, spanned by orbitals $\{\phi^{\text{emb}}\} = \{\phi^x\} \oplus \{\phi^{\text{bath}}\}$. The simple way in which the environment degrees of freedom enter into $|\Phi\rangle$ is a property of the Schmidt decomposition, and this guarantees that the number of bath orbitals $\{\phi^{\text{bath}}\}$ is at most the number of fragment orbitals. Mathematically, $\{\phi^{\text{bath}}\}$ are obtained from the singular-value decomposition (SVD) of the fragment/environment block of the coefficient matrix, or eigenvectors of the environment block of the one-particle reduced density matrix, of $|\Phi\rangle$ (*41, 55*).

The DMET impurity problem is defined by next projecting the interacting problem for the whole crystal into the embedded space $\{\phi^{\text{emb}}\} \otimes |\Phi^{x,\text{core}}\rangle$. Using the projector $P$, we define an embedded Hamiltonian

$$H^{\text{emb}} = PHP, \tag{S14}$$

where $H$ is the Hamiltonian of the crystal, and the high-level DMET solution is the ground-state of $H^{\text{emb}}$. The form of the many-body ground-state is now $|\Psi^{\text{corr}}\rangle = |\Psi^{x,\text{emb}}\rangle \otimes |\Phi^{x,\text{core}}\rangle$, where $|\Psi^{x,\text{emb}}\rangle$ is now in general a correlated state rather than a Slater determinant. Since the embedded Hilbert space is very small (spanned by only twice the number of fragment orbitals) this is a great reduction in complexity, and we can solve for the many-body ground-state $|\Psi^{\text{corr}}\rangle$ using sophisticated (and even exact) ground-state solvers. Note that the mean-field state is an eigenstate of $P$, thus the embedding is exact at a mean-field level. However, the bath orbitals defined from $|\Phi\rangle$ only fully capture the entanglement between the fragment and its environment in the mean-field state.

To improve the bath orbitals, we can update the mean-field wavefunction using the information from the correlated solution $|\Psi^{\text{corr}}\rangle$. Since $|\Phi\rangle$ is the eigenstate of a mean-field Hamiltonian $h^{\text{mf}}$, we can adjust $|\Phi\rangle$ by adding a correlation potential,

$$h^{\text{mf}}_{ij} = h^{\text{core}}_{ij} + v^{\text{eff}}_{ij} + u_{ij}, \tag{S15}$$

where $h^{\text{core}}$ is the bare one-particle Hamiltonian, $v^{\text{eff}}$ is the mean-field effective potential from the electron-electron interaction (e.g., $v_J - v_K$ in Hartree-Fock, $v_J + v_{\text{xc}}$ in density functional theory), and $u$ is the correlation potential ($ij$ denote indices defined on the fragments). The correlation potential $u$ is then chosen to minimize the deviation of the high-level ground-state $|\Psi^{\text{corr}}\rangle$ and $|\Phi\rangle$. In practice, we minimize the difference in the 1-particle reduced density matrix $\gamma$ in a least-squares sense

$$u = \underset{u}{\operatorname{argmin}}\, w(u) = \underset{u}{\operatorname{argmin}} \sum_{ij} \left\{ \gamma^{\text{mf}}_{ij}[h^{\text{mf}}(u)] - \gamma^{\text{corr}}_{ij} \right\}^2. \tag{S16}$$

In large impurities, this minimization can be numerically difficult, and we restrict $ij$ to selected active orbitals in the fragment (see below for details).

DMET includes the interactions inside the chosen fragment exactly, but captures interactions between the fragment and environment only via the bath. The replacement of the environment by an effective bath determined self-consistently resembles the formalism of DMFT, but note that the bath degrees of freedom in DMET are not fictitious (as they are in DMFT) but corresponding to a projection from the environment orbitals into a smaller space. Further, unlike in DMFT, the bath is designed to capture ground-state entanglement, rather than the local Green's function or density of states. For the current work, this leads to several crucial computational simplifications: (1) the theory is frequency-independent, thus for ground-state properties one does not need to compute Green's functions, significantly reducing the cost; (2) the number of bath orbitals is at most the number of impurity sites. In DMFT a formally infinite size bath is needed when it is represented explicitly. There are thus no bath discretization errors in DMET, and it is possible to scale to large fragments even with solvers that require explicit baths; (3) unlike for Green's functions, there are a large of number of many-body solvers that can compute ground-states of Hamiltonians in *ab initio* settings. These characteristics help facilitate the high-level *ab initio* treatment of large fragments that is necessary for the systematic study in this work.

### 1.2.2 Ab initio formulation

The current work uses the full cell (all-electron) quantum embedding scheme, recently discussed for both DMET and DMFT in periodic systems (*10–12*). We briefly review some of the formalism and additional technical innovations that are used in this work.

*Local orbitals.* To carry out DMET, we require an orthogonal local orbital (LO) basis. We transform the Gaussian AOs into periodic intrinsic atomic orbitals (IAOs) (*10, 60*) and projected atomic orbitals (PAOs) (*61*), and use these as our local orbitals. These orbitals can be viewed as a series of atom-centered projected Wannier functions, and therefore, no numerical optimization is required during their construction. In particular, the periodic IAOs are based on the projection to a set of predefined valence AO orbitals (the so-called IAO reference functions), whose number is smaller than the computational AO basis and do not include polarization or diffuse components. See Ref. (*10, 60*) for their construction. In this work, we use atomic spherically averaged Hartree-Fock orbitals as the IAO reference functions, because the segmented Gaussian basis functions that we use (such as def2-SVP) do not individually possess meaningful AO character. The IAOs represent the valence space (occupied + virtuals of valence character) of the materials, while the PAOs represent the remaining virtual space,

$$\left|\phi_p^{\text{PAO},\mathbf{k}}\right\rangle = \sum_i \left(1 - \left|\phi_i^{\text{IAO},\mathbf{k}}\right\rangle\left\langle\phi_i^{\text{IAO},\mathbf{k}}\right|\right) \left|\chi_p^{\text{AO},\mathbf{k}}\right\rangle. \tag{S17}$$

The union of the two sets spans the full orbital space. The coefficient matrix $C^{\text{LO}}$ defines the transformation from the computational AO basis to the LO basis.

*Bath construction.* We determine the bath orbitals from the valence (IAO, not PAO) part of the one-particle density matrix. We assume below that the impurity corresponds to a reference cell $\mathbf{R} = 0$, thus the bath orbitals live in the cells $\mathbf{R} \neq 0$. The off-diagonal block of the density matrix of the whole crystal ("lattice") is computed directly from the Fourier transform of the $\mathbf{k}$-space density matrix obtained in the mean-field calculation (*10, 56*),

$$\gamma_{ij}^{\mathbf{R}\neq\mathbf{0}} = \frac{1}{N_\mathbf{k}} \sum_\mathbf{k} e^{i\mathbf{k}\cdot\mathbf{R}} \gamma_{ij}^\mathbf{k}. \tag{S18}$$

Constraining $i$, $j$ to be IAO (i.e., valence) indices, the valence bath is obtained from an SVD of the off-diagonal block,

$$\gamma_{ij}^{\mathbf{R}\neq\mathbf{0}} = \sum_{ik} B_{ik}^{\mathbf{R}\neq\mathbf{0}} \Lambda_k V_{kj}^\dagger, \tag{S19}$$

where $\Lambda$ measures the entanglement between the bath and impurity orbitals and $B$ is the coefficient matrix of the (orthogonalized) bath orbitals,

$$\left|\phi_j^{\text{bath}}\right\rangle = \sum_{\mathbf{R}\neq\mathbf{0},i} \left|\phi_i^{\text{LO},\mathbf{R}}\right\rangle B_{ij}^\mathbf{R}. \tag{S20}$$

The overall embedding orbital (EO) space is spanned by impurity orbitals (in the reference cell $\mathbf{R} = 0$) and the above bath orbitals,

$$C_{ij}^{\text{EO},\mathbf{R}} = \begin{bmatrix} \mathbf{1} & \mathbf{0} \\ \mathbf{0} & \mathbf{B}^{\mathbf{R}\neq\mathbf{0}} \end{bmatrix}. \tag{S21}$$

For subsequent integral transformations (see below), it is more convenient to Fourier transform the embedding orbitals to the $\mathbf{k}$-space,

$$C_{ij}^{\text{EO},\mathbf{k}} = \sum_\mathbf{R} e^{-i\mathbf{k}\cdot\mathbf{R}} C_{ij}^{\text{EO},\mathbf{R}}. \tag{S22}$$

*Integral transformation.* The construction of the embedding Hamiltonian is equivalent to a set of integral transformations using the coefficient matrix $C$ of the embedding basis (*10, 11*).

The one-body part of the embedding Hamiltonian can be directly evaluated using the projection Eq. (S14),

$$H_{ij}^{\text{emb}} = \frac{1}{N_\mathbf{k}} \sum_\mathbf{k} C_{ip}^{\mathbf{k}\dagger} [h_{pq}^{\text{core},\mathbf{k}} + v_{pq}^{\text{eff},\mathbf{k}}] C_{qj}^\mathbf{k} - v_{ij}^{\text{eff,loc}} - \mu\delta_{ij}, \tag{S23}$$

where we have included in the definition $\mu$, a chemical potential that adjusts the electron density on the fragment such that each cell has the correct number of electrons. $v_{ij}^{\text{eff,loc}}$ is the effective potential in the embedding space originating from the density matrix of the embedded space,

$$v_{ij}^{\text{eff,loc},\sigma} = \left(\sum_{\sigma'} V_{ijkl}^{\text{emb},\sigma\sigma'} \gamma_{lk}^{\text{emb},\sigma'}\right) - V_{iklj}^{\text{emb},\sigma\sigma} \gamma_{kl}^{\text{emb},\sigma}, \tag{S24}$$

where $V^{\text{emb}}$ is the embedding two-body hamiltonian (see below for its construction) and $\sigma$ is a spin label.

The two-body part of the embedding hamiltonian must be constructed appropriately to minimize computational cost. With density fitting, ERIs in the embedding space can be evaluated from 3-center integrals in the reference cell,

$$W_{Lij}^{\mathbf{k}_L 00} = \frac{1}{N_\mathbf{k}} \sum_{\mathbf{k}_p \mathbf{k}_q}{}' C_{ip}^{\mathbf{k}_p\dagger} W_{Lpq}^{\mathbf{k}_p \mathbf{k}_q} C_{qj}^{\mathbf{k}_q}, \quad (S25)$$

where $'$ indicates the summation is over momentum conserving crystal momenta $\mathbf{k}_L = \mathbf{k}_p - \mathbf{k}_q + n\mathbf{b}$. The cost of this step scales as $\mathcal{O}(n_\mathbf{k}^2 n_{\text{bas}}^4)$. The final embedding ERI is a contraction which scales as $\mathcal{O}(n_\mathbf{k} n_{\text{bas}}^5)$,

$$V_{ijkl}^{\text{emb}} = \frac{1}{N_\mathbf{k}} \sum_{\mathbf{k}_L L} W_{Lij}^{\mathbf{k}_L 00 *} W_{Lkl}^{\mathbf{k}_L 00}. \quad (S26)$$

Note that time reversal symmetry of the integrals and coefficients can be used to reduce the computational cost. For example, time reversal symmetry over $\mathbf{k}_L$ effectively reduces costs by about a factor of 2, as we only need consider the non-negative $\mathbf{k}_L$.

For each pair $(\mathbf{k}_p, \mathbf{k}_q)$, there will be another pair $\bar{\mathbf{k}}_q = -\mathbf{k}_q$ and $\bar{\mathbf{k}}_p = -\mathbf{k}_p$ that are related,

$$\begin{aligned} W_{Lji}^{\bar{\mathbf{k}}_q \bar{\mathbf{k}}_p} &= C_{jq}^{\bar{\mathbf{k}}_q\dagger} W_{Lqp}^{\bar{\mathbf{k}}_q \bar{\mathbf{k}}_p} C_{pi}^{\bar{\mathbf{k}}_p} \\ &= C_{jq}^{\mathbf{k}_q T} W_{Lpq}^{\mathbf{k}_p \mathbf{k}_q} C_{pi}^{\mathbf{k}_p *} \\ &= W_{Lij}^{\mathbf{k}_p \mathbf{k}_q T}, \end{aligned} \quad (S27)$$

where we have used the relation $C^{\mathbf{k}} = C^{\bar{\mathbf{k}}*}$. This relation further gives a factor of 2 cost reduction in the transformation.

Finally, we note that after the summation over $\mathbf{k}$, the resulting embedding 3-center integrals and the final embedding 4-center integrals have permutation symmetry over the orbital indices. In fact, the embedding ERI is real and has 8-fold symmetry: $V_{ijkl}^{\text{emb}} = V_{jikl}^{\text{emb}} = V_{ijlk}^{\text{emb}} = V_{klij}^{\text{emb}} = \cdots$. This relation gives another factor of 4 during the contraction step.

### 1.2.3 Multi-fragment extension

The above *ab initio* DMET formulation assumes we are embedding a full crystal cell (which may be a supercell of primitive cells) in the environment of other cells. However, for complicated crystal structures, e.g., in the multi-layer compounds, or for inhomogeneous systems and defect calculations, the full cell calculation is prohibitively expensive. New techniques are thus required to further reduce the impurity size. Here we have developed and implemented a multi-fragment extension of the *ab initio* DMET. This allows a further decomposition of the full cell impurity into fragments while retaining the periodicity among different cells.

In this scheme, the reference impurity cell is divided into fragments which are each embedded in the bath of the other fragments and other cells. For example, for the double-layer compound Hg-1212, the cell is sliced into 3 fragments, the first one involving the bottom layer of the CuO$_2$ plane and the corresponding apical oxygen; the second, the upper layer of copper and oxygen; and the third fragment, the other buffer-layer atoms, i.e., Hg, Ba and Ca [see Fig. 1(a)].

The total energy is defined as the sum of all fragment energies, $E = \sum_x E^x$. Implementing the DMET democratic partitioning formula (which defines how to reassemble expectation values from each fragment), each fragment energy is the expectation value of a scaled Hamiltonian,

$$E^x = \sum_{ij} \left\langle \tilde{H}_{ij}^x a_i^\dagger a_j \right\rangle + \frac{1}{2} \sum_{ijkl} \left\langle \tilde{V}_{ijkl}^x a_i^\dagger a_k^\dagger a_l a_j \right\rangle = \sum_{ij} \tilde{H}_{ij}^x \gamma_{ji}^x + \frac{1}{2} \sum_{ijkl} \tilde{V}_{ijkl}^x \Gamma_{ijkl}^x, \quad (S28)$$

where $\tilde{H}$ and $\tilde{V}$ are scaled Hamiltonians, and $\gamma$ and $\Gamma$ are 1-body and 2-body density matrices respectively. $\tilde{H}$ is defined as,

$$\tilde{H}_{ij} = w_{ij} \left( h_{ij}^{\text{core}} + \frac{1}{2} v_{ij}^{\text{eff}} - \frac{1}{2} v_{ij}^{\text{eff,loc}} \right). \quad (S29)$$

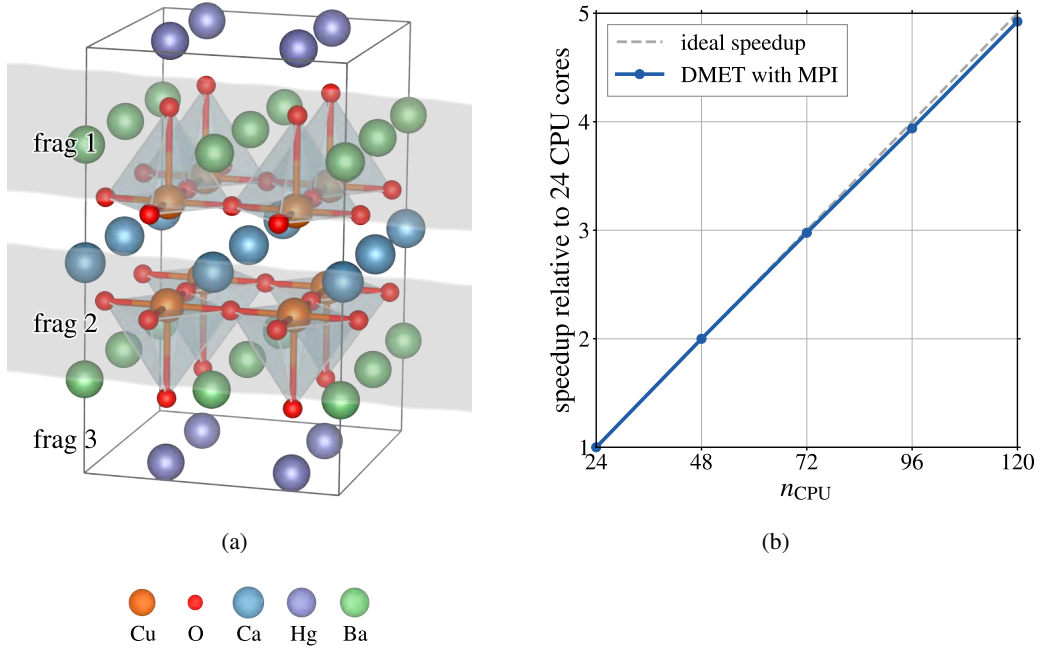

Figure S1: (a) Illustration of the multi-fragmentation scheme in the multi-layer cuprate Hg-1212. The system is divided into 3 pieces: the fragments 1 and 2 involve the two Cu-O layers and fragment 3 contains all other ions in the cell. (b) MPI efficiency of the multi-fragment implementation of a h-BN crystal.

Note the $\frac{1}{2}$ factor in the Coulomb energy. The scaling weight is defined by the fraction of indices in the local orbitals of fragment $x$,

$$w_{ij} = \begin{cases} 1, & \text{if } ij \in x, \\ \frac{1}{2}, & \text{if } i \in x \text{ or } j \in x, \\ 0, & \text{if } i \notin x \text{ and } j \notin x. \end{cases} \quad (S30)$$

Similarly, the 2-body part of the Hamiltonian,

$$\tilde{V}_{ijkl} = w_{ijkl} V_{ijkl}^{\text{emb}}, \quad (S31)$$

includes a weight factor $w_{ijkl}$ to correctly account for the number of fragment indices.

$v^{\text{eff}}$ in Eq. (S29) is re-evaluated using the DMET global density matrix $\gamma^{\text{glob}}$ (62),

$$\gamma^{\text{glob},\mathbf{R}} = \frac{1}{2}\left(C^{\mathbf{R}} \gamma^{\text{emb}} C^{0\dagger} + C^{0} \gamma^{\text{emb}} C^{\mathbf{R}\dagger}\right), \quad (S32)$$

where $C^{\mathbf{R}}$ is the coefficient of the embedding basis of the $\mathbf{R}^{\text{th}}$ unit cell. This ensures a consistent Fock potential is used in all fragments.

The multi-fragment calculations in DMET are easily parallelized e.g., using MPI. The evaluation of the bath orbitals, construction of the embedding Hamiltonian, high-level solver calculations, energy computation and correlation potential fitting are all independent of each other. The communication only happens when (1) determining the chemical potential [communication cost $\mathcal{O}(1)$]; (2) constructing the global density matrix [cost $\mathcal{O}(N)$], where $N$ is the number of embedding orbitals; (3) combining subblocks of the correlation potential [cost $\mathcal{O}(N)$]. We illustrate the MPI efficiency in Fig. 1(b), in which the fragments are pairs of atoms (i.e., of equal size) in a 2D boron nitride crystal. The speedup is very close to ideal.

### 1.2.4 Solver

*Coupled cluster.* The main solver we use in this work is coupled cluster singles and doubles (CCSD) (*43*), which can be easily applied to *ab initio* Hamiltonians with hundreds of correlated orbitals. It is based on a wavefunction ansatz of the form

$$|\Psi\rangle = e^{\hat{T}_1+\hat{T}_2} |\Phi\rangle, \tag{S33}$$

where $|\Phi\rangle$ is a reference Slater determinant and the cluster excitation operators read,

$$\hat{T}_1 = \sum_{ia} t_i^a a_a^\dagger a_i, \tag{S34}$$

$$\hat{T}_2 = \sum_{ijab} t_{ij}^{ab} a_a^\dagger a_b^\dagger a_j a_i. \tag{S35}$$

CC approximations have a number of important properties. First, they are exact for all products of correlations involving a finite number of particles. For example, the CCSD approximation is exact for any product of two-particle correlations, which allows for the accurate description of correlated singlet-like physics. Second, they are extensive, which means that the approximation does not deteriorate simply from increasing system size. Third, they are in principle systematically improvable, by increasing the excitation level (although the cost also increases exponentially with excitation level). Finally, they are especially accurate for gapped and ordered states. This describes the AFM parent state and magnetic configurations considered in this work. In ordered states, one chooses the reference $|\Phi\rangle$ to break the appropriate symmetry. Here, we break $S^2$ symmetry and choose an unrestricted (spin-polarized) Hartree-Fock determinant as our reference state, solving the unrestricted CC equations (UCCSD). Note that Hartree-Fock is truly a mean-field theory of the bare Coulomb interaction. Thus, all fluctuations observed in this work are due to the CC correlations.

The DMET energy expression requires the reduced density matrices. The CC density matrices are obtained from the CC Λ equations (*63*).

Because CCSD is an approximate method, it is always important to benchmark its accuracy for the phenomenon of interest. In molecular quantum chemistry, CCSD(T) (coupled cluster singles and doubles with perturbative triples) is often regarded as the "gold standard" because it achieves high accuracy for ordered or gapped reference states (so-called "single reference" states). In this work, we do not extensively use the triples correction because our implementation of the Λ equations is efficient only at the singles and doubles level. However, we can verify the accuracy of the unrestricted CCSD solver against unrestricted CCSD(T) in a smaller subset of examples. We also benchmark against the *ab initio* DMRG solver. Since DMRG works well away from ordered states (i.e., for multi-reference correlations), this test allows us to verify the basic assumption underlying the accuracy of CC approximations in these systems. We describe these benchmarks further below.

Since DMET involves a self-consistency loop and a search over the chemical potential, it is necessary to solve the quantum impurity problem many times. To do this efficiently, the solver can be approximately restarted from the previous solution by matching the embedding basis. Say $C_1$ and $C_2$ are the embedding bases in the first and the second cycles of a DMET calculation. The bases can be approximately matched using the orbital overlap matrix and an SVD,

$$C_1^\dagger S C_2 = U \Sigma V^\dagger, \tag{S36}$$

where $S$ is the overlap matrix in the computational basis (here this is the AO overlap matrix), and

$$\tilde{C}_1 = C_1 R = C_1 U V^\dagger \tag{S37}$$

defines the closest orbitals to $C_2$ in the Frobenius norm sense and $R = UV^\dagger$ is a unitary rotation matrix. The wavefunction from the first cycle can then be transformed with the rotation matrix, i.e., the one- and two-body amplitudes in the CCSD equations are rotated as,

$$\tilde{t}_k^c = \sum_{ia} R_{ca}^\dagger t_i^a R_{ik}, \tag{S38}$$

$$\tilde{t}_{kl}^{cd} = \sum_{ijab} R_{ca}^\dagger R_{db}^\dagger t_{ij}^{ab} R_{ik} R_{jl}. \tag{S39}$$

This restart scheme greatly reduces the total cost spent in the many-body solver. Typically, the CCSD amplitude equations converge in $< 5$ iterations after the second DMET iteration.

*Density matrix renormalization group.* The *ab initio* density matrix renormalization group (DMRG) (*64–66*) uses a matrix product state (MPS) defined on a 1-dimensional ordering of the orbitals,

$$|\Psi\rangle = \sum_{n_1, \cdots, n_L} \mathbf{A}^{n_1} \mathbf{A}^{n_2} \cdots \mathbf{A}^{n_L} |n_1 n_2 \cdots n_L\rangle, \tag{S40}$$

where $L$ is the number of orbitals, $n$ is the occupation number of an orbital and the $\mathbf{A}$'s are $M \times M$ matrices. The accuracy of DMRG is controlled by the so-called bond-dimension $M$ and as $M \to \infty$, DMRG becomes exact. In the large bond dimension regime, the energy has a linear relation with respect to the DMRG discarded weight $\delta$ (*67–69*), which allows for stable extrapolation to the exact limit.

To carry out *ab initio* DMRG calculations using our embedding Hamiltonian, we follow the strategy described in Ref. (*70*). We first define an orthogonal local basis as the orbitals in DMRG. In particular, we use split localized unrestricted MP2 natural orbitals, where orbitals of occupied and virtual character are separately localized by the Edmiston-Ruedenberg (ER) method (*71*). Using local natural orbitals improves the convergence of the DMRG with respect to bond dimension.

## 1.3 Analysis methods

### 1.3.1 Charge and spin population analysis

Since we allow $S^2$ symmetry breaking in our calculations, charge and spin order can be analyzed using the spin-resolved one-particle reduced density matrix ($\sigma = \alpha, \beta$),

$$\gamma_{ij}^\sigma = \langle a_{j\sigma}^\dagger a_{i\sigma} \rangle. \tag{S41}$$

In particular, the charge of orbital $i$ reads,

$$n_i = \gamma_{ii}^\alpha + \gamma_{ii}^\beta, \tag{S42}$$

and the local magnetic moment of orbital $i$ reads,

$$m_i = \gamma_{ii}^\alpha - \gamma_{ii}^\beta. \tag{S43}$$

In principle, the charge (spin) populations depend on the choice of atom centered local orbitals $\{\phi_i\}$. This typically has a strong basis set dependence if the population analysis is carried using the computational AO basis. However, the basis dependence can be largely removed by measuring the population in the IAO basis (*60*), which is what we do here.

### 1.3.2 Bonding analysis

To analyze bonding in the system in a straightforward way, we can use the atom centered local orbitals, i.e., the IAOs + PAOs used in the population analysis above, and evaluate bond orders, which measure the off-diagonal density matrix element between two local orbitals. In this work, we use the 2-center Mayer bond order (*72*) which, for atoms $A$ and $B$ (or two subsets of orbitals) is defined as,

$$b_{AB} = \sum_\sigma b_{AB}^\sigma = 2 \sum_\sigma \sum_{i \in A} \sum_{j \in B} (\gamma^\sigma S)_{ji} (\gamma^\sigma S)_{ij}, \tag{S44}$$

where $\gamma^\sigma$ is the one-particle density matrix with spin $\sigma$ and $S$ is the overlap matrix of the local basis. Since we use IAOs + PAOs as our basis, $S$ is the identity matrix. For non-polarized covalent bonds, the Mayer bond order typically agrees very well with chemical intuition (e.g., H$_2$ and N$_2$ roughly have bond orders 1 and 3 in calculations). For strong polarized covalent bonds or even ionic bonds, the Mayer bond order is generally qualitatively reasonable.

As an alternative to the bond order, we also use the electron density $\rho(\mathbf{r})$ and electron localization function ELF$^\sigma(\mathbf{r})$ (*73, 74*) as real space indicators of the bonding. ELF was originally proposed to measure the localization of electrons and helps reveal atomic shell structure, bonding, and lone electron pairs. ELF values lie in [0, 1]. When ELF = 1, the electron is completely localized while ELF = $\frac{1}{2}$ suggests that the electron behaves like it does in the electron gas of the given density at that position. Since ELF is defined in real space it is less sensitive to the choice of basis set. Typically, large ELF values indicate a core region, a lone pair of electrons, and covalent bonding. Thus, the ELF is a useful tool to distinguish between covalent and non-covalent (such as ionic) bonding.

A third way to understand the bonding is to examine the individual localized orbitals in the occupied and virtual spaces, which reveals the bonds and antibonds of the system. Here, we localized the occupied and virtual embedding orbitals via Pipek-Mezey (PM) localization (*75*), which maximizes the population charges on the atoms,

$$U = \underset{U}{\mathrm{argmax}}\, w(U) = \underset{U}{\mathrm{argmax}} \sum_{I}^{\mathrm{atoms}} \sum_{i} |q_i^I|^2, \tag{S45}$$

and $q_i^I$ is the charge of the $i^{\mathrm{th}}$ orbital of atom $I$ (IAOs + PAOs are used as the population method, in order to reduce the basis set dependence). The resulting unitary rotation $U$ defines a set of localized orbitals,

$$|\phi_i^{\mathrm{PM}}\rangle = \sum_m |\psi_m\rangle U_{mi}. \tag{S46}$$

Note that PM localization preserves the separation between $\sigma$ and $\pi$ orbitals.

### 1.3.3 Spin-spin correlation function

The ($S_z$ component) spin-spin correlation function $\langle S_z(0) S_z(r) \rangle$ reflects the spin-spin correlation between Cu in the reference cell (0) and another Cu at position ($r$). If the correlation function does not decay to 0 at large $r$, the system has long-range order.

Using the spin operator of local orbital $i$

$$\hat{S}_i^z = \frac{1}{2}\left(a_{i\alpha}^\dagger a_{i\alpha} - a_{i\beta}^\dagger a_{i\beta}\right), \tag{S47}$$

we express the correlation function as a contraction of the reduced 1-particle $\gamma_{ij}^\sigma$ and 2-particle $\Gamma_{ijkl}^{\sigma\tau} \equiv \langle a_{i\sigma}^\dagger a_{k\tau}^\dagger a_{l\tau} a_{j\sigma} \rangle$ density matrices,

$$\begin{aligned}
\langle S_z(0) S_z(r) \rangle &= \sum_{i \in \mathrm{Cu}(0)} \sum_{j \in \mathrm{Cu}(r)} \langle \hat{S}_i^z \hat{S}_j^z \rangle \\
&= \frac{1}{4} \sum_{i \in \mathrm{Cu}(0)} \sum_{j \in \mathrm{Cu}(r)} \left(\gamma_{ij}^\alpha \delta_{ij} + \Gamma_{iijj}^{\alpha\alpha} - \Gamma_{iijj}^{\alpha\beta} - \Gamma_{jjii}^{\alpha\beta} + \gamma_{ij}^\beta \delta_{ij} + \Gamma_{iijj}^{\beta\beta}\right).
\end{aligned} \tag{S48}$$

where the summation is constrained to the local orbitals of Cu. Note that we do not consider the oxygen contribution to the correlation function.

### 1.3.4 Natural orbital analysis

Spin-traced natural orbitals can be obtained by diagonalizing the (spin-traced) density matrix $\gamma$,

$$\gamma_{qp}^{\mathbf{k}} C_{pi}^{\mathbf{k}} = C_{qi}^{\mathbf{k}} \Lambda_i^{\mathbf{k}}, \tag{S49}$$

where $\Lambda_p$ is a natural occupation number (between 0 to 2) and $C_{pi}$ are the natural orbital coefficients. If the density matrix originates from a pure state with $S = 0$, then the further the natural occupation is away from 0 or 2 (a single

Slater determinant), the more correlated an orbital is. We can define the *half-filling index* to summarize the contribution of local orbitals $\{p\}$, to half-filled natural orbitals,

$$f_p^{\text{half}} = \frac{1}{N_{\mathbf{k}} N_p} \sum_{i\mathbf{k}} |C_{pi}^{\mathbf{k}}|^2 \min(\Lambda_i^{\mathbf{k}}, 2 - \Lambda_i^{\mathbf{k}}), \tag{S50}$$

where $N_p$ is the number of orbitals in an orbital group (e.g., Cu $3d_{x^2-y^2}$, O $2p$). The situation is a bit more subtle for a state with $S \neq 0$ or a symmetry broken state. In a symmetry broken state, the degree of half-filling measures both the fluctuations as well as the degree of spatial symmetry breaking. For example, in a symmetry broken Slater determinant with overall low-spin ($S_z = 0$) (which has no fluctuations), $\Lambda_p \to 1$ means that there are spin-orbitals of opposite spin with no spatial overlap. Nonetheless, since it is important to include spin fluctuations between such spin-orbitals, the half-filling index remains a useful indicator of the most important local orbitals to include in a minimal atomic model.

The spin-resolved natural orbitals can be obtained by diagonalizing the spin-resolved density matrix $\gamma^{\sigma}$. Now $\Lambda_p$ ranges from 0 to 1. In a symmetry broken state, it is the deviation of the *spin-resolved* occupancies $\Lambda_p$ from their extremal values that measures the importance of dynamical fluctuations. If all spin-resolved occupancies are 0 or 1, then the state is exactly of mean-field character and all single-particle lifetimes are infinite (no dynamical effects). Deviation from this occupancy pattern indicates correlation, and very strong deviation indicates strong correlation [e.g., in a system far from a Fermi liquid where the quasiparticle picture breaks down such as a Luttinger liquid, the occupancy (or momentum distribution function) no longer shows a jump between values close to 0 and values close to 1]. In this work, we estimate if there are strong dynamical effects by examining if all the natural occupancies are close to 0 or 1.

Also, although DMET does not provide direct access to the single-particle energy spectrum, we can use the occupancies of the spin-resolved natural orbitals as proxies for proximity to the top edge of the valence band/bottom edge of the conduction band. In particular, the highest occupied natural orbital (occupancy $> 1/2$ but furthest from 1) is a pseudo-valence-band maximum; while the lowest occupied natural orbital (occupancy $< 1/2$ but furthest from 0) is a pseudo-conduction band minimum. We analyze the character of these states in the main text.

## 1.4 Single-particle methods

We have carried out calculations with 3 kinds of single-particle approaches. The first is Hartree-Fock (HF) which forms the low-level mean-field method that is the starting point for the *ab initio* density matrix embedding. We have also carried out DFT (hybrid functional) and DFT+$U$ calculations. These latter calculations represent the typical approach to electronic structure in the cuprates, and in some sense, represent methods we aim to supersede with an explicit many-body approach. The primary drawback of DFT and DFT+$U$ is the level of empiricism that enters in the choice of functional, $U$, and double-counting correction, which means that the errors are not simply improvable. We provide DFT and DFT+$U$ results in this work purely for comparison; the DFT quantities do not enter into the many-body and quantum embedding calculations.

Our Hartree-Fock and DFT implementations within the PYSCF package have been described elsewhere [See Ref. (*14*) and references within]. Since our DFT+$U$ implementation has not, and since the definition of the method involves some implementation specific details, we briefly describe it below.

In the PYSCF program, we have implemented Dudarev's rotationally invariant DFT+$U$ formulation (*76*) in a periodic Gaussian basis. Suppose we have a set of local orbitals $\{\phi\}$ expanded in a crystalline AO basis $\{\chi\}$,

$$|\phi_i^{\mathbf{k}}\rangle = \sum_p |\chi_p^{\mathbf{k}}\rangle C_{pi}^{\text{LO},\mathbf{k}} \tag{S51}$$

and a set of molecular orbitals,

$$|\psi_m^{\mathbf{k}\sigma}\rangle = \sum_p |\chi_p^{\mathbf{k}}\rangle C_{pm}^{\text{MO},\mathbf{k}\sigma}. \tag{S52}$$

The default localized orbitals are atomic orbitals in the Gaussian basis set, whose projector is defined as,

$$\begin{aligned} \gamma_{ij}^{\mathbf{k}\sigma} &= \sum_m \langle \phi_i^{\mathbf{k}} | \psi_m^{\mathbf{k}\sigma} \rangle f_m^{\mathbf{k}\sigma} \langle \psi_m^{\mathbf{k}\sigma} | \phi_j^{\mathbf{k}} \rangle \\ &= \sum_{pqrs} C_{ip}^{\text{LO},\mathbf{k}\dagger} S_{pq}^{\mathbf{k}} \gamma_{qr}^{\mathbf{k}\sigma} S_{rs}^{\mathbf{k}} C_{rj}^{\text{LO},\mathbf{k}}, \end{aligned} \tag{S53}$$

where $S^{\mathbf{k}}_{pq}$ is the AO overlap matrix and $\gamma^{\mathbf{k}\sigma}_{pq} = \sum_m C^{MO,\mathbf{k}\sigma}_{pm} f^{\mathbf{k}\sigma}_m C^{MO,\mathbf{k}\sigma\dagger}_{mq}$ is the reduced one-particle density matrix in the AO basis where $f^{\mathbf{k}\sigma}_m$ denotes the occupancy of molecular orbital $\psi^{\mathbf{k}\sigma}_m(\mathbf{r})$. The DFT+$U$ Hamiltonian is then obtained from a partial derivative,

$$h^{DFT+U,\mathbf{k}\sigma}_{pq} = \frac{\partial E^{DFT}}{\partial \gamma^{\mathbf{k}\sigma}_{qp}} + \frac{\partial(E^U - E^{loc})}{\partial \gamma^{\mathbf{k}\sigma}_{qp}}$$
$$= h^{DFT,\mathbf{k}\sigma}_{pq} + \sum_I \frac{U^I - J^I}{2} \sum_{rs} S^{\mathbf{k}}_{pr} \sum_{ij} C^{LO,I\mathbf{k}}_{ri} \left(\delta^I_{ij} - 2\gamma^{I\mathbf{k}\sigma}_{ij}\right) C^{LO,I\mathbf{k}\dagger}_{js} S^{\mathbf{k}}_{sq},$$ (S54)

where $I$ labels the atom whose $d$ or $f$ orbitals are corrected by the $U - J$ term.

Although conceptually very simple, the detailed implementation of DFT+$U$ can differ in different computer programs due to two factors: (1) several versions of DFT+$U$ exist, including various choices of double counting corrections (77, 78). (2) different choices of local orbital projectors. For example, in the PySCF program, we implement $U$ in a standard Gaussian basis, while in plane-wave codes, the local orbital is often chosen to be a pseudopotential atomic orbital (79) or a localized Wannier orbital (80, 81).

## 1.5 Computational details

### 1.5.1 System

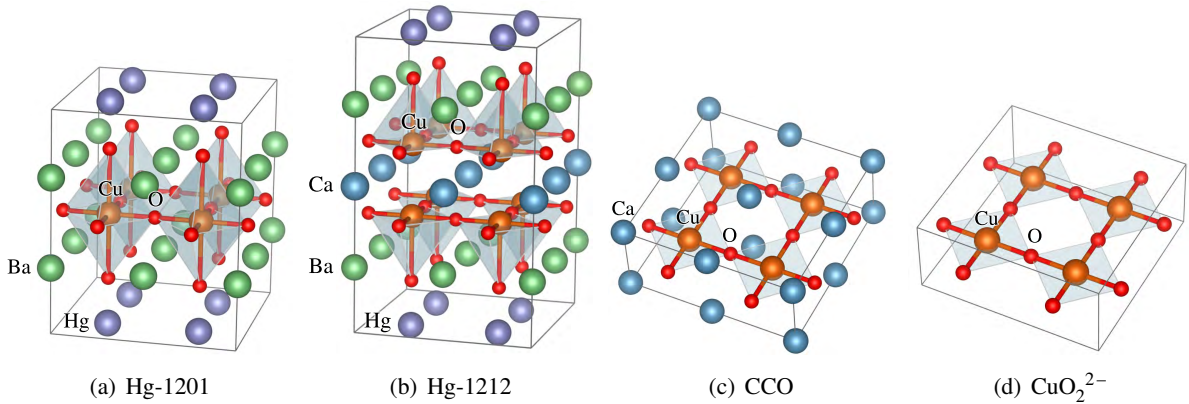

(a) Hg-1201  (b) Hg-1212  (c) CCO  (d) CuO$_2^{2-}$

Figure S2: Crystal structures of HgBa$_2$CuO$_4$ (Hg-1201), HgBa$_2$CaCu$_2$O$_6$ (Hg-1212), CaCuO$_2$ (CCO) and CuO$_2^{2-}$.

Table S1: Crystal structures of HgBa$_2$CuO$_4$ (Hg-1201), HgBa$_2$CaCu$_2$O$_6$ (Hg-1212), CaCuO$_2$ (CCO) and CuO$_2^{2-}$.

| Compound | $a$ [Å] | $c$ [Å] | ∠ Cu-O-Cu [°] | apical $\delta^z_{Cu-O}$ [Å] |
|---|---|---|---|---|
| Hg-1201 [a] | 3.8714 | 9.5023 | 180.0 | 2.767 |
| Hg-1212 [b] | 3.8630 | 12.6978 | 179.5 | 2.822 |
| CCO [c] | 3.8556 | 3.1805 | 180.0 | |
| CuO$_2^{2-}$ | 3.8556 | 3.1805 | 180.0 | |

[a] From Ref. (82).  [b] From Ref. (83).  [c] From Ref. (84).

We primarily consider 4 compounds in this work: (a) the single-layer compound HgBa$_2$CuO$_4$ (Hg-1201), (b) the double-layer compound HgBa$_2$CaCu$_2$O$_6$ (Hg-1212), (c) the infinite-layer compound CaCuO$_2$ (CCO), and (d) a hypothetical CuO$_2^{2-}$ layer (repeated in the vertical direction) (see Fig. S2). The lattice parameters are summarized in Table S1.

We use two types ($\sqrt{2}\times\sqrt{2}$ and $2\times 2$) of supercells in this work to accommodate different magnetic configurations (see Sec. 1.5.2 for details). Their crystal structure files can be found in Appendix 3.1.

### 1.5.2 Magnetic configurations and model mapping

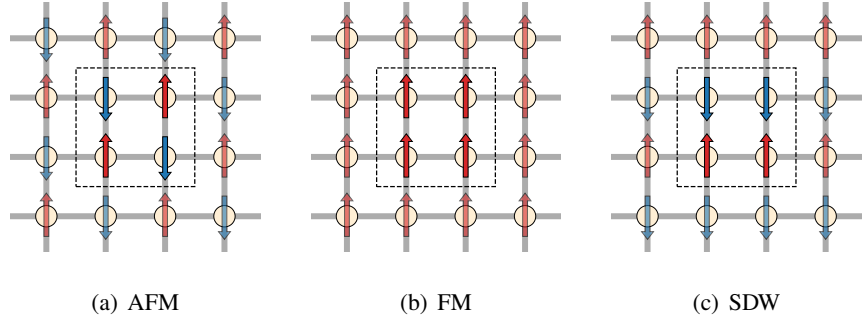

(a) AFM  (b) FM  (c) SDW

Figure S3: Magnetic configurations considered in this work. Only Cu atoms are shown in the figure and the two flavors of spin are represented by up and down arrows respectively.

In this work, we consider 3 magnetic configurations for the in-plane exchange coupling in all cuprate calculations, namely AFM, FM, SDW states (see Fig. S3). The spins in the AFM state [Fig. 3(a)] are arranged in a checkerboard pattern while in the FM [Fig. 3(b)] state the spins are all aligned in the same direction. In the SDW phase [Fig. 3(c)], spins are aligned along the $x$ direction, but are anti-parallel along the $y$ direction.

*Heisenberg model.* In this work, we consider a nearest-neighbor (NN) Heisenberg spin Hamiltonian with the nearest ($J_1$) neighbor coupling parameter,

$$H = J_1 \sum_{\langle ij \rangle} \mathbf{S}_i \cdot \mathbf{S}_j, \tag{S55}$$

where $\langle \cdots \rangle$ denotes the nearest neighbors. Within the Heisenberg model, the energies of the 3 magnetic states can be expressed as,

$$\begin{aligned} E^{\text{AFM}} &= E_0 - J_1 N Z_1 S^2, \\ E^{\text{FM}} &= E_0 + J_1 N Z_1 S^2, \\ E^{\text{SDW}} &= E_0, \end{aligned} \tag{S56}$$

where $Z_n$ denotes the average number of $n^{\text{th}}$ nearest neighbors (here $Z_1 = Z_2 = 2$), $N$ is the number of Cu atoms per cell (here $N = 4$ for the $2 \times 2$ cell) and $S = 1/2$.

*1-band Hubbard model.* Using the fourth order perturbation theory of the one-band Hubbard model (with hopping $t$ and onsite interaction $U$) around the $U = \infty$ limit, we obtain a spin Hamiltonian with 4 terms (*85*),

$$\begin{aligned} H =& J_1 \sum_{\langle ij \rangle} \mathbf{S}_i \cdot \mathbf{S}_j + J_2 \sum_{\langle\langle ij \rangle\rangle} \mathbf{S}_i \cdot \mathbf{S}_j + J_3 \sum_{\langle\langle\langle ij \rangle\rangle\rangle} \mathbf{S}_i \cdot \mathbf{S}_j \\ &+ J_c \sum_{\langle ijkl \rangle} (\mathbf{S}_i \cdot \mathbf{S}_j)(\mathbf{S}_k \cdot \mathbf{S}_l) + (\mathbf{S}_i \cdot \mathbf{S}_l)(\mathbf{S}_k \cdot \mathbf{S}_j) - (\mathbf{S}_i \cdot \mathbf{S}_k)(\mathbf{S}_j \cdot \mathbf{S}_l), \end{aligned} \tag{S57}$$

where

$$J_1 = 4\frac{t^2}{U} - 24\frac{t^4}{U^3}, \tag{S58}$$

and

$$J_c = 80\frac{t^4}{U^3} \tag{S59}$$

is the cyclic magnetic coupling parameter that measures the exchange pathway around the plaquette of the four Cu's.

$$J_2 = J_3 = \frac{J_c}{20} = 4\frac{t^4}{U^3}. \tag{S60}$$

Within this expansion of the Hubbard model, the energies of the 3 magnetic states can be expressed as,

$$\begin{aligned}
E^{\text{AFM}} &= E_0 - J_1 N Z_1 S^2 + J_2 N Z_2 S^2 + J_3 N Z_3 S^2 + J_c N Z_c S^4, \\
E^{\text{FM}} &= E_0 + J_1 N Z_1 S^2 + J_2 N Z_2 S^2 + J_3 N Z_3 S^2 + J_c N Z_c S^4, \\
E^{\text{SDW}} &= E_0 - J_2 N Z_2 S^2 + J_3 N Z_3 S^2 + J_c N Z_c S^4,
\end{aligned} \tag{S61}$$

where $Z_1 = Z_2 = Z_3 = 2$, $Z_c = 1$. Equivalently, the energies can be expressed in terms of the two independent Hubbard parameters, $t$ and $U$,

$$\begin{aligned}
E^{\text{AFM}} &= E_0 - 8\frac{t^2}{U} + 84\frac{t^4}{U^3}, \\
E^{\text{FM}} &= E_0 + 8\frac{t^2}{U} - 12\frac{t^4}{U^3}, \\
E^{\text{SDW}} &= E_0 + 20\frac{t^4}{U^3}.
\end{aligned} \tag{S62}$$

These parameters can be determined from the least-squares solution of the above equations.

*Effective 3J (multi-J) Heisenberg model.* The magnetic couplings from the 1-band Hubbard model can be renormalized into a multi-$J$ Heisenberg model with couplings $J_1$, $J_2$ and $J_3$ (*47, 50*).

$$H = J_1^{\text{eff}} \sum_{\langle ij \rangle} \mathbf{S}_i \cdot \mathbf{S}_j + J_2^{\text{eff}} \sum_{\langle\langle ij \rangle\rangle} \mathbf{S}_i \cdot \mathbf{S}_j + J_3^{\text{eff}} \sum_{\langle\langle\langle ij \rangle\rangle\rangle} \mathbf{S}_i \cdot \mathbf{S}_j. \tag{S63}$$

The effective $J$'s are related to the previous 1-band Hubbard $J$'s,

$$\begin{aligned}
J_1^{\text{eff}} &= J_1 - 2 J_c S^2, \\
J_2^{\text{eff}} &= J_2 - J_c S^2, \\
J_3^{\text{eff}} &= J_3.
\end{aligned} \tag{S64}$$

*Inter-layer coupling $J_\perp$.* We consider both inter-layer AFM and FM coupled configurations for CCO to evaluate the inter-layer coupling $J_\perp$,

$$J_\perp = \frac{E_{\text{FM}} - E_{\text{AFM}}}{2 N_{\text{Cu}} Z_\perp S^2}, \tag{S65}$$

where the perpendicular coordination number $Z_\perp = 1$. (Note: the individual cuprate layers in CCO are AFM coupled; FM above refers only to the inter-layer, or layer-layer, coupling).

The inter-layer magnetic order of the double-layer compound Hg-1212 is fixed to be AFM coupled; we do not evaluate $J_\perp$ for this compound since it is not required for the spin-wave spectrum at $k_z = 0$, the setting for Hg-1212. The inter-layer couplings in the other compounds are very weak and are thus neglected.

*Spin wave spectrum.* Once the spin model parameters are determined from the *ab initio* calculation, the spin wave spectrum can be obtained from linear spin wave theory, which converts a spin problem to a quadratic bosonic problem (*86, 87*). We use the SPINW program (*87*) to generate the spin wave dispersions of the three compounds and we compare them to data from resonant inelastic X-ray scattering (RIXS). We present spin-wave spectra for the NN Heisenberg and $3J^{\text{eff}}$ model. Following typical experimental conventions (which allows us to compare directly to the experimental couplings), we do not use a quantum renormalization factor for the NN Heisenberg spin-wave spectrum, but use a quantum renormalization factor of $Z_c = 1.219$ (*47*) for the multi-$J$ model.

### 1.5.3 Single-particle method settings

The single particle mean-field (SCF) calculations (HF, DFT, DFT+$U$) were carried out in crystalline Gaussian bases using the PYSCF package (*14, 88*), and were cross checked with plane wave basis calculations using the VASP package (*89–93*).

For CCO and CuO$_2{}^{2-}$, we used the minimal basis GTH-SVP-MOLOPT-SR for various benchmarks. This consists of $1s1p1d$ shells for Cu, $1s1p$ for O, $2s1p$ for Ca, and uses the GTH pseudopotential for the core electrons (*94, 95*). GDF was used to compute the two-electron integrals. We used an even-tempered Gaussian basis as the density fitting auxiliary basis ($n_{\text{aux}} \sim 10 n_{\text{AO}}$).

For the more realistic calculations, we used an all-electron basis of polarized double-zeta (split-valence) quality, def2-SVP (*46*) for all elements (consisting of $5s3p2d1f$ shells for Cu, $3s2p1d$ for O, $4s2p1d$ for Ca, $5s2p2d1f$ for Hg, $3s2p1d$ for Ba, and $5s2p2d1f$ for La). The sufficiency of the basis was further checked with a larger polarized triple-zeta basis set def2-TZVP (*46*) as well as plane-wave basis calculations. For Hg, Ba and La, an effective core potential (ECP) was used to handle the core electrons and scalar relativistic effects (*96, 97*). For the Hg, Ba, Ca and La bases, small exponent Gaussians ($< 0.05$) were dropped to remove linear dependencies and to ensure numerical stability. GDF was also used for the two-electron integrals. We used the density fitting auxiliary basis def2-SVP-RI (*98, 99*), which is specially optimized for correlated calculations with the def2-SVP basis ($n_{\text{aux}} \sim 5 n_{\text{AO}}$).

For the plane wave basis calculations, a projector augmented wave (PAW) (*93, 100*) representation was used to treat the core electrons and we used a plane wave kinetic energy cutoff of 500 eV.

We sampled the Brillouin zone with a $\Gamma$-centered **k** mesh: $6 \times 6 \times 2$ for the $\sqrt{2} \times \sqrt{2}$ cell of the single layer compounds CuO$_2{}^{2-}$, CCO and Hg-1201; $6 \times 6 \times 1$ for the $\sqrt{2} \times \sqrt{2}$ cell of the double layer compound Hg-1212; $4 \times 4 \times 2$ for the $2 \times 2$ supercell of the single layer compounds CuO$_2$, CCO and Hg-1201; $4 \times 4 \times 1$ for the $2 \times 2$ supercell of the double layer compound Hg-1212. All mean-field calculations were converged to an accuracy of better than $10^{-8}$ a.u. per unit cell.

We used the Perdew-Burke-Ernzerhof (PBE) functional (*101*) in the DFT+$U$ calculations and also used the PBE0 (*102*) hybrid functional. PBE+$U$ calculations were performed using Dudarev's approach with a $U$ value of 7.5 eV for the Cu 3d AOs. We also refer to additional DFT data using other functionals from the literature (see below).

### 1.5.4 DMET settings

All DMET routines, including the bath construction, integral transformation, solver interface, chemical potential and correlation potential fitting, are implemented in the LIBDMET package (*10, 103*). To remove core orbitals, which make the bath construction unstable and increases computational cost, we froze the lowest mean-field bands ($1s2s2p3s3p$ bands for Cu and Ca, $1s2s$ bands for O, $5s5p$ bands for Hg, Ba and La). We added the correlation potential $u$ to all Cu and O orbitals and fit the three-band orbital blocks of the density matrices, which avoids any instabilities in the DMET self-consistency. The convergence criterion on the DMET self-consistency was chosen such that the maximal change of an element in $u$ was less than $5 \times 10^{-5}$ a.u., which corresponds roughly to an energy accuracy of better than $1 \times 10^{-5}$ a.u.

### 1.5.5 Solver settings

We used the UCCSD and UCCSD(T) methods implemented in PYSCF as solvers. The CCSD energy and $\Lambda$ equations were converged to an energy of better than $10^{-6}$ a.u.

The DMRG impurity solver used the BLOCK2 program (*8, 68, 104–106*). We used the standard DMRG sweep settings and a genetic algorithm for orbital ordering. The tolerance of the DMRG sweep energy was set to $10^{-6}$ a.u., the largest bond dimension was chosen to be 5000 and extrapolation of the DMET energy was performed (see below).

The largest embedding problem we treated using the UCCSD solver was of size (364o, 168e), with multiple such size fragments solved in parallel in the multi-fragment embedding formalism. For the UCCSD(T) and DMRG solvers, the largest problems treated were of size (122o, 60e) and (60o, 60e) respectively.

# 2 Data

## 2.1 Benchmarks

### 2.1.1 Finite size effects

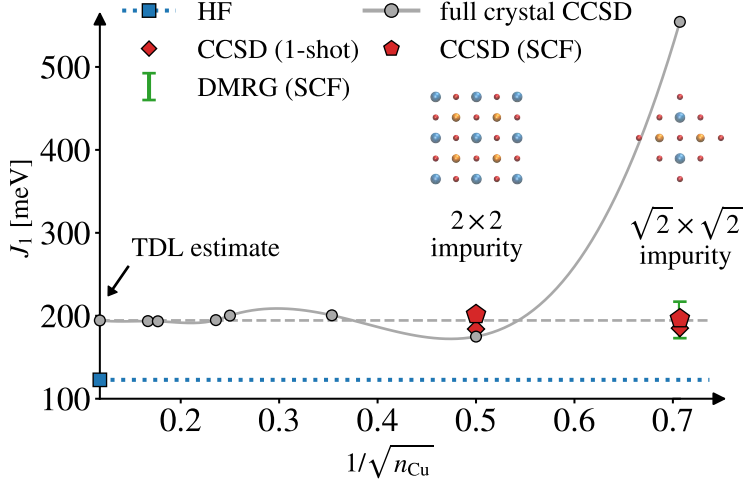

Figure S4: Benchmark with **k**-CCSD. See the caption of Fig. 1(f) in the main text. Here, we additionally show DMET 1-shot results and HF results.

We first benchmark the finite impurity size error of the DMET calculations. We extract the nearest-neighbor coupling $J_1$ in a large periodic lattice from a **k**-CCSD solver, and compare that to $J_1$ extracted from impurity calculations of different sizes, also with the CCSD solver. The difference between these results is the finite size error. The largest periodic lattice used for this purpose was a $6 \times 6 \times 1$ lattice of the AFM cell of CCO (72 primitive unit cells). In Fig. S4, we show the convergence of the $J_1$ values from **k**-CCSD calculations and different **k** meshes (cluster sizes) (note that the mean-field finite size error is always corrected by the result from the largest mean-field calculation, so the data is showing the convergence with respect to correlation effects only).

For the impurity, we use two cluster shapes ($\sqrt{2} \times \sqrt{2}$ and $2 \times 2$ cells of CCO, see Appendix 3.1). Even in the very small $\sqrt{2} \times \sqrt{2}$ impurity (the smallest magnetic supercell), the embedding calculation gives a very accurate $J_1$. Importantly, the DMET calculations show significantly less finite size error compared to CCSD on periodic clusters of the same size, showing the effectiveness of the embedding. Another feature we observe is that the 1-shot DMET calculation (1$^{st}$ iteration) gives similar results to the self-consistent one. This indicates that the initial mean-field, which breaks $S^2$ symmetry, is already close to the final one corrected by the correlation potential. This is, however, not true in the larger basis set calculations, where correlations produce larger corrections and self-consistency is important.

### 2.1.2 Multi-fragment scheme

We next test the accuracy of the multi-fragment scheme. From Table S2, we see that the error in $J_1$ from the multi-fragment treatment is less than $\sim 10$ meV and indeed energies in all the schemes are very close to the TDL **k**-CCSD value. The multi-fragment scheme also does not affect the local magnetic moments. Since the multi-fragment scheme does not introduce significant errors in the minimal basis but greatly reduces the computation cost, we use it in all following calculations.

### 2.1.3 Basis set completeness

We first check basis set convergence for the mean-field (single-particle) methods. Cross-checks between def2-SVP and a plane-wave basis are summarized in Appendix 3.3. The data in Tables S15-S18 clearly show that the relative energies in both the HF and DFT calculations are well converged. The error is 5 meV or less in HF, and 10 meV or less in PBE0. The deviation in PBE+$U$ is larger, primarily due to the different choices of local projector, but not the basis set completeness. The basis convergence in HF is also shown in Table S3, where the difference among def2-SVP (the main basis used in this work), def2-TZVP and plane wave is less than 1 meV.

Table S2: Local magnetic moment and nearest exchange coupling parameter of CCO with a minimal basis set.

| Method | $m_{\text{AFM}}$ [$\mu_B$] | $m_{\text{FM}}$ [$\mu_B$] | $J_1$ [meV] |
| --- | --- | --- | --- |
| multi-frag ($\sqrt{2} \times \sqrt{2}$ cell, 1-shot) | 0.62 | 0.76 | 178 |
| multi-frag ($\sqrt{2} \times \sqrt{2}$ cell, SCF) | 0.61 | 0.76 | 191 |
| full-cell ($\sqrt{2} \times \sqrt{2}$ cell, 1-shot) | 0.63 | 0.77 | 185 |
| full-cell ($\sqrt{2} \times \sqrt{2}$ cell, SCF) | 0.62 | 0.77 | 197 |
| multi-frag ($2 \times 2$ cell, 1-shot) | 0.63 | 0.77 | 183 |
| multi-frag ($2 \times 2$ cell, SCF) | 0.63 | 0.77 | 211 |
| full-cell ($2 \times 2$ cell, 1-shot) | 0.63 | 0.78 | 184 |
| full-cell ($2 \times 2$ cell, SCF) | 0.63 | 0.78 | 202 |
| UCCSD (extrap.) | | | 194 |

Table S3: Basis set size convergence. Both mean-field (HF) and correlated (DMET with $\sqrt{2} \times \sqrt{2}$ impurity cell size) calculations of CCO are shown.

| Basis set | method | $m_{\text{AFM}}$ [$\mu_B$] | $m_{\text{FM}}$ [$\mu_B$] | $J_1$ [meV] |
| --- | --- | --- | --- | --- |
| def2-SVP | | | | |
| | HF | 0.81 | 0.87 | 38.0 |
| | DMET (CCSD) | 0.68 | 0.77 | 122 |
| def2-TZVP | | | | |
| | HF | 0.81 | 0.86 | 37.5 |
| | DMET (CCSD) | 0.68 | 0.77 | 117 |
| plane wave | | | | |
| | HF | | | 37.1 |

Converging the correlation parts of the energy is in principle more challenging, requiring bases with more valence, polarization, and diffuse functions. We assess the basis set completeness in the small impurity ($\sqrt{2} \times \sqrt{2}$ cell) DMET calculations in Table S3. Compared to a larger basis def2-TZVP, the magnetic moments from def2-SVP agree well, and the difference in the derived NN magnetic coupling $J_1$ is only 5 meV. This suggests that def2-SVP basis provides a satisfactory balance between accuracy and efficiency for this study, allowing for reasonably converged energy scales while enabling larger impurity sizes ($2 \times 2$ supercell) in the following realistic calculations.

### 2.1.4 Solver accuracy

We further check the accuracy of the solver (CCSD) against more accurate solvers [CCSD(T) and DMRG]. In Fig. 5(a), we extrapolate the DMET energy from the DMRG solver to zero discarded weight $\delta \to 0$ [infinite bond dimension ($M \to \infty$)] for CCO. Both the AFM and FM states energies exhibit good linearity with respect to the discarded weight (the AFM state energy is slightly more linear). Despite the small size of the energy difference, the extrapolated $J_1$ agrees very well with the CCSD solver. This illustrates the accuracy of CCSD for the AFM ordered state starting from the symmetry broken mean-field reference. Similarly, the magnetic moments are also very close.

CCSD(T) includes more dynamical correlation than CCSD and this becomes important in larger basis sets. From Table S4, one can see that for the minimal basis, CCSD(T) gives a very small correction of 4 meV in $J_1$. For the larger basis def2-SVP, as expected, it gives a slightly larger correction of 10 meV. The change in the magnetic moment is about 0.01 $\mu_B$. We also show HF reference magnetic moments. The CC results are significantly different from the Hartree-Fock reference, showing the magnitude of magnetic fluctuations. In summary, for the parent state, CCSD yields good accuracy in the magnetic properties and its error mainly comes from the neglect of some dynamical correlation (about 10 meV in $J$), rather than any breakdown of the CC approximation due to multi-reference effects.

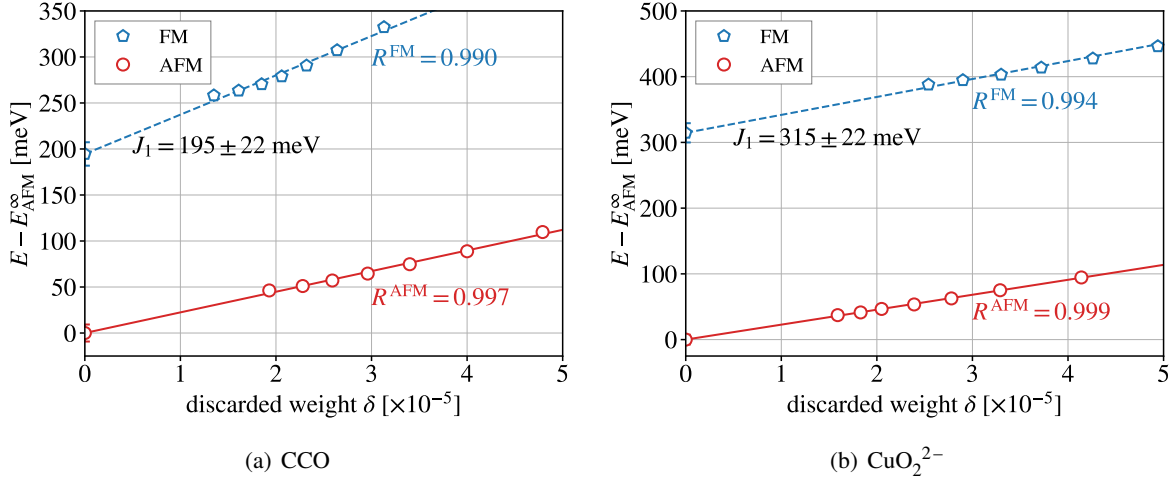

Figure S5: Linear extrapolation of the DMET energy using a DMRG solver for (a) CCO and (b) $CuO_2^{2-}$. The energies are generated by reverse sweeps of the DMRG calculation from bond dimension $M = 5000$. We use $M = 1500, 2000, 2500, 3000, 3500, 4000, 4500$ to perform the linear energy extrapolation with respect to the discarded weight. The energy zero is taken as the extrapolated AFM energy (per Cu). The error bar is estimated as 1/5 of the extrapolation distance, i.e., $[E(M = 4500) - E(M = \infty)]/5$.

Table S4: DMET solver benchmark. The results use the embedding Hamiltonian from the last DMET self-consistent iteration in CCO ($\sqrt{2} \times \sqrt{2}$ cell) with the minimal and def2-SVP basis sets.

| Basis set | method | $m_{AFM}$ [$\mu_B$] | $m_{FM}$ [$\mu_B$] | $J_1$ [meV] |
| --- | --- | --- | --- | --- |
| minimal basis | | | | |
| | CCSD solver | 0.61 | 0.76 | 191 |
| | CCSD(T) solver | 0.61 | 0.75 | 195 |
| | DMRG solver ($M = 1000$) | 0.61 | 0.76 | 231 |
| | DMRG solver ($M = 5000$) | 0.61 | 0.75 | 212 |
| | DMRG solver (extrap.) | 0.61 | 0.75 | $195 \pm 22$ |
| def2-SVP | | | | |
| | HF | 0.81 | 0.87 | 38 |
| | CCSD solver | 0.68 | 0.77 | 122 |
| | CCSD(T) solver | 0.67 | 0.76 | 132 |

We also benchmarked the artificial $CuO_2^{2-}$ material using the same strategy as above. The results are summarized in Fig. 5(b) and Table S5. The DMRG extrapolation shows a similar degree of linearity and energy uncertainty as in CCO, and the conclusions about the accuracy of CCSD in comparison to CCSD(T) and DMRG are unaltered.

### 2.1.5 Additional data for $La_2CuO_4$

As a more realistic benchmark example, we applied our methods to $La_2CuO_4$, which has been extensively studied both experimentally and theoretically. There are two commonly studied structural phases of $La_2CuO_4$, namely the high-temperature tetragonal (HTT) and low-temperature orthorhombic (LTO, stabilized below 520 K) phases. In the

Table S5: Same as caption of Table S4, but for $CuO_2^{2-}$.

| Basis set | method | $m_{AFM}$ [$\mu_B$] | $m_{FM}$ [$\mu_B$] | $J_1$ [meV] |
|---|---|---|---|---|
| minimal basis | | | | |
| | CCSD solver | 0.50 | 0.69 | 294 |
| | CCSD(T) solver | 0.50 | 0.69 | 297 |
| | DMRG solver ($M = 1000$) | 0.50 | 0.70 | 395 |
| | DMRG solver ($M = 5000$) | 0.49 | 0.69 | 349 |
| | DMRG solver (extrap.) | 0.49 | 0.69 | 315 ± 22 |

HTT phase, the $CuO_6$ octahedra are perfectly aligned along the $z$ axis and all Cu's are equivalent while in the LTO phase, the octahedra are distorted and the Cu-O-Cu angle is no longer 180° (see Fig. S6). We computed the exchange coupling parameters in the two phases in Table S6 and plot their spin wave dispersions in Fig. S6. When fitted to the NN Heisenberg model, the $J_1$ results of the two phases are similar and agree well with experimentally derived parameters, which reflects the fact that the local chemical environments of Cu are similar. However, the long-range parameters ($J_2$, $J_3$ and $J_c$) in the two phases are different and larger in the HTT phase. It is likely that the distortion among the $CuO_6$ octahedra is harmful for the long-range exchange process due to weaker overlap of orbitals. In general, the LTO phase spin-wave spectrum agrees better with the experimentally measured spectrum. Away from the $\Gamma$ point and the Brillouin zone boundary, the error compared to the experimental spectrum is larger; this can be traced to the smaller value of $J_1$ compared to experiment.

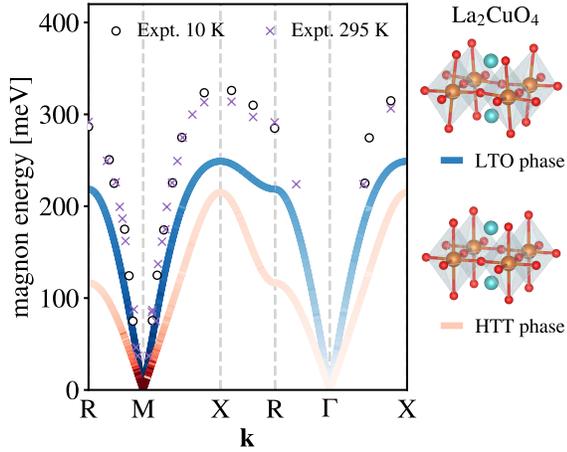

Figure S6: Spin wave spectrum of $La_2CuO_4$. Experimental data is taken from Ref. (85). Special points in Brillouin zone: R ($\frac{3}{4}, \frac{1}{4}$), M ($\frac{1}{2}, \frac{1}{2}$), X ($\frac{1}{2}, 0$), $\Gamma$ (0, 0).

Table S6: Magnetic exchange coupling parameters [in meV] of $La_2CuO_4$ fitted from *ab initio* DMET and experiments. [a] from Ref. (85), inelastic neutron scattering data fitted to the Heisenberg and 1-band Hubbard models.

| Method | Heisenberg | 1-band Hubbard | | |
|---|---|---|---|---|
| | $J_1$ | $J_1$ | $J_2, J_3$ | $J_c$ |
| DMET (HTT) | 102 | 102 | 7 | 135 |
| DMET (LTO) | 106 | 106 | 2 | 42 |
| Expt.[a] | 112 | 138 | 2 | 39 |

## 2.2 Multi-orbital electronic structure

### 2.2.1 Population analysis

Table S7: Population analysis. $n$: number of electron; $m$: magnetization.

| Element | orbital | Hg-1201 | | Hg-1212 | | CCO | | CuO$_2^{2-}$ | |
|---|---|---|---|---|---|---|---|---|---|
| | | $n$ | $m$ | $n$ | $m$ | $n$ | $m$ | $n$ | $m$ |
| **Cu** | | | | | | | | | |
| | $4s$ | 0.18 | -0.02 | 0.21 | -0.02 | 0.24 | -0.02 | 0.27 | -0.03 |
| | $4p_x$ | 0.24 | -0.02 | 0.25 | -0.02 | 0.26 | -0.02 | 0.28 | -0.03 |
| | $4p_y$ | 0.25 | -0.02 | 0.26 | -0.02 | 0.27 | -0.02 | 0.28 | -0.03 |
| | $4p_z$ | 0.18 | 0.00 | 0.19 | 0.00 | 0.18 | 0.00 | 0.25 | 0.00 |
| | $3d_{xy}$ | 1.99 | 0.00 | 1.99 | 0.00 | 1.99 | 0.00 | 1.99 | 0.00 |
| | $3d_{yz}$ | 1.98 | 0.00 | 1.98 | 0.00 | 1.98 | 0.00 | 1.98 | 0.00 |
| | $3d_{zx}$ | 1.98 | 0.00 | 1.98 | 0.00 | 1.98 | 0.00 | 1.98 | 0.00 |
| | $3d_{z^2}$ | 1.95 | 0.00 | 1.93 | 0.01 | 1.91 | 0.00 | 1.90 | 0.00 |
| | $3d_{x^2-y^2}$ | 1.20 | 0.76 | 1.21 | 0.74 | 1.22 | 0.72 | 1.29 | 0.63 |
| | $5s$ | 0.00 | 0.00 | 0.00 | 0.00 | 0.00 | 0.00 | 0.00 | 0.00 |
| | $4d_{xy}$ | 0.01 | 0.00 | 0.01 | 0.00 | 0.01 | 0.00 | 0.01 | 0.00 |
| | $4d_{yz}$ | 0.01 | 0.00 | 0.01 | 0.00 | 0.01 | 0.00 | 0.01 | 0.00 |
| | $4d_{z^2}$ | 0.01 | 0.00 | 0.01 | 0.00 | 0.01 | 0.00 | 0.01 | 0.00 |
| | $4d_{xz}$ | 0.01 | 0.00 | 0.01 | 0.00 | 0.01 | 0.00 | 0.01 | 0.00 |
| | $4d_{x^2-y^2}$ | 0.01 | 0.00 | 0.01 | 0.00 | 0.01 | 0.00 | 0.01 | 0.00 |
| | $4f(\times 7)$ | 0.00 | 0.00 | 0.00 | 0.00 | 0.00 | 0.00 | 0.00 | 0.00 |
| | **total** | **10.01** | **0.71** | **10.06** | **0.69** | **10.10** | **0.67** | **10.30** | **0.55** |
| **O in-plane** | | | | | | | | | |
| | $2p_x$ | 1.60 | 0.00 | 1.58 | 0.00 | 1.55 | 0.00 | 1.50 | 0.00 |
| | $2p_y$ | 1.92 | 0.00 | 1.92 | 0.00 | 1.92 | 0.00 | 1.94 | 0.00 |
| | $2p_z$ | 1.90 | 0.00 | 1.87 | 0.00 | 1.84 | 0.00 | 1.85 | 0.00 |
| | $3s$ | 0.00 | 0.00 | 0.00 | 0.00 | 0.00 | 0.00 | 0.00 | 0.00 |
| | $3p_x$ | 0.00 | 0.00 | 0.00 | 0.00 | 0.00 | 0.00 | 0.00 | 0.00 |
| | $3p_y$ | 0.01 | 0.00 | 0.01 | 0.00 | 0.01 | 0.00 | 0.01 | 0.00 |
| | $3p_z$ | 0.01 | 0.00 | 0.01 | 0.00 | 0.01 | 0.00 | 0.01 | 0.00 |
| | $3d(\times 5)$ | 0.00 | 0.00 | 0.00 | 0.00 | 0.00 | 0.00 | 0.00 | 0.00 |
| | **total** | **5.46** | **0.00** | **5.40** | **0.00** | **5.35** | **0.00** | **5.33** | **0.00** |

Table S7: Population analysis (cont.).

| Element | orbital | Hg-1201 | | Hg-1212 | | CCO | | CuO$_2^{2-}$ | |
|---|---|---|---|---|---|---|---|---|---|
| | | $n$ | $m$ | $n$ | $m$ | $n$ | $m$ | $n$ | $m$ |
| **O apical** | | | | | | | | | |
| | $2p_x$ | 1.95 | 0.00 | 1.95 | 0.00 | | | | |
| | $2p_y$ | 1.95 | 0.00 | 1.95 | 0.00 | | | | |
| | $2p_z$ | 1.63 | 0.00 | 1.65 | 0.00 | | | | |
| | $3s$ | 0.00 | 0.00 | 0.00 | 0.00 | | | | |
| | $3p_x$ | 0.01 | 0.00 | 0.01 | 0.00 | | | | |
| | $3p_y$ | 0.01 | 0.00 | 0.01 | 0.00 | | | | |
| | $3p_z$ | 0.00 | 0.00 | 0.00 | 0.00 | | | | |
| | $3d(\times 5)$ | 0.00 | 0.00 | 0.00 | 0.00 | | | | |
| | **total** | **5.57** | **0.00** | **5.58** | **0.00** | | | | |
| **Ca** | | | | | | | | | |
| | $4s$ | | | 0.15 | 0.00 | 0.16 | 0.00 | | |
| | $3d(\times 5)$ | | | 0.00 | 0.00 | 0.00 | 0.00 | | |
| | **total** | | | **0.15** | **0.00** | **0.16** | **0.00** | | |
| **Ba** | | | | | | | | | |
| | $6s$ | 0.16 | 0.00 | 0.16 | 0.00 | | | | |
| | $5d(\times 5)$ | 0.00 | 0.00 | 0.00 | 0.00 | | | | |
| | $6p(\times 3)$ | 0.00 | 0.00 | 0.00 | 0.00 | | | | |
| | $7s$ | 0.00 | 0.00 | 0.00 | 0.00 | | | | |
| | **total** | **0.16** | **0.00** | **0.16** | **0.00** | | | | |
| **Hg** | | | | | | | | | |
| | $6s$ | 1.03 | 0.00 | 1.03 | 0.00 | | | | |
| | $5d_{xy}$ | 1.98 | 0.00 | 1.98 | 0.00 | | | | |
| | $5d_{yz}$ | 1.98 | 0.00 | 1.98 | 0.00 | | | | |
| | $5d_{zx}$ | 1.98 | 0.00 | 1.98 | 0.00 | | | | |
| | $5d_{z^2}$ | 1.55 | 0.00 | 1.52 | 0.00 | | | | |
| | $5d_{x^2-y^2}$ | 1.98 | 0.00 | 1.98 | 0.00 | | | | |
| | $6p_x$ | 0.01 | 0.00 | 0.01 | 0.00 | | | | |
| | $6p_y$ | 0.01 | 0.00 | 0.01 | 0.00 | | | | |
| | $6p_z$ | 0.00 | 0.00 | 0.00 | 0.00 | | | | |
| | $5f(\times 7)$ | 0.01 | 0.00 | 0.01 | 0.00 | | | | |
| | $6d(\times 5)$ | 0.00 | 0.00 | 0.00 | 0.00 | | | | |
| | $7s, 8s, 9s$ | 0.00 | 0.00 | 0.00 | 0.00 | | | | |
| | **total** | **10.59** | **0.00** | **10.57** | **0.00** | | | | |

### 2.2.2 Bonding analysis

Table S8: Bonding analysis of different compounds. Both the total bond order $b$ and orbital specific bond order are shown.

| Bond type | Hg-1201 | | Hg-1212 | | CCO | | $CuO_2^{2-}$ | |
|---|---|---|---|---|---|---|---|---|
| | length | order | length | order | length | order | length | order |
| Cu-O (in plane) | 1.936 | 0.320 | 1.932 | 0.352 | 1.928 | 0.377 | 1.928 | 0.430 |
| Cu-O (apical) | 2.767 | 0.060 | 2.824 | 0.055 | | | | |
| Cu-Cu (intra-layer) | 3.871 | 0.044 | 3.863 | 0.052 | 3.856 | 0.060 | 3.856 | 0.073 |
| Cu-Cu (inter-layer) | 9.502 | 0.000 | 3.119 | 0.025 | 3.180 | 0.023 | 3.180 | 0.047 |
| O-O (nearest) | 2.737 | 0.005 | 2.732 | 0.007 | 2.726 | 0.008 | 2.726 | 0.010 |
| O-O (next nearest) | 3.871 | 0.002 | 3.863 | 0.002 | 3.856 | 0.001 | 3.856 | 0.001 |
| Ca-Cu | | | 3.145 | 0.004 | 3.156 | 0.004 | | |
| Ca-O | | | 2.488 | 0.030 | 2.499 | 0.033 | | |
| Ba-Cu | 3.340 | 0.002 | 3.351 | 0.003 | | | | |
| Ba-O | 2.722 | 0.035 | 2.732 | 0.035 | | | | |
| Hg-O (apical) | 1.984 | 0.393 | 1.966 | 0.386 | | | | |
| Orbital specific bond order | | | | | | | | |
| Cu $3d_{x^2-y^2}$ - O $2p_x$ | | 0.073 | | 0.075 | | 0.077 | | 0.092 |
| Cu $4s$ - O $2p_x$ ($\sigma$) | | 0.043 | | 0.047 | | 0.054 | | 0.057 |
| Cu $4p_x$ - O $2p_x$ ($\sigma$) | | 0.157 | | 0.160 | | 0.162 | | 0.161 |
| Cu $4p_y$ - O $2p_y$ ($\pi$) | | 0.009 | | 0.012 | | 0.015 | | 0.024 |
| Cu $4p_z$ - O $2p_z$ ($\pi$) | | 0.038 | | 0.055 | | 0.068 | | 0.093 |
| **total** | | **0.320** | | **0.349** | | **0.376** | | **0.427** |
| Cu $3d_{z^2}$ - O $2p_z$ (apical) | | 0.000 | | 0.000 | | | | |
| Cu $4s$ - O $2p_z$ (apical) | | 0.009 | | 0.008 | | | | |
| Cu $4p_z$ - O $2p_z$ (apical) | | 0.051 | | 0.047 | | | | |
| **total** | | **0.060** | | **0.055** | | | | |

### 2.2.3 Comparison to DFT population

Table S9: DFT charge, magnetic moment, bond order of CCO compared to DMET.

| Method | $n_{CuO_2}$ | $m_{Cu}$ | $b_{Cu-O}$ |
|---|---|---|---|
| PBE | 15.58 | 0.00 | 0.442 |
| PBE0 | 15.54 | 0.54 | 0.476 |
| DMET | 15.45 | 0.67 | 0.377 |
| HF | 15.37 | 0.81 | 0.356 |

We compare the population from different DFT functionals to DMET and HF in Table S9. The semi-local PBE functional, as expected, completely fails in describing magnetism ($m_{Cu} = 0$). HF is in another limit, where the electrons are over-localized and the magnetic moment is overestimated. PBE0, due to the mixing of 25% HF exchange in the functional, is between the two limits and predicts reasonable charge and magnetic moment similar to DMET. As described in the main text, hybrid functionals like PBE0, gives qualitatively correct results for a single compound.

However, hybrid DFT or DFT+$U$ may fail in predicting systematical trends among different compounds, especially for the subtle influence of the buffer layer.

### 2.2.4 Real space density and ELF analysis

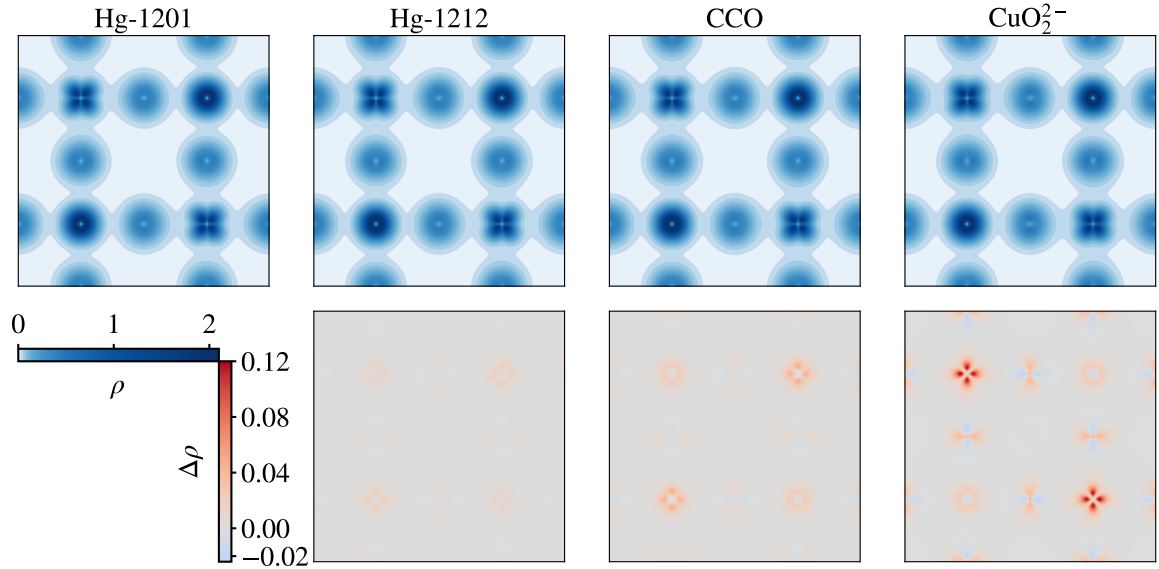

Figure S7: Electron density contour in the $CuO_2$ plane of different compounds (From left to right: Hg-1201, Hg-1212, CCO, $CuO_2^{2-}$). Only the valence electron $\rho^\alpha(\mathbf{r})$ of the AFM state is shown. The second row shows the density difference between compounds $X$ and the reference Hg-1201, i.e., $\Delta\rho = \rho(X) - \rho(\text{Hg-1201})$.

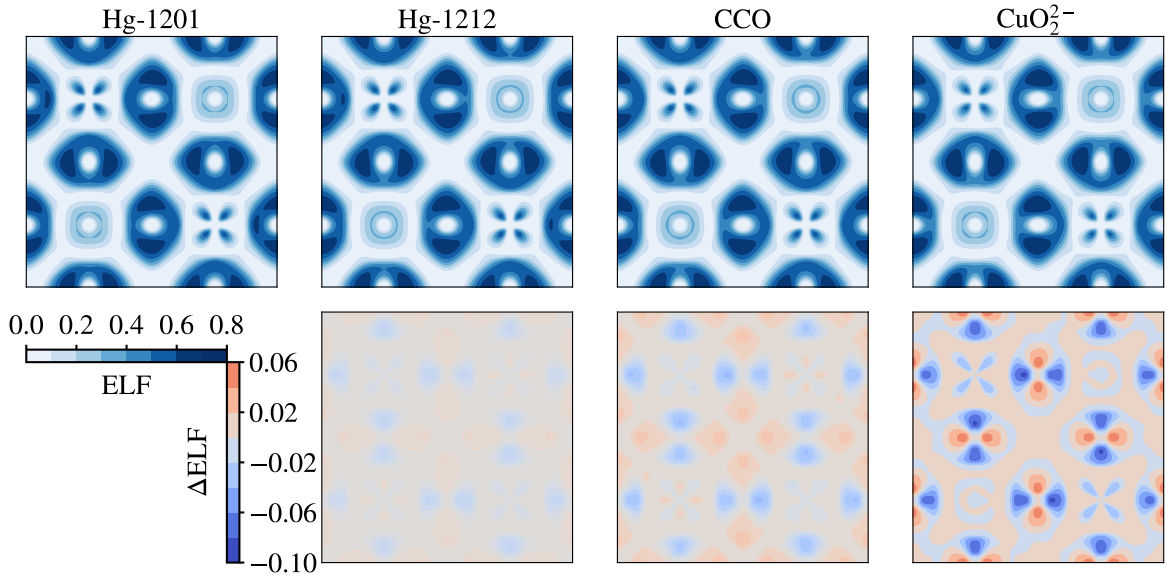

Figure S8: Electron localization function (ELF) contour in the $CuO_2$ plane of different compounds (From left to right: Hg-1201, Hg-1212, CCO, $CuO_2^{2-}$). Only valence electrons $ELF^\alpha(\mathbf{r})$ of the AFM state are shown. The second row shows the difference of ELF between compounds $X$ and the reference Hg-1201, i.e., $\Delta ELF = ELF(X) - ELF(\text{Hg-1201})$.

For a real space description of the charge and bonding in the $xy$ plane, we have analyzed the electron density $\rho(\mathbf{r})$ in Fig. S7 and the electron localization function (ELF) in Fig. S8. In general, the three compounds have very similar plots of the density and ELF. From the density plot (Fig. S7), the differential density shows that the Cu electron density increases from Hg-1201 to $CuO_2^{2-}$, consistent with the previous population analysis. In the ELF plots, we clearly see the lone electron pairs on the non-3-band oxygen $2p$'s. Between Cu and O, although there is a maximum in the ELF function, it does not show a very typical covalent bonding pattern. This suggests the bonding between Cu and O is more ionic than covalent. The differential ELF plots show that the electron is less likely to localize around the core region of oxygen moving from Hg-1201 to $CuO_2^{2-}$, i.e., the covalent bonding in the inter-atomic bonding region is increasing and the ionic character becomes weaker.

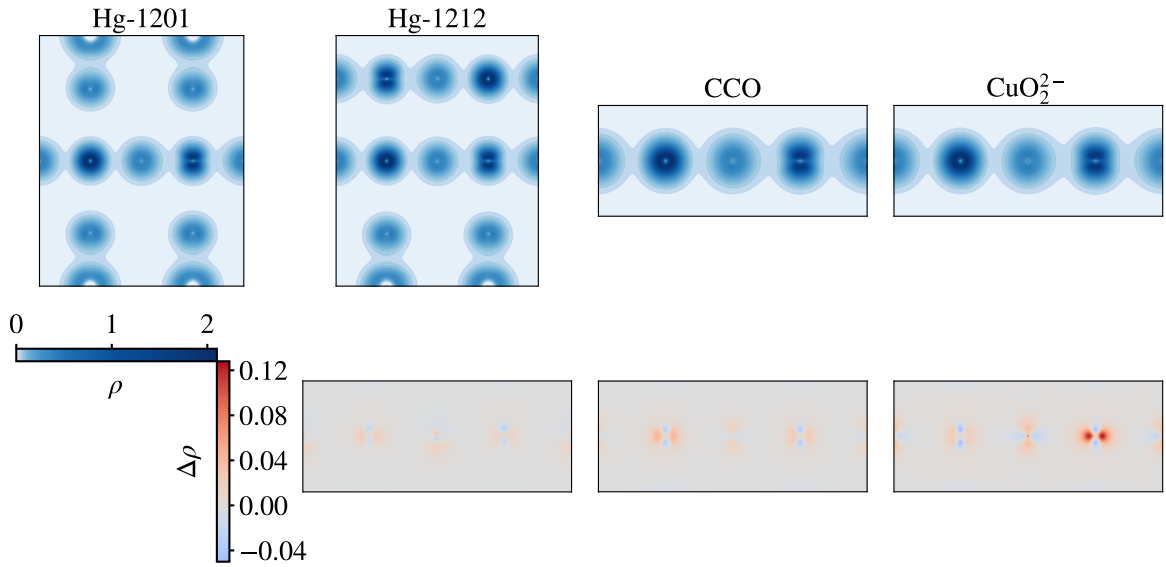

Figure S9: Same caption as Fig. S9, but on the $xz$ plane.

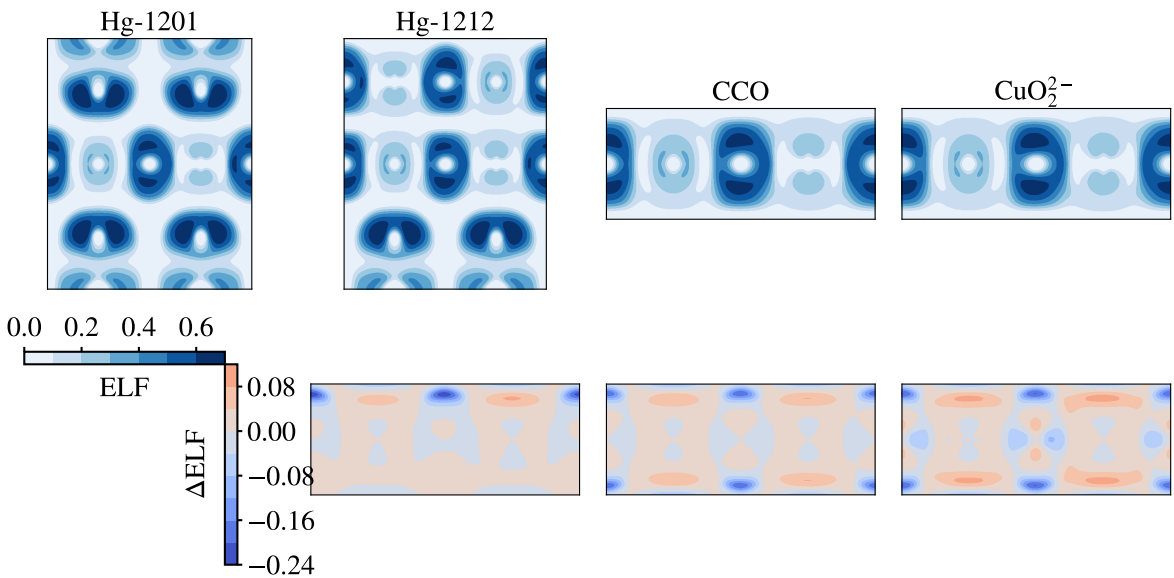

Figure S10: Same caption as Fig. S10, but in the $xz$ plane.

A similar analysis can be done for the $xz$ plane, see Fig. S9 and S10. In the differential density plot, we see that the electron density around Cu increases in the $xy$ plane, but decreases in the $xz$ plane. This can be interpreted as a change in bond order. When the system has apical oxygens, it has two effects: the first is to form bonds in the $z$ direction with Cu. Since the total valence of Cu (the ability to form covalent bonds) is finite, Cu then has less ability to form covalent bonds in the CuO$_2$ plane. The other effect is to make Cu's change more positive and the whole system becomes more ionic and the overall covalent bond order is then decreased. In the language of electronic bands, the inclusion of apical oxygen enlarges the orbital energy gap between the Cu $3d$ and O $2p$ bands (c.f. $\Delta_{pd} = \epsilon_p - \epsilon_d$ is an important parameter in the 3-band model) and makes the hybridization weaker. Also, the density / ELF plots show covalency between Hg and the apical O, which has been discussed in the bond order analysis.

### 2.2.5 Spin-traced natural orbitals around Fermi level

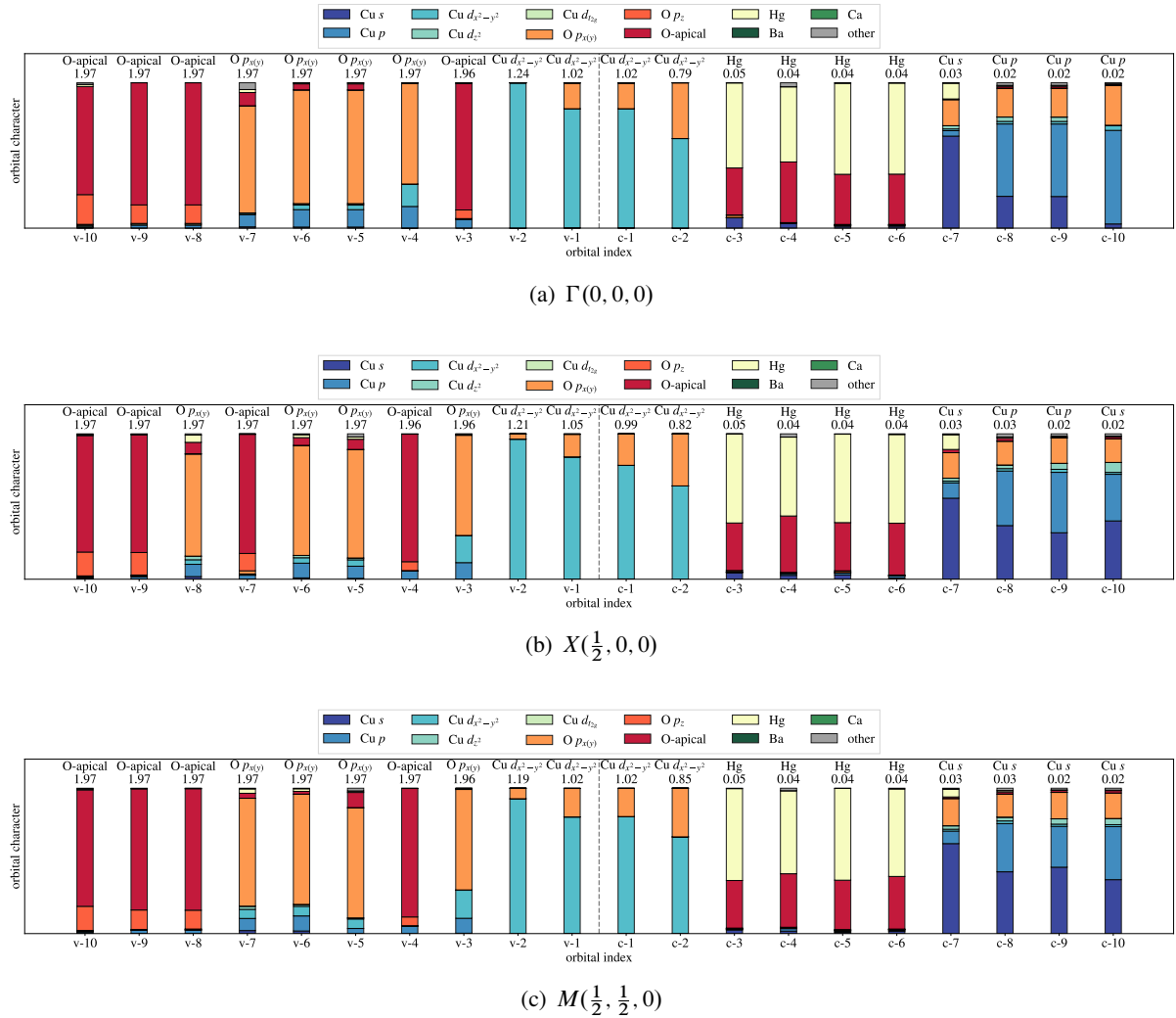

Figure S11: Spin-traced natural orbitals of Hg-1201 from DMET around the Fermi level (dash line) at different **k** points. The main orbital character and the occupancy are labelled.

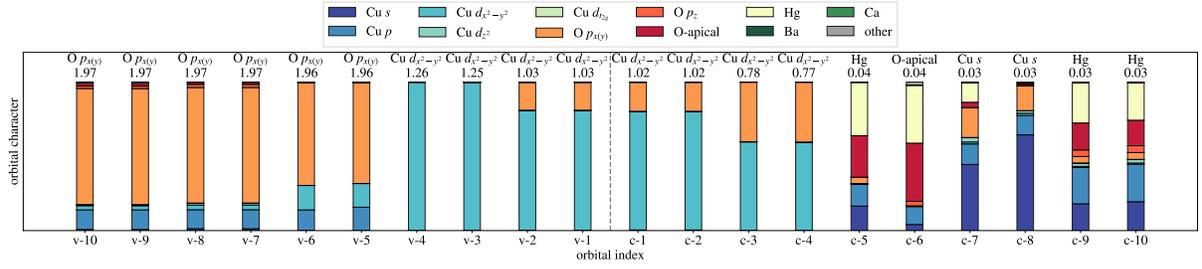

(a) $\Gamma(0, 0, 0)$

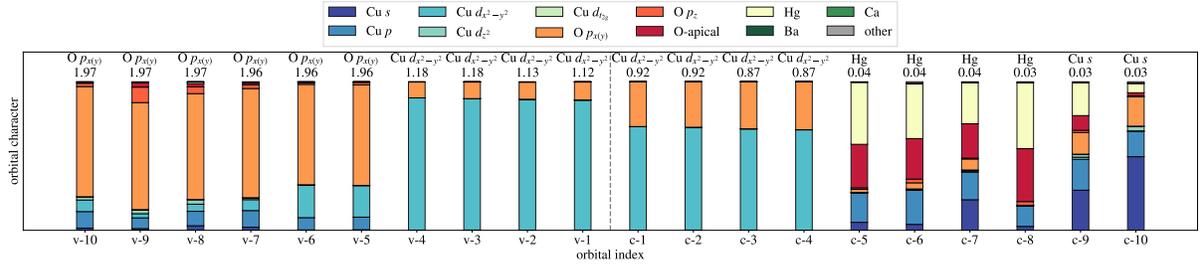

(b) $X(\frac{1}{2}, 0, 0)$

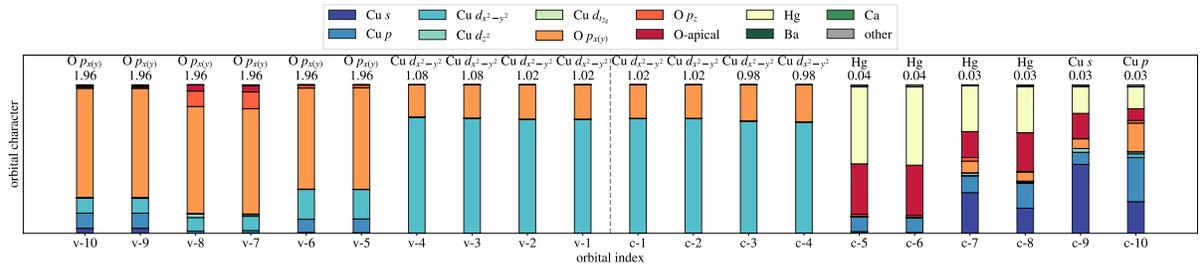

(c) $M(\frac{1}{2}, \frac{1}{2}, 0)$

Figure S12: Spin-traced natural orbitals of Hg-1212 from DMET around the Fermi level (dashed line) at different **k** points. The main orbital character and occupancy are labelled.

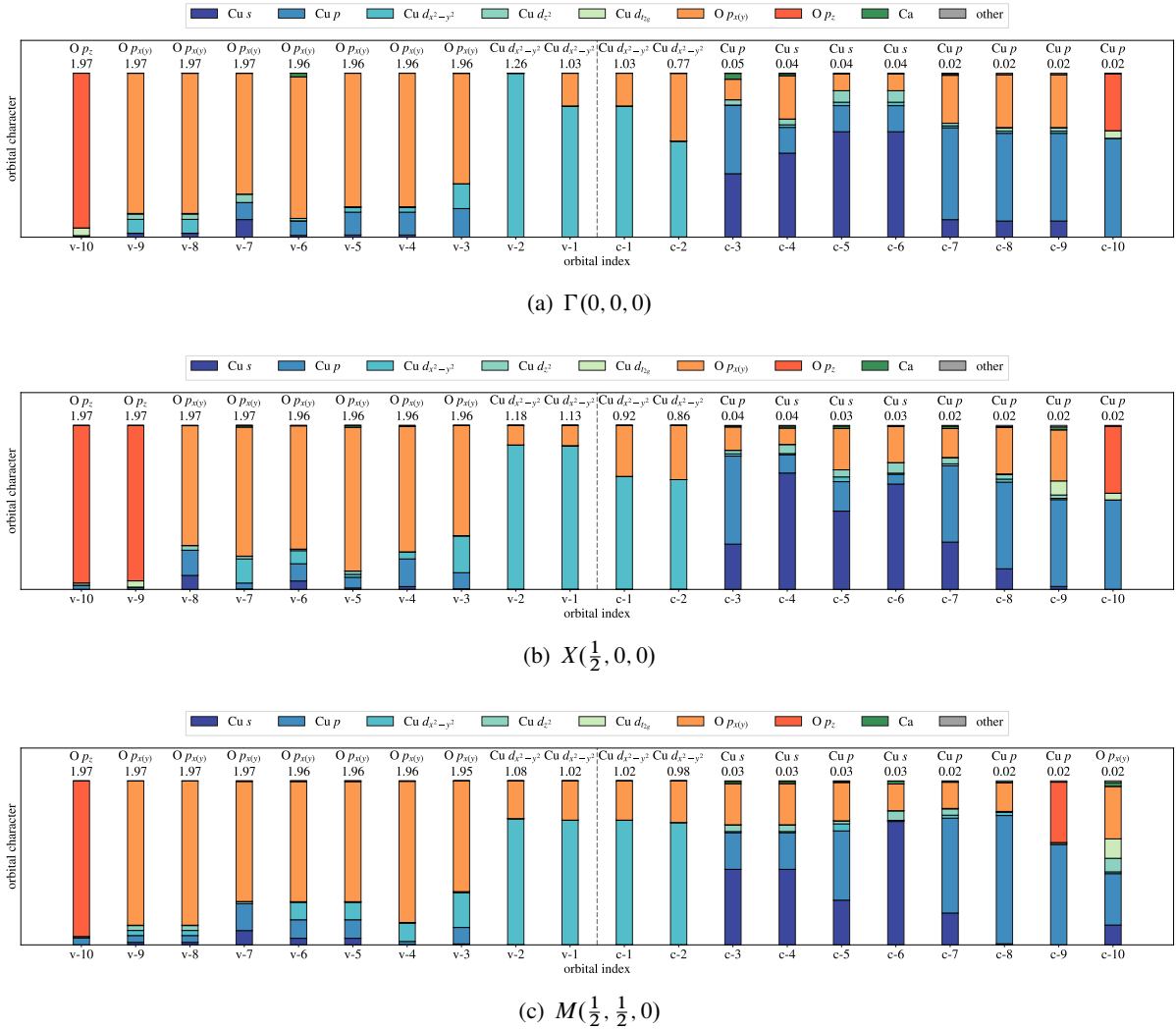

Figure S13: Spin-traced natural orbitals of CCO from DMET around the Fermi level (dashed line) at different **k** points. The main orbital character and occupancy are labelled.

Although the correlated band structure is not currently available in our DMET calculations, one can analyze the orbitals around the Fermi level through the natural orbitals, see Figs. S11, S12 and S13. Starting with the CCO natural orbitals, we can see that around the Fermi level, the valence bands mainly have O $p$ character and the conduction bands have a mixture of Cu $d$ and Cu $s$ characters, which is typical for a charge transfer insulator. There is some dispersion along the different **k** points, but it does not change either the orbital character or natural occupancy significantly. The natural occupancies are not very far from 1 and 0, which means the system is not far from a symmetry-breaking single reference system (and this is why UCCSD gives a very accurate description).

Compared to the CCO natural orbitals, Hg-1212 has some Hg-apical O bands among the low-lying virtual bands (even more appear for Hg-1201). This feature has also been observed in the band structures using hybrid functionals (*107*), i.e., as the number of Hg-O layers increases, the system CBM is dominated by the Hg-O bands and the band gap approaches zero. This plays a role in the layer effects on the superexchange constants, as discussed in the main text and further below.

## 2.2.6 Spin-resolved natural orbitals around Fermi level

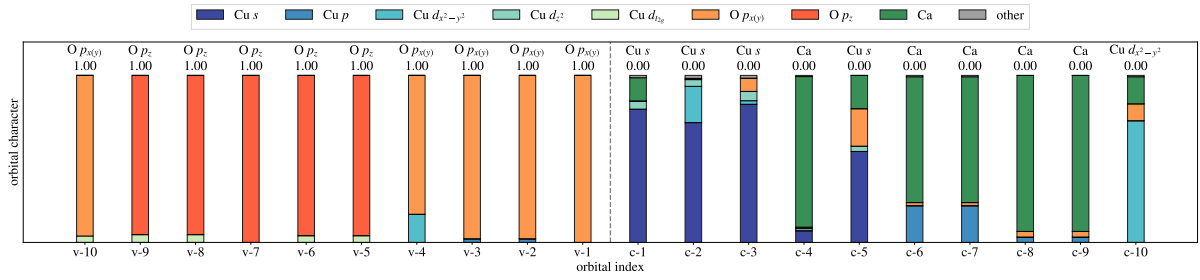

(a) $\Gamma(0, 0, 0)$

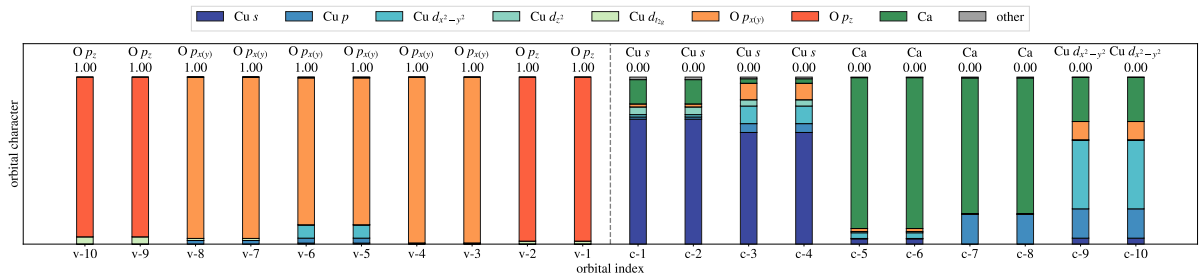

(b) $X(\frac{1}{2}, 0, 0)$

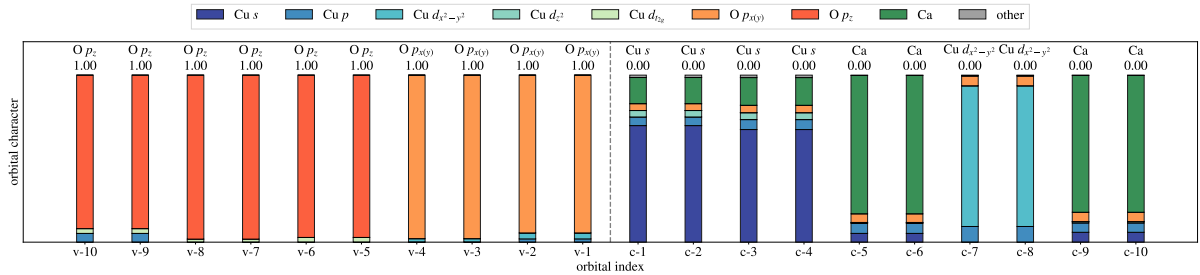

(c) $M(\frac{1}{2}, \frac{1}{2}, 0)$

Figure S14: Spin-resolved HF orbitals of CCO around Fermi level (dash line) at different **k** points. The main orbital character and the occupancy are labelled.

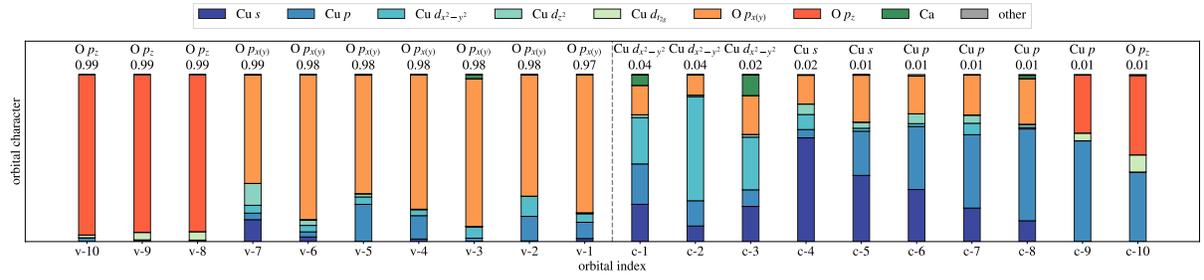

(a) $\Gamma(0, 0, 0)$

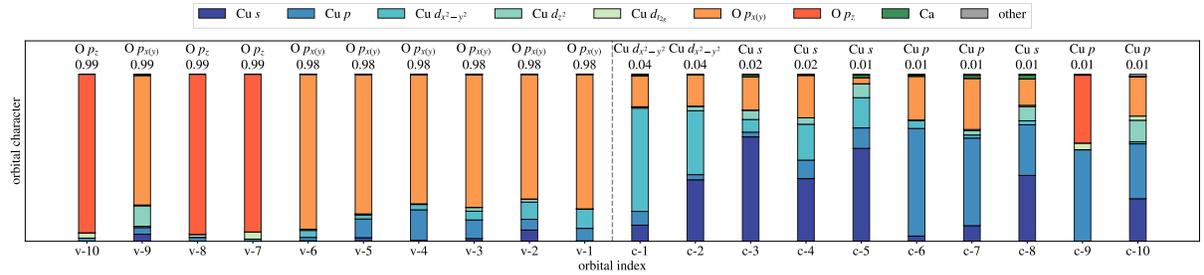

(b) $X(\frac{1}{2}, 0, 0)$

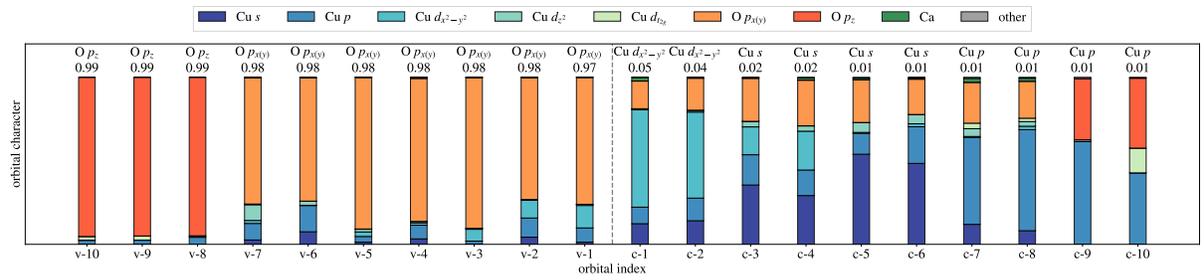

(c) $M(\frac{1}{2}, \frac{1}{2}, 0)$

Figure S15: Spin-resolved natural orbitals of CCO from DMET around the Fermi level (dash line) at different **k** points. The main orbital character and the occupancy are labelled.

## 2.3 Magnetic trends across the cuprates

We present additional data from the literature and mean-field methods on the magnetic exchange coupling parameters in this section. (Tables S10, S11, S12, S13 and Figs S16, S17, S18).

### 2.3.1 Hg-1201

Table S10: Magnetic exchange coupling parameters (in meV) of Hg-1201 calculated from different methods, fitted to different spin models.

| Method | Heisenberg | 1-band Hubbard | | | | $3J^{\text{eff}}$ Heisenberg | | |
|---|---|---|---|---|---|---|---|---|
| | $J_1$ | $J_1$ | $J_2, J_3$ | $J_c$ | $U/t$ | $J_1^{\text{eff}}$ | $J_2^{\text{eff}}$ | $J_3^{\text{eff}}$ |
| PBE+$U$ | 149.7 | 149.7 | 9.7 | 194.7 | 4.6 | 52.4 | -38.9 | 9.7 |
| PBE0 | 198.8 | 198.8 | 11.3 | 225.4 | 4.9 | 86.1 | -45.1 | 11.3 |
| HF | 33.7 | 33.7 | 1.0 | 20.2 | 6.3 | 23.6 | -4.0 | 1.0 |
| DMET | 103.8 | 103.8 | 2.4 | 48.4 | 7.0 | 79.6 | -9.7 | 2.4 |
| HSE06 | 204 | | | | | | | |
| B3LYP | 235 | | | | | | | |
| DDCI | 136.2 [a], 141 [b] | | | | | | | |
| CASPT2 | 123 [c] | | | | | | | |
| Expt. | 123 [d], 135 [e] | | | | | | | |

[a] From Ref. (26), difference dedicated configuration interaction (DDCI) (molecular model) calculation fitted to the Heisenberg model.

[b] From Ref. (108), difference dedicated configuration interaction (DDCI) (molecular model) calculation fitted to the Heisenberg model.

[c] From Ref. (108), complete active space second-order perturbation theory (CASPT2) (molecular model) calculation fitted to the Heisenberg model.

[d] From Ref. (49), resonant inelastic X-ray-scattering data fitted to the Heisenberg model.

[e] From Ref. (48), resonant inelastic X-ray-scattering data fitted to the Heisenberg model.

### 2.3.2 Hg-1212

Table S11: Magnetic exchange coupling parameters (in meV) of Hg-1212 calculated from different methods, fitted to different spin models.

| Method | Heisenberg | 1-band Hubbard | | | | $3J^{\text{eff}}$ Heisenberg | | |
|---|---|---|---|---|---|---|---|---|
| | $J_1$ | $J_1$ | $J_2, J_3$ | $J_c$ | $U/t$ | $J_1^{\text{eff}}$ | $J_2^{\text{eff}}$ | $J_3^{\text{eff}}$ |
| PBE+$U$ | 159.6 | 159.6 | 11.0 | 220.9 | 4.5 | 49.1 | -44.2 | 11.0 |
| PBE0 | 210.2 | 210.2 | 12.5 | 250.2 | 4.8 | 85.1 | -50.0 | 12.5 |
| HF | 36.3 | 36.3 | 1.2 | 23.4 | 6.1 | 24.6 | -4.7 | 1.2 |
| DMET | 122.1 | 122.1 | 5.3 | 106.7 | 5.4 | 68.7 | -21.3 | 5.3 |
| HSE06 | 215 | | | | | | | |
| B3LYP | 224 | | | | | | | |
| DDCI | 153.8 [a] | | | | | | | |
| Expt. | 176 [b] | | | | | | | |

[a] From Ref. (26), DDCI (molecular model) calculation fitted to the Heisenberg model.

[b] From Ref. (48), resonant inelastic X-ray-scattering data fitted to the Heisenberg model.

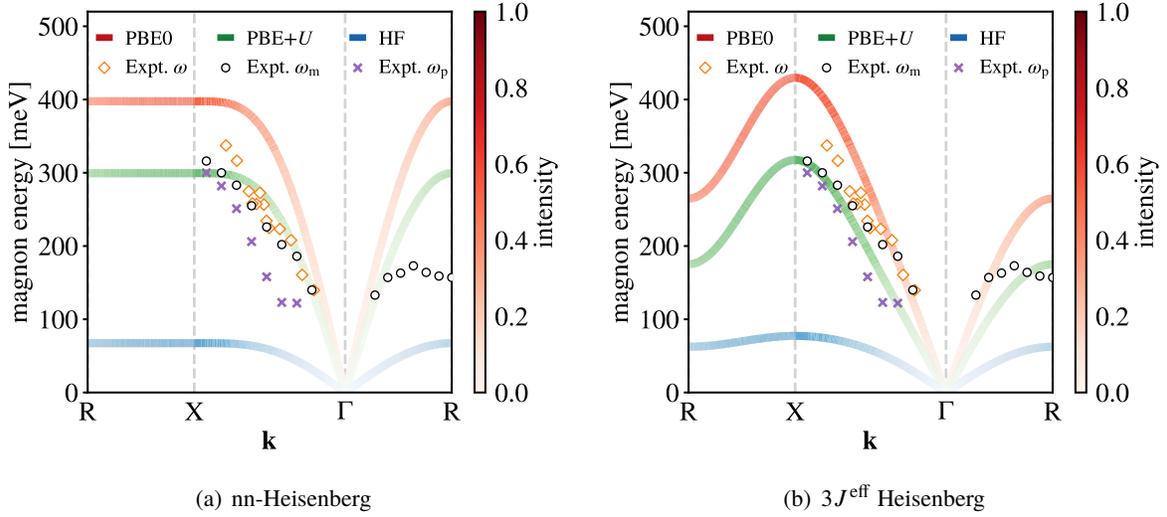

Figure S16: Spin wave dispersion of Hg-1201 from different effective single-particle approaches (PBE+$U$, PBE0 and HF) in the 2D magnetic Brillouin zone [$\Gamma$: (0, 0), X: ($\frac{1}{2}$, 0), R: ($\frac{1}{4}$, $\frac{1}{4}$), $k_z$ is fixed as 0]. Results from both (a) the nearest neighbor Heisenberg model and (b) the $3J^{\text{eff}}$ model parameters are shown. The experimental RIXS data (48, 49) is also shown. $\omega_{\text{m}}$ and $\omega_{\text{p}}$ denote the maximal-intensity magnon energy and the paramagnon energy respectively.

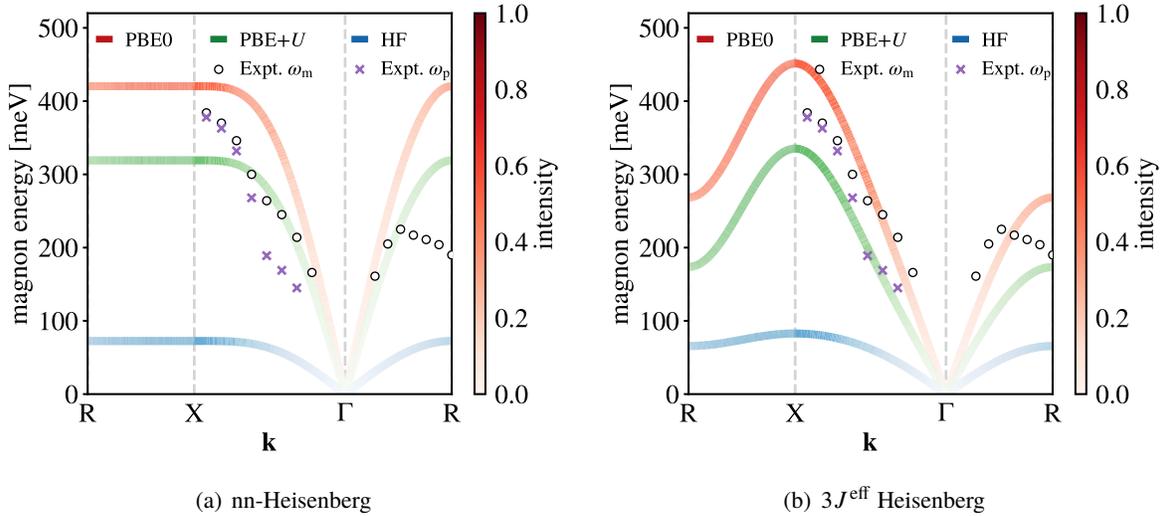

Figure S17: Spin wave dispersion of Hg-1212 from different effective single-particle approaches (PBE+$U$, PBE0 and HF) in the 2D magnetic Brillouin zone [$\Gamma$: (0, 0), X: ($\frac{1}{2}$, 0), R: ($\frac{1}{4}$, $\frac{1}{4}$), $k_z$ is fixed as 0]. Results from both (a) the nearest neighbor Heisenberg model and (b) the $3J^{\text{eff}}$ model parameters are shown. The experimental RIXS data (48) is also shown. $\omega_{\text{m}}$ and $\omega_{\text{p}}$ denote the maximal-intensity magnon energy and the paramagnon energy respectively.

### 2.3.3 CCO

Table S12: Magnetic exchange coupling parameters (in meV) of CCO calculated from different methods, fitted to different spin models.

| Method | Heisenberg | 1-band Hubbard | | | | $3J^{\text{eff}}$ Heisenberg | | | $J_\perp$ |
|---|---|---|---|---|---|---|---|---|---|
| | $J_1$ | $J_1$ | $J_2, J_3$ | $J_c$ | $U/t$ | $J_1^{\text{eff}}$ | $J_2^{\text{eff}}$ | $J_3^{\text{eff}}$ | |
| PBE+$U$ | 168.9 | 168.9 | 14.0 | 279.0 | 4.3 | 29.4 | -55.8 | 14.0 | 10.4 |
| PBE0 | 213.9 | 213.9 | 13.4 | 267.2 | 4.7 | 80.3 | -53.4 | 13.4 | 12.0 |
| HF | 38.0 | 38.0 | 1.4 | 27.0 | 5.8 | 24.5 | -5.4 | 1.4 | 2.7 |
| DMET | 155.4 | 155.4 | 9.7 | 194.4 | 4.7 | 58.2 | -38.9 | 9.7 | 8.9 |
| QMC | 142 [a] | | | | | | | | |
| Expt. | 142 [b], 158 [c] | 182 [d] | 10.3 [d] | 205.6 [d] | 4.9 [d] | 79.5 [e] | -41.1 [e] | 10.3 [e] | 6.5 [e] |

[a] From Ref. (*23*), fixed-node diffusion Monte Carlo (FN-DMC) (crystal) calculation fitted to the Heisenberg model.
[b] From Ref. (*51*), Raman spectrum data fitted to the Heisenberg model.
[c] From Ref. (*50*), RIXS data fitted to the Heisenberg model.
[d] From Ref. (*50*), RIXS data fitted to the 1-band Hubbard model.
[e] From Ref. (*50*), RIXS data fitted to the $3J^{\text{eff}}$ Heisenberg model.

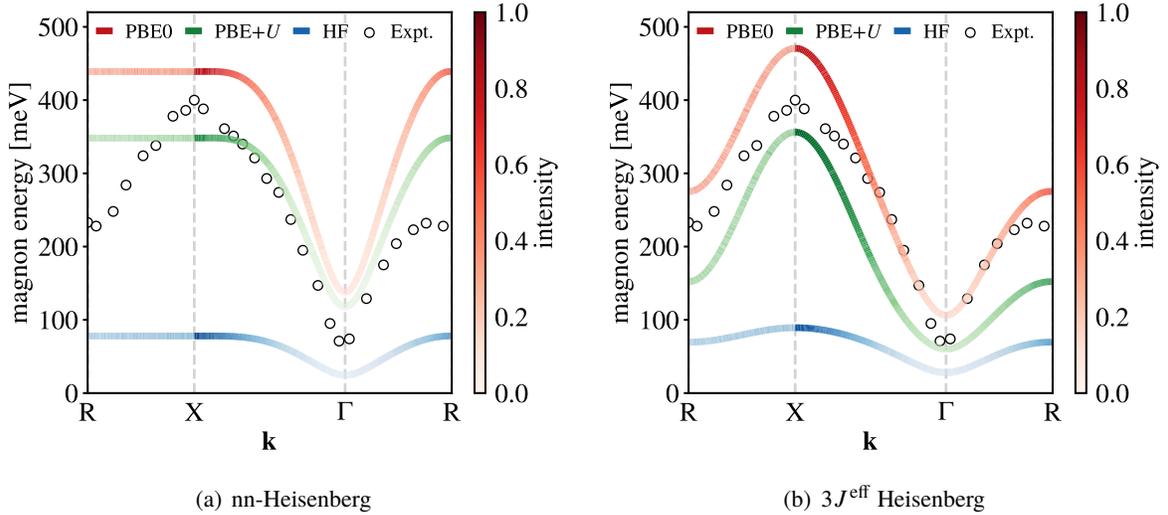

(a) nn-Heisenberg  (b) $3J^{\text{eff}}$ Heisenberg

Figure S18: Spin wave dispersion of CCO from different effective single-particle approaches (PBE+$U$, PBE0 and HF) in the 2D magnetic Brillouin zone [Γ: (0, 0), X: ($\frac{1}{2}$, 0), R: ($\frac{1}{4}$, $\frac{1}{4}$), $k_z$ is fixed as 0.46 to match the experimental condition]. Results from both (a) the nearest neighbor Heisenberg model and (b) the $3J^{\text{eff}}$ model parameters are shown. The experimental RIXS data (*50*) is also shown.

### 2.3.4 $CuO_2^{2-}$

Table S13: Magnetic exchange coupling parameters (in meV) of $CuO_2^{2-}$ calculated from different methods, fitted to different spin models. $U = 7.5$ eV is added to the Cu $3d$ orbitals in the PBE+$U$ method.

| Method | Heisenberg | 1-band Hubbard | | | | $3J^{\text{eff}}$ Heisenberg | | |
|---|---|---|---|---|---|---|---|---|
| | $J_1$ | $J_1$ | $J_2, J_3$ | $J_c$ | $U/t$ | $J_1^{\text{eff}}$ | $J_2^{\text{eff}}$ | $J_3^{\text{eff}}$ |
| PBE+$U$ | 165.5 | 165.5 | 26.0 | 520.4 | 3.5 | -94.7 | -104.1 | 26.0 |
| PBE0 | 269.8 | 269.8 | 11.9 | 238.4 | 5.4 | 150.6 | -47.7 | 11.9 |
| HF | 55.5 | 55.5 | 2.1 | 41.9 | 5.7 | 34.6 | -8.4 | 2.1 |
| DMET | 205.5 | 205.5 | 14.0 | 279.6 | 4.6 | 65.8 | -55.9 | 14.0 |
| QMC | 241 [a] | | | | | | | |

[a] From Ref. (*23*), fixed-node diffusion Monte Carlo (FN-DMC) (crystal) calculation fitted to the Heisenberg model.

### 2.3.5 Remarks on sign and error of exchange coupling parameters

We note that the sign of the 2$^{\text{nd}}$ neighbor exchange coupling parameter $J_2$ is related to the specific spin model that is being fit. When fitting to a $J_1$-$J_2$ Heisenberg model, $J_2$ is positive, indicating an antiferromagnetic coupling. On the other hand, in the effective $3J$ model, after absorbing the cyclic exchange $J_c$, $J_2$ becomes negative, i.e., ferromagnetic coupling.

When comparing to the experimental spectra, there are several possible sources of error: (i) Finite-size effects: As the largest cluster size we used in the embedding calculation is a $2 \times 2$ supercell, it is more likely that the long-range parameters $J_2$, $J_3$, $J_c$ have larger error. Errors in these parameters can typically be seen in the spin-wave dispersion away from the $\Gamma$ point (e.g. the X point in Fig. 4 in the main text, where the curvature is dominated by $J_c$). Also because of the current mean-field Hartree-Fock treatment of long-range Coulomb interactions outside of the computational cell (and given that Hartree-Fock underestimates the Heisenberg exchange parameters in this system) we expect that if the cluster size is further enlarged, the derived $J$'s will only increase, further improving agreement with with experiment. (ii) Model error: the current spin-wave spectrum is derived by fitting energies to spin models and then applying linear spin-wave theory. It is possible that the chosen spin models do not fully capture the high-energy part of the spin wave dispersion. Also, the mapping from the ab initio energies to the spin model assumes that the chosen electronic energies relate to Ising-like effective spins, but there is some ambiguity in this mapping. For instance, the current mapping assumes that $\langle S_z \rangle = \pm\frac{1}{2}$; however, due to the charge fluctuation of $3d_{x^2-y^2}$ orbitals, their $|\langle S_z \rangle| < \frac{1}{2}$ and fitting to such $S_z$ will make the $J$ values larger. (iii) Experimental uncertainty: the mercury-barium cuprate samples are typically doped and the spin-wave dispersion can differ from that of the undoped parent state. The doping dependence of the spin-wave dispersion in $La_2CuO_4$ has been studied and the dispersion along $\Gamma$ to R was found to be softened (to lower energy) compared to the parent state. On the other hand, the dispersion along $\Gamma$ to X was insensitive to the doping (*109*).

## 2.4 Untangling layer effects

### 2.4.1 Freezing out-of-plane orbitals

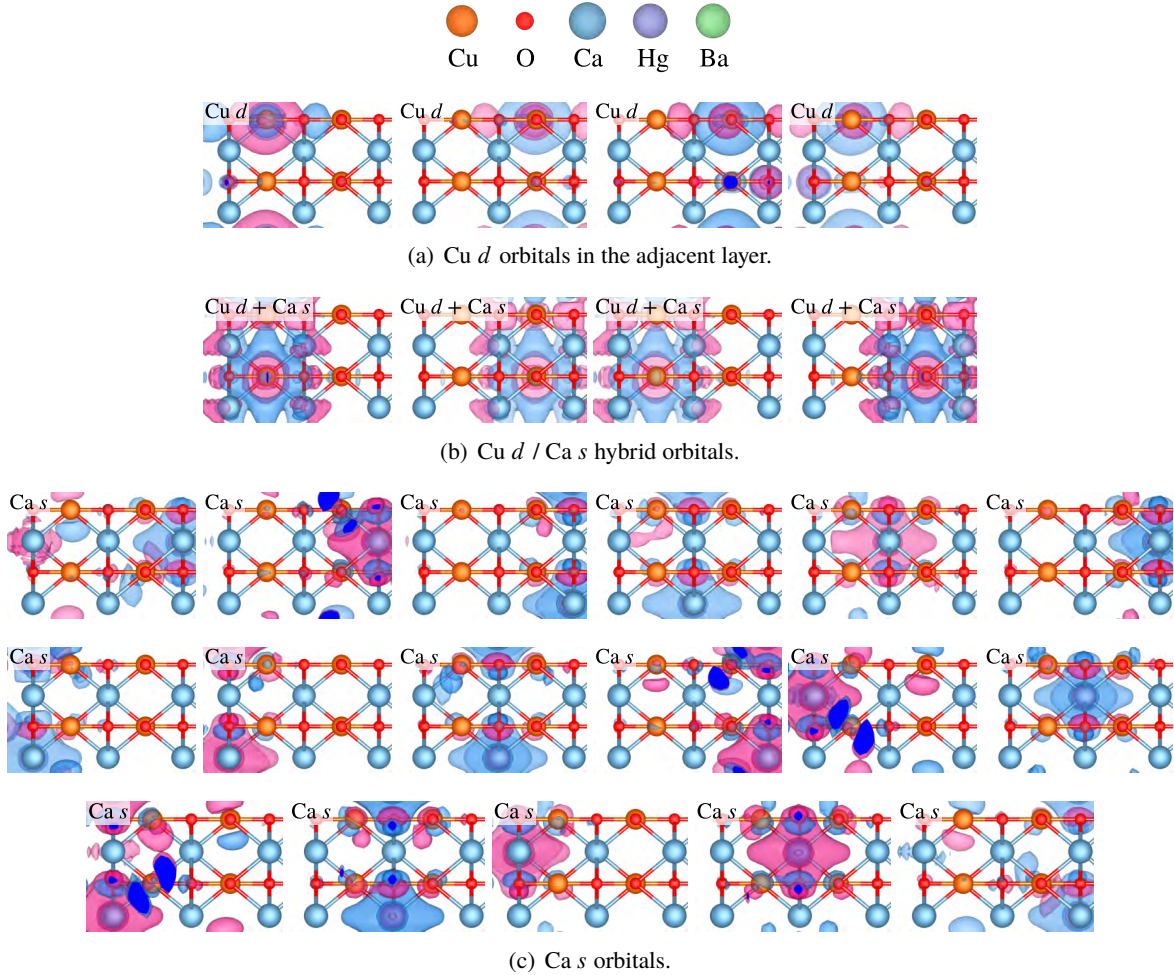

(a) Cu $d$ orbitals in the adjacent layer.

(b) Cu $d$ / Ca $s$ hybrid orbitals.

(c) Ca $s$ orbitals.

Figure S19: Out-of-plane localized embedding orbitals (isosurfaces) of CCO (view along $x$ axis). The main character of each orbital is labeled.

Table S14: Effect of freezing orbitals on the magnetic exchange coupling parameters of CCO and Hg-1201 (in meV).

| Compound | Heisenberg | 1-band Hubbard | | | | $3J^{\text{eff}}$ Heisenberg | | |
|---|---|---|---|---|---|---|---|---|
| | $J_1$ | $J_1$ | $J_2, J_3$ | $J_c$ | $U/t$ | $J_1^{\text{eff}}$ | $J_2^{\text{eff}}$ | $J_3^{\text{eff}}$ |
| CCO (1-shot) | 114.6 | 114.6 | 3.3 | 67.0 | 6.3 | 81.1 | -13.4 | 3.3 |
| CCO (frozen out-of-plane) | 105.7 | 105.7 | 1.0 | 19.7 | 10.6 | 95.8 | -3.9 | 1.0 |
| change | -8% | | | -71% | +68% | | | |
| Hg-1201 (1-shot) | 92.3 | 92.3 | 1.1 | 22.0 | 9.5 | 81.3 | -4.4 | 1.1 |
| Hg-1201 (frozen out-of-plane) | 90.3 | 90.3 | 0.8 | 16.9 | 10.6 | 81.9 | -3.4 | 0.8 |
| change | -2% | | | -23% | +12% | | | |

To understand the effects of the buffer layers (including the apical oxygens), we first localized the embedding orbitals using PM localization (see Fig. S19 for CCO and S20 for Hg-1201). We see that most of the out-of-plane orbitals are part of the virtual bands, except for some of the apical oxygen orbitals. The two compounds are similar w.r.t. Ca and Ba centered orbitals. The CCO bath also has some additional orbitals that come from the Cu $d$ of the adjacent layers, while Hg-1201 has additional apical O and Hg orbitals.

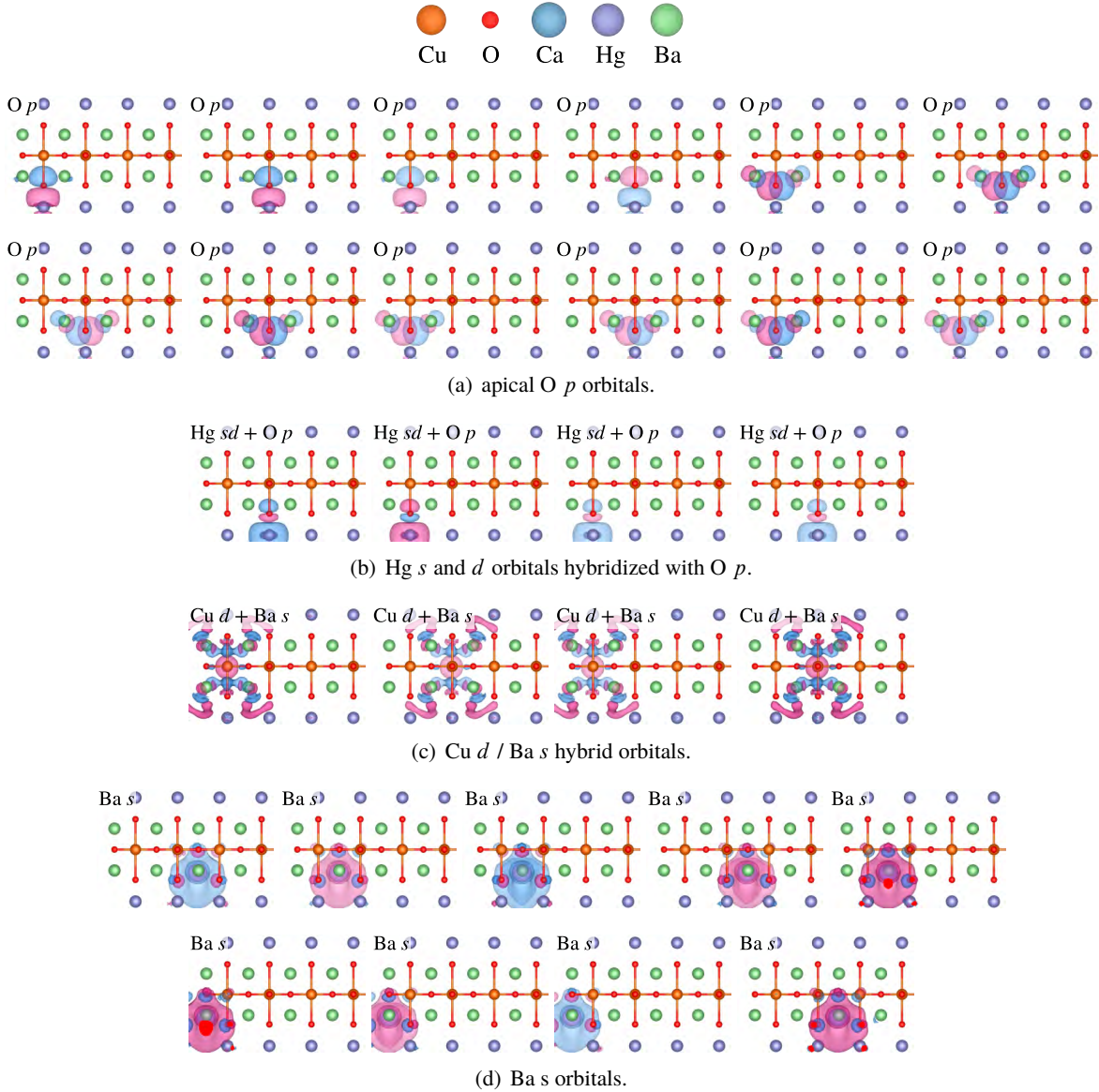

Figure S20: Out-of-plane localized embedding orbitals (isosurfaces) of Hg-1201 (view along $x$ axis). The main character of each orbital is labeled. Only the bottom buffer layer orbitals are shown.

We then freeze the out-of-plane orbitals in CCO and Hg-1201 and recompute the (1-shot) DMET impurity wavefunctions. Concretely, the buffer and its coupling to the $CuO_2$ layer are treated by HF in the impurity solver; then the freezing procedure forbids the excitation/de-excitation process (excitation = particle-hole excitations, including multiple particle-hole channels) from the $CuO_2$ layer to buffer layers. Thus the correlated impurity wavefunction, when formally expanded in singles, doubles, etc. excitations relative to the Hartree-Fock determinant, is missing those specific excited configurations. The resulting $J$ values are shown in Table S14. One sees that $J_1$ decreases by 8 %. $J_c$ is very strongly influenced by the freezing of the buffer layer orbitals and decreases by 71%. This suggest that $J_1$ is a relatively local property and is less influenced by freezing exchange pathways that involve the buffer layers; $J_c$ is a long-range property and its value is more strongly controlled by excitations to / from buffer layers. In Hg-1201, $J_1$ is almost unchanged and the magnitude of the change in $J_c$ is significantly smaller than in CCO. After freezing the buffer layer, the exchange couplings in CCO and Hg-1201 become very similar, highlighting the importance of explicit excitations involving the buffer in differentiating the physics.

### 2.4.2 Wavefunction excitation analysis

Additional insight into the type of excitations involving the buffer layer that affect the magnetic physics can be obtained by explicitly analyzing the CC wavefunction in the impurity solution. This is discussed below.

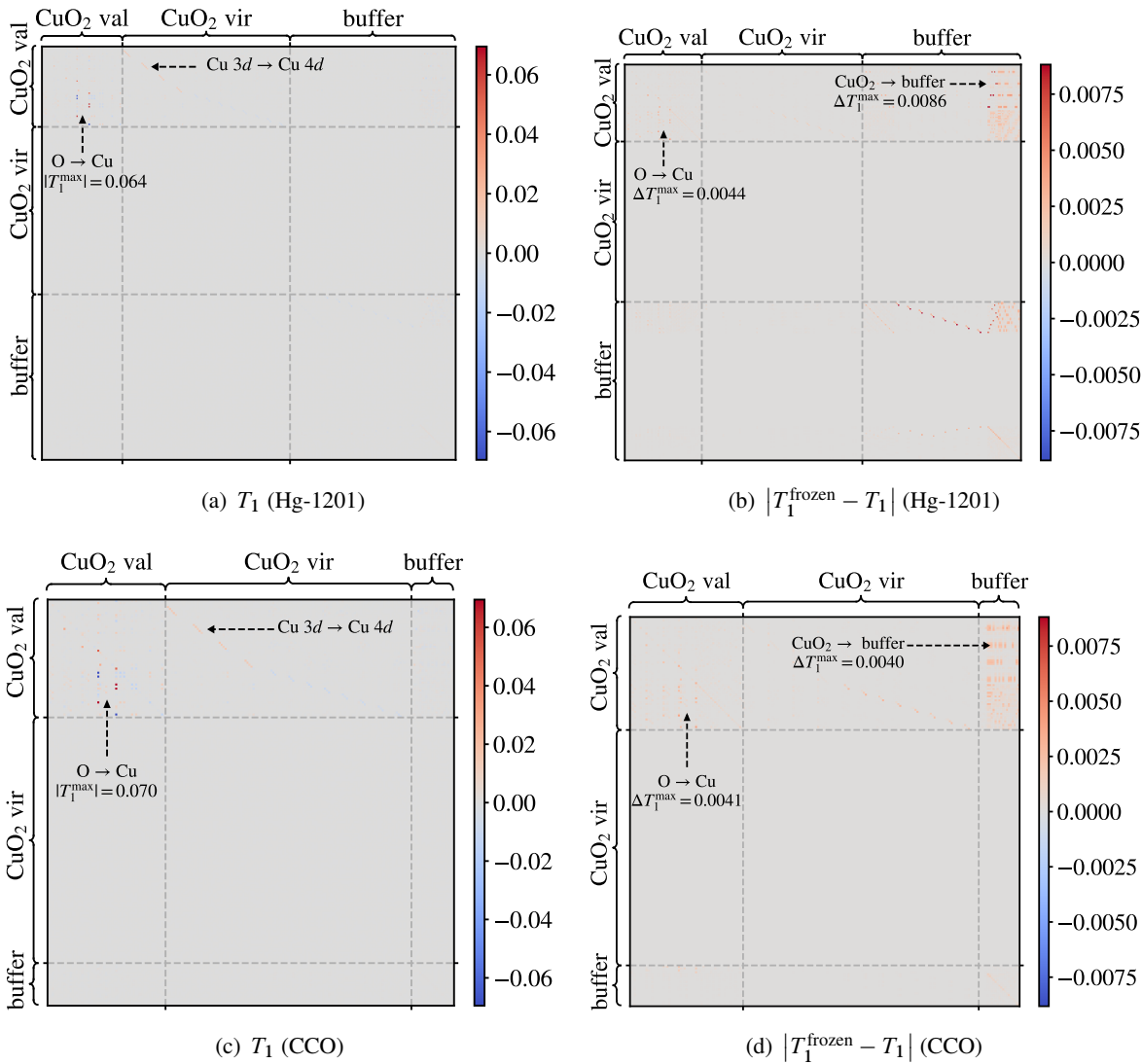

Figure S21: Visualization of the coupled-cluster $T_1$ amplitude in a local orbital basis. The row indices are transformed from the occupied molecular orbitals and the column indices are transformed from the virtual orbitals. (a) $T_1$ amplitude of Hg-1201, where the important orbital hybridization excitations are labeled: O - Cu and Cu $3d$ - $4d$. (b) Difference between $T_1$ before and after freezing out-of-plane (buffer) orbitals of Hg-1201, where the primary changes are labeled: O - Cu and Cu - buffer. (c), (d): Same as (a), (b), but for CCO.

We first transform the CCSD $T_1$ amplitudes to the local orbital basis

$$T_p^q = \sum_{ij} C_{pi} t_i^a C_{qa}^\dagger, \tag{S66}$$

and this quantity is plotted in Fig. S21. The $T_1$ amplitude carries information on the single-particle excitations that correct the Hartree-Fock solution. It thus describes the change in orbital character (rehybridization) driven by fluctuations. Visualizing this quantity (a matrix) in the local orbital basis allows us to describe the rehybridization in terms of the atomic orbitals. The difference in the $T_1$ amplitude on freezing the orbitals thus identifies the change in fluctuation driven hybridization, where the fluctuations involve the buffer degrees of freedom.

The basic feature seen in the $T_1$ amplitude is a strong excitation from O $2p$ to Cu $3d$ and hybridization between Cu $3d$ and $4d$. The former is slightly stronger in CCO (0.070) than in Hg-1201 (0.065), reflecting the stronger super-exchange in CCO, which has a larger $J_1$ than Hg-1201. The latter has also been observed in some recent CASPT2 calculations (*110*).

We next focus on the amplitude change after freezing the buffer orbitals [see Fig. S21 (b), (d)]. It is clear that freezing has three significant effects: (i) excitations within the buffer layer are prohibited (bottom right corner); (ii) excitations from the $CuO_2$ plane to the buffer layer are blocked (upper right corner); (iii) since screening effects from the buffer are also removed (which increases the charge-transfer gap), there is a change in the in-plane excitations, in particular the in-plane O → Cu excitation (upper left corner). (i) does not directly affect the in-plane magnetism, as it is limited to rehybridization of the buffer orbitals themselves. The change in (iii) is similar in the two compounds. However, the change in (ii) is almost two times larger in Hg-1201 than in CCO, due to much stronger $CuO_2$ → buffer (Hg and apical O) excitation.

The effect of these processes on the resulting super-exchange can be understood to come from several effects. First, the in-plane O→Cu excitations directly lead to increased superexchange (as this is part of the superexchange mechanism). Second, longer range exchange (including ring-like exchange $J_c$) can be connected to non-local hopping between oxygen orbitals facilitated by a diffuse orbital on Cu. The strong excitation into the buffer layer changes the character of this orbital, reducing its effective mixing with the oxygen orbitals in the virtual hopping process. [This is similar to the mechanism envisioned in Ref. (*17, 111*)]. Third, excitations from the ground-configuration to other non-super-exchange configurations overall renormalizes all the exchange constants. The first and second effects are the likely the largest ones and they act in opposite directions in Hg-1201, leading to the overall insensitivity of the couplings to freezing/unfreezing the buffer orbitals.

The $T_2$ amplitudes contain information on the connected two-particle excitations (see Fig. S22 for the largest 2000 elements in $T_2$ and $\Delta T_2$). Again, the amplitudes are transformed to the local orbital basis and partitioned into 4 types: pure in-plane excitations; coupled and double excitation terms involving indices in both the buffer and the $CuO_2$ plane (double refers to two holes/two-particles in the buffer); pure buffer-buffer excitations. We find that CCO has a larger change in the coupling component of the two-particle excitations than Hg-1201. Note that this change in the connected two-particle excitation reflects a fluctuation that cannot be renormalized into an effective static picture, and is thus not contained in earlier arguments that rely on such a picture, e.g., Ref. (*17, 111*). Although we have not carefully derived the influence on superexchange of this dynamical effect, it seems likely that the larger coupled layer-buffer excitation can couple into longer-range exchange processes in CCO, further increasing $J_c$ relative to Hg-1201.

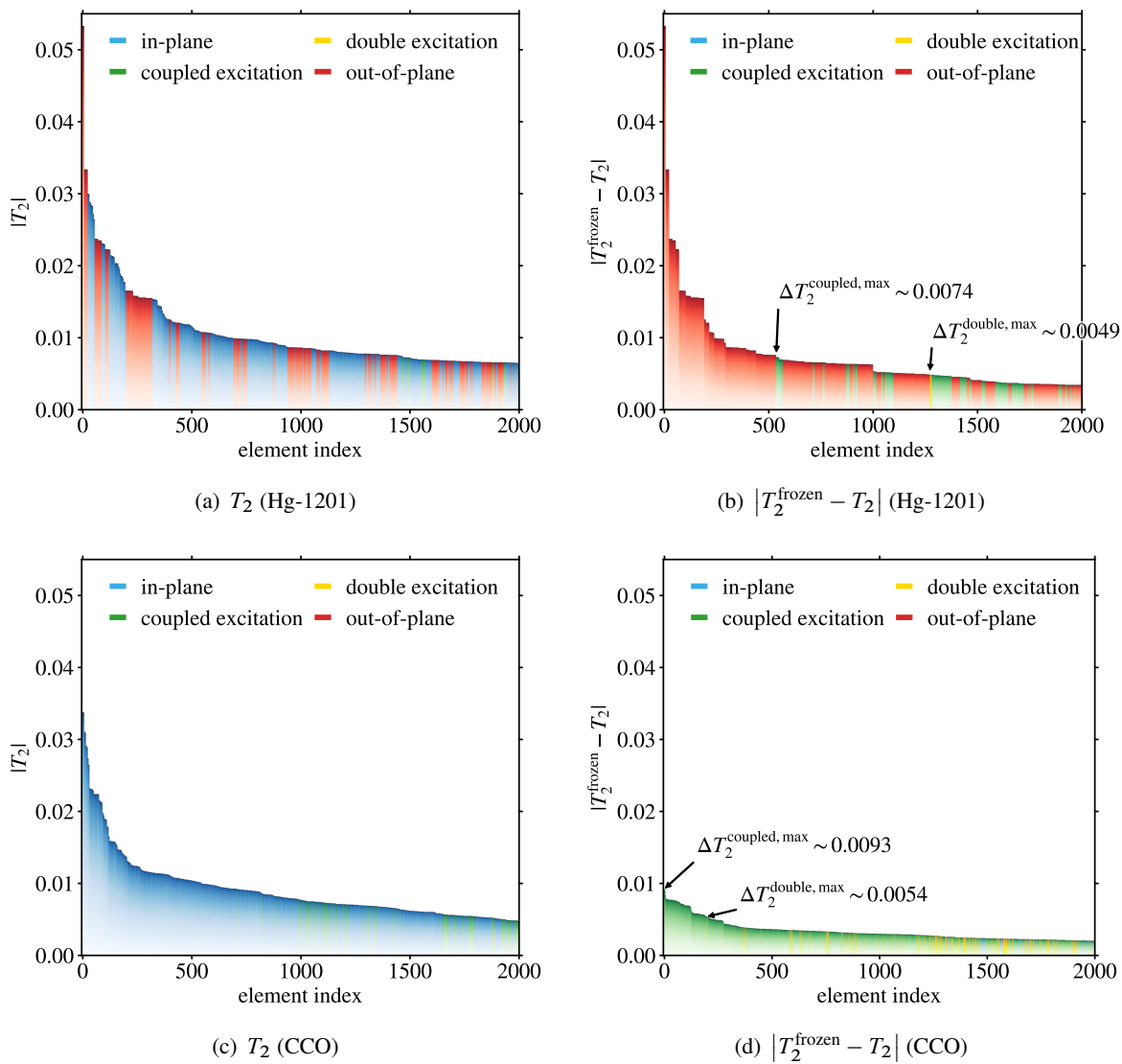

Figure S22: Visualization of the coupled-cluster $T_2$ amplitude in a local orbital basis. The largest 2000 elements of $(T_2)_{pqrs}$ are labelled as in-plane (all 4 indices belong to the in-plane orbitals), coupled/double (some indices are in-plane and some are out-of-plane) and out-of-plane (all 4 indices are for out-of-plane orbitals). (a) $T_2$ of Hg-1201. (b) Difference between $T_2$ before and after freezing out-of-plane (buffer) orbitals of Hg-1201. (c), (d): Same as (a), (b), but for CCO.

### 2.4.3 Effect of shifting apical oxygen

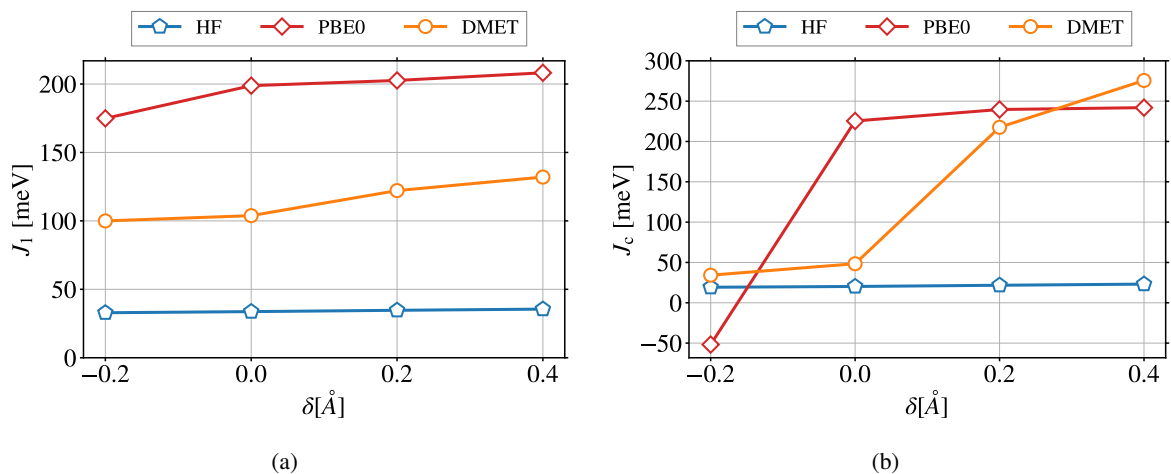

Figure S23: Effect of apical oxygen distance $\delta$ on (a) $J_1$ and (b) $J_c$ of Hg-1201.

We studied the influence of apical oxygen in more detail by shifting the apical oxygen closer or further away from the $CuO_2$ plane of Hg-1201. Its influence on $J_1$ and $J_c$ is shown in Fig. S23.

# 3 Appendix

## 3.1 Crystal structures, input scripts and data files

The crystal structures of $CuO_2^{2-}$, CCO, Hg-1201, Hg-1212; input scripts for different types of calculations; computational results data can be found at the GitHub repository https://github.com/zhcui/cuprate_parent_state_data .

## 3.2 $U$ dependence in DFT+$U$

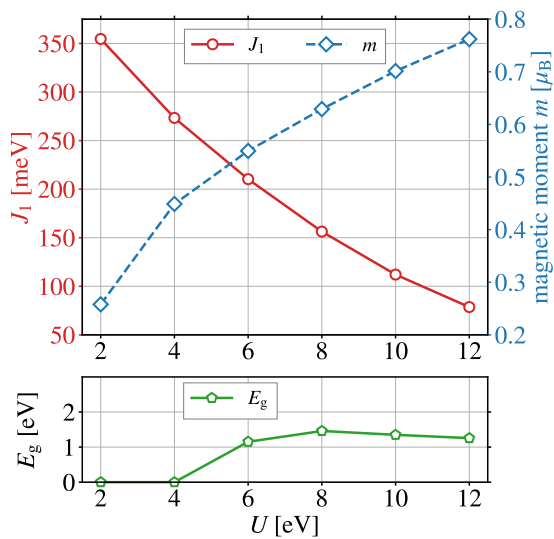

Figure S24: PBE+$U$ calculation in CCO: Upper panel: The nearest-neighbor exchange coupling $J_1$ (left axis) and the local magnetic moment $m$ (right axis) of the AFM phase as a function of $U$; Lower panel: band gap $E_g$ as a function of $U$.

Here, we discuss the DFT+$U$ approximation which is commonly used to study cuprates, and in particular the $U$ dependence of the magnetic coupling parameter $J$ and gap (see Fig. S24). As expected from second order perturbation theory,

$$J = \frac{4t^2}{U}, \tag{S67}$$

thus $J$ continuously decreases from 350 meV to 70 meV as $U$ increases from 2 eV to 12 eV. Meanwhile, the local magnetic moment monotonically increases when $U$ increases. If no $U$ is added to the system (not shown), i.e., when using the pure PBE functional, the ground state of the system is predicted to be non-magnetic, which is qualitatively wrong. A reasonable on-site $U$ then helps the system to stabilize the local moment. Another area where PBE+$U$ gives improvement is for the band gap, which is incorrectly predicted to be zero by the pure PBE functional. A small $U$ value ($< 4$ eV) is not sufficient to open a gap. For $U \geqslant 6$ eV, a gap opens, however it does not increase monotonically with $U$. The maximum of the gap occurs at $U = 8$ eV and it slightly decreases after that. Overall, quantities such as the magnetic coupling parameter $J$, local moment $m$ and band gap $E_g$ strongly depend on the size of $U$. Considering all three physical quantities, a reasonable $U$ appears to lie in the range of 6 - 9 eV.

Furthermore, as we mentioned in Sec. 1.4, the results have a dependence on the definition of the local projector. This effect can be clearly seen in Appendix 3.3, where the same calculation is performed using pseudopotential AOs as the projectors and the difference between the two software implementations is much larger than for the other methods. For instance, in Table S17, $J_1$ from PBE+$U$ using PYSCF is 168.9 meV which is quite different from the value from VASP, 199.6 meV.

## 3.3 Single-particle method cross-checks

Table S15: Comparison of the single-particle approach results for Hg-1201 from PySCF and VASP. Some long-range parameters are left blank since the SDW state in the plane wave basis converges to a paramagnetic state.

| Method | software | Heisenberg | 1-band Hubbard | | | | $3J^{\text{eff}}$ Heisenberg | | |
|---|---|---|---|---|---|---|---|---|---|
| | | $J_1$ | $J_1$ | $J_2, J_3$ | $J_c$ | $U/t$ | $J_1^{\text{eff}}$ | $J_2^{\text{eff}}$ | $J_3^{\text{eff}}$ |
| PBE+$U$ | PySCF | 149.7 | 149.7 | 9.7 | 194.7 | 4.6 | 52.4 | -38.9 | 9.7 |
| | VASP | 175.2 | 175.2 | 8.0 | 159.0 | 5.3 | 95.7 | -31.8 | 8.0 |
| PBE0 | PySCF | 198.8 | 198.8 | 11.3 | 225.4 | 4.9 | 86.1 | -45.1 | 11.3 |
| | VASP | 206.0 | 206.0 | 13.9 | 278.2 | 4.6 | 66.9 | -55.6 | 13.9 |
| HF | PySCF | 33.7 | 33.7 | 1.0 | 20.2 | 6.3 | 23.6 | -4.0 | 1.0 |
| | VASP | 34.2 | | | | | | | |

Table S16: Comparison of the single-particle approach results for Hg-1212 from PySCF and VASP.

| Method | software | Heisenberg | 1-band Hubbard | | | | $3J^{\text{eff}}$ Heisenberg | | |
|---|---|---|---|---|---|---|---|---|---|
| | | $J_1$ | $J_1$ | $J_2, J_3$ | $J_c$ | $U/t$ | $J_1^{\text{eff}}$ | $J_2^{\text{eff}}$ | $J_3^{\text{eff}}$ |
| PBE+$U$ | PySCF | 159.6 | 159.6 | 11.0 | 220.9 | 4.5 | 49.1 | -44.2 | 11.0 |
| | VASP | 181.5 | 181.5 | 7.5 | 149.5 | 5.5 | 106.8 | -29.9 | 7.5 |
| PBE0 | PySCF | 210.2 | 210.2 | 12.5 | 250.2 | 4.8 | 85.1 | -50.0 | 12.5 |
| | VASP | 214.1 | 214.1 | 14.6 | 291.5 | 4.5 | 68.3 | -58.3 | 14.6 |
| HF | PySCF | 36.3 | 36.3 | 1.2 | 23.4 | 6.1 | 24.6 | -4.7 | 1.2 |
| | VASP | 36.1 | | | | | | | |

Table S17: Comparison of the single-particle approach results for CCO from PySCF and VASP.

| Method | software | Heisenberg | 1-band Hubbard | | | | $3J^{\text{eff}}$ Heisenberg | | |
|---|---|---|---|---|---|---|---|---|---|
| | | $J_1$ | $J_1$ | $J_2, J_3$ | $J_c$ | $U/t$ | $J_1^{\text{eff}}$ | $J_2^{\text{eff}}$ | $J_3^{\text{eff}}$ |
| PBE+$U$ | PySCF | 168.9 | 168.9 | 14.0 | 279.0 | 4.3 | 29.4 | -55.8 | 14.0 |
| | VASP | 199.6 | 199.6 | 14.0 | 279.7 | 4.5 | 59.7 | -55.9 | 14.0 |
| PBE0 | PySCF | 213.9 | 213.9 | 13.4 | 267.2 | 4.7 | 80.3 | -53.4 | 13.4 |
| | VASP | 217.2 | 217.2 | 16.0 | 319.0 | 4.4 | 57.6 | -63.8 | 16.0 |
| HF | PySCF | 38.0 | 38.0 | 1.4 | 27.0 | 5.8 | 24.5 | -5.4 | 1.4 |
| | VASP | 37.1 | 37.1 | 1.7 | 33.2 | 5.3 | 20.5 | -6.6 | 1.7 |

Table S18: Comparison of the single-particle approach results for $CuO_2^{2-}$ from PySCF and VASP. Some long-range parameters are left blank since the SDW state in the plane wave basis converges to a paramagnetic state.

| Method | software | Heisenberg | 1-band Hubbard | | | | $3J^{\text{eff}}$ Heisenberg | | |
|---|---|---|---|---|---|---|---|---|---|
| | | $J_1$ | $J_1$ | $J_2, J_3$ | $J_c$ | $U/t$ | $J_1^{\text{eff}}$ | $J_2^{\text{eff}}$ | $J_3^{\text{eff}}$ |
| PBE+$U$ | PySCF | 165.5 | 165.5 | 26.0 | 520.4 | 3.5 | -94.7 | -104.1 | 26.0 |
| | VASP | 194.8 | | | | | | | |
| PBE0 | PySCF | 269.8 | 269.8 | 11.9 | 238.4 | 5.4 | 150.6 | -47.7 | 11.9 |
| | VASP | 280.9 | | | | | | | |
| HF | PySCF | 55.5 | 55.5 | 2.1 | 41.9 | 5.7 | 34.6 | -8.4 | 2.1 |
| | VASP | 50.9 | | | | | | | |